\documentclass[journal,onecolumn,draftclsnofoot]{IEEEtran}
\usepackage{setspace}
\usepackage{enumitem}
\usepackage{graphicx}
\usepackage{epstopdf}
\usepackage{csquotes}
\usepackage{subfig}
\usepackage{mathtools}
\usepackage{amssymb, amsmath,amsthm, amsfonts,mathrsfs}
\usepackage{accents}
\usepackage{ifthen}
\usepackage{tikz}
\usepackage{cite}
\usepackage{verbatim}
\usepackage{pifont}
\usepackage{bm}  
\usepackage{stackengine}
\usepackage{bbm}

\usepackage{kantlipsum}
\allowdisplaybreaks

\newcommand{\RN}[1]{%
  \textup{\uppercase\expandafter{\romannumeral#1}}%
}
\newcommand{\kl}[2]{D\left(#1||#2\right)}

\DeclareMathOperator*{\argmin}{arg\,min}
\DeclareSymbolFont{bbold}{U}{bbold}{m}{n}
\DeclareSymbolFontAlphabet{\mathbbold}{bbold}

\newtheorem{theorem}{Theorem}
\newtheorem{cor}{Corollary}
\newtheorem{lemma}{Lemma}
\newtheorem{prop}{Proposition}

\newtheorem{remark}{Remark}
\newtheorem{definition}{Definition}
\newtheorem{example}{Example}

\newcommand{\ind}{\mathbbm 1}
\newcommand{\Y}{\mathcal{Y}}
\newcommand{\Ucal}{\mathcal{U}}
\newcommand{\W}{\mathcal{W}}
\newcommand{\V}{\mathcal{V}}
\newcommand{\X}{\mathcal{X}}
\newcommand{\M}{\mathcal{M}}

\newcommand{\Z}{\mathcal{Z}}

\usepackage{xpatch}
\makeatletter
\xpatchcmd{\@thm}{\thm@headpunct{.}}{\thm@headpunct{}}{}{}
\makeatother

\newcommand\blfootnote[1]{%
	\begingroup
	\renewcommand\thefootnote{}\footnote{#1}%
	\addtocounter{footnote}{-1}%
	\endgroup
}

\def\h2{\bar \Pi}

\def\hm1{\hat h_{-1}}

\begin{document}

\title{Distributed Hypothesis Testing over a Noisy Channel: Error-exponents Trade-off} 
\author{Sreejith~Sreekumar and
Deniz G\"und\"uz}

\maketitle

\begin{abstract}
A two-terminal distributed binary hypothesis testing problem over a noisy channel is studied. The two terminals, called the observer and the decision maker, each has access to $n$ independent and identically  distributed  samples, denoted by $\mathbf{U}$ and $\mathbf{V}$, respectively. The observer communicates to the decision maker over a discrete memoryless channel, and the decision maker performs a binary hypothesis test on the joint probability distribution of $(\mathbf{U},\mathbf{V})$ based on $\mathbf{V}$ and the  noisy information received from the observer. The trade-off between the exponents of the type I and type II error probabilities  is investigated. Two inner bounds  are obtained, one using a separation-based scheme that involves type-based compression and unequal error-protection channel coding, and the other using a joint scheme that incorporates  type-based hybrid coding. The separation-based scheme is shown to recover the inner bound obtained by Han and Kobayashi for the special case of a rate-limited noiseless channel, and also the one obtained by the authors previously for a corner point of the trade-off.  Finally, we show via an example that the joint scheme achieves a strictly tighter bound than the  separation-based scheme for some points of the error-exponents trade-off.
\end{abstract}


\section{Introduction}

\blfootnote{This  work  is  supported  in  part  by  the  European  Research  Council  (ERC) through Starting Grant BEACON (agreement \#677854).
S. Sreekumar was with the Department of Electrical and Electronic Engineering, Imperial College London, at the time of this work. He is now with the School  of Electrical and Computer Engineering, Cornell University, Ithaca, NY 14850, USA (email: sreejithsreekumar@cornell.edu). D. G\"{u}nd\"{u}z is with the Department of Electrical and Electronic Engineering, Imperial College London, London SW72AZ, UK (e-mail: d.gunduz@imperial.ac.uk).}
Hypothesis testing (HT), which refers to the problem of choosing between one or more alternatives based on available data,  plays a central role in statistics and information theory. Distributed HT (DHT) problems arise in situations where the test data are scattered across multiple terminals, and need to be communicated to a central terminal, called the \textit{decision maker}, which performs the hypothesis test. The need to jointly optimize the communication scheme and the hypothesis test makes DHT problems much more challenging than their centralized counterparts. Indeed, while an efficient characterization of the optimal hypothesis test and its asymptotic performance is well known in the centralized setting, thanks to~\cite{NP-1933,Chernoff-1952,Hoeff-1965,Blahut-1974,Tuncel-2005}, the same problem  in even the simplest distributed setting remains open, except for some special cases (see~\cite{Ahlswede-Csiszar, Han, Shimokawa,Rahman-Wagner, SD_2020, Sadaf-Wigger-HTN}).
 
 In this work, we consider a DHT problem with two parties, an \textit{observer} and a decision maker, such that the former communicates to the latter over a noisy channel. The observer and the decision maker each has access to independent and identically distributed (i.i.d.) samples, denoted by $\mathbf{U}$ and $\mathbf{V}$, respectively. Based on the information received from the observer and its own observations $\mathbf{V}$,  the decision maker performs a binary hypothesis test  on the joint distribution of $(\mathbf{U},\mathbf{V})$. 
 Our goal is to characterize the trade-off between the best achievable  rate of decay (or exponent) of the type I and type II error probabilities with respect to the sample size. We will refer to this problem as \textit{DHT over a noisy channel}, and its special instance with the noisy channel replaced by a rate-limited noiseless channel as \textit{DHT over a noiseless channel}.
\subsection{Background}
Distributed statistical inference problems were first conceived in~\cite{Berger_1979} and the information-theoretic study of   
 DHT over a noiseless channel was first investigated in~\cite{Ahlswede-Csiszar},  
where the objective is to characterize the Stein's exponent $\kappa_{\mathrm{se}}(\epsilon)$, i.e., the optimal \textit{type II error-exponent} 
subject to the  type I error probability constrained to be at most $\epsilon \in (0,1)$. The authors therein established  a multi-letter characterization of this quantity including a strong converse, which shows that $\kappa_{\mathrm{se}}(\epsilon)$  is independent of $\epsilon$. Furthermore, a single-letter characterization of $\kappa_{\mathrm{se}}(\epsilon)$ is obtained for a special case of HT known as \textit{testing against independence} (TAI), in which the joint distribution factors as a product of the marginal distributions under the alternative hypothesis.  Improved lower bounds on $\kappa_{\mathrm{se}}(\epsilon)$ were subsequently obtained  in~\cite{Han,Shimokawa}, respectively,  and the strong converse was extended to zero-rate  settings~\cite{Shalaby-pap}.  While all the aforementioned works focus on $\kappa_{\mathrm{se}}(\epsilon)$, the trade-off between the exponents of both the type I and type II error probabilities  in the same setting was first explored in~\cite{HK-1989}.

In the recent years, there has been a renewed interest in distributed statistical inference problems motivated by emerging machine learning applications to be served at the wireless edge, particularly in the context of semantic communications in 5G/6G communication systems~\cite{Gunduz:ComMag:20, Gunduz:JSAC:23}. Several  extensions of the DHT over a noiseless channel  problem have been studied, such as generalizations to multi-terminal settings~\cite{Rahman-Wagner,Zhao-Lai,Wigger-Timo,Sadaf-Wigger-Li,Inaki-Zaidi-2019, Zaidi-2023}, DHT under security or privacy constraints~\cite{Maggie-Pablo,SD-10-security,SCG-2019,GSSV-2018}, DHT with lossy compression~\cite{Katz-estdetjourn}, interactive settings~\cite{Xiang-Kim-2,Xiang-Kim-1}, successive refinement models~\cite{Tian-Chen-2008}, and more.  Improved bounds have been obtained on the type I and type II error-exponents region~\cite{Haim-Kochman,WKJ-2017}, and on $\kappa_{\mathrm{se}}(\epsilon)$ for  testing correlation between  bivariate  standard normal distributions~\cite{Hadar-Liu-Polyansky-2019}. In the simpler zero-rate communication setting, there has been some progress in terms of second-order optimal  schemes~\cite{Watanabe-18}, geometric interpretation of type I and type II error-exponent region~\cite{Xu-2022}, and characterization of $\kappa_{\mathrm{se}}(\epsilon)$ for sequential HT~\cite{Sadaf-Tan-2021}. DHT over noisy communication channels with the goal of characterizing  $\kappa_{\mathrm{se}}(\epsilon)$ has been considered in~\cite{SD_2020,SD_ISIT2020,Sadaf-Wigger-HTN,Sadaf-2020-VLC}.

 \subsection{Contributions}
In this work, our objective is to explore the trade-off between the type I and type II error-exponents  for DHT over a noisy channel. This problem is a generalization of~\cite{HK-1989} from noiseless rate-limited channels to noisy channels, and also  of~\cite{SD_2020,Sadaf-Wigger-HTN} from a  type I error probability constraint to a  positive type I error-exponent constraint. 

Our main contributions  can be summarized as follows:
\begin{enumerate}[label=(\roman*)]
    \item 
    We obtain an inner bound (Theorem \ref{lbbinningts}) on the error-exponents trade-off by using a \textit{separate HT and channel coding} scheme (SHTCC) that is a combination of a type-based (type here  refers to the empirical probability distribution of a sequence, see~\cite{Csiszar-Korner}) quantize-bin strategy and unequal error-protection scheme of~\cite{Borade-09}.  This result is shown to recover the bounds established in~\cite{HK-1989,SD_2020}. 
  Furthermore, we evaluate Theorem  \ref{lbbinningts} for two important instances of DHT, namely TAI and its opposite, i.e.,  \textit{testing against dependence} (TAD) in which the joint distribution under the null hypothesis factors as a product of marginal distributions.  
    \item We also obtain a second inner bound (Theorem \ref{jhtccthm}) on the error-exponents trade-off by using a \textit{joint HT and channel coding scheme} (JHTCC) based on \textit{hybrid coding}~\cite{Lim-minero-kim-2015}.   
    Subsequently, we show via an example that the JHTCC scheme strictly outperforms the SHTCC scheme for some points on the error-exponent trade-off.
\end{enumerate}
While the above schemes are inspired from those in~\cite{SD_2020}, which have been proposed with the goal of maximizing the type II error-exponent, novel modifications in its design and analysis are required when considering  both of  the error-exponents. More specifically, the schemes presented here perform separate quantization-binning or hybrid coding on each individual source sequence type at the observer/encoder (as opposed to a typical ball in~\cite{SD_2020}) with the corresponding reverse operation implemented at the decision-maker/decoder. This necessitates a different analysis to compute the probabilities of the various error events contributing to the overall error-exponents. We finally mention that the DHT problem considered here was recently investigated  in~\cite{WKW_isit19}, where an  inner bound on the error-exponents trade-off (Theorem 2 in \cite{WKW_isit19}) is obtained using a combination of a type-based quantization scheme and unequal error protection scheme of~\cite{Csiszar-1982} with two special messages.  A qualitative  comparison between  Theorem \ref{jhtccthm} and~Theorem 2 in~\cite{WKW_isit19}  seems to suggest that  the JHTCC scheme here  uses a stronger decoding rule depending jointly on the  source-channel statistics. In comparison,  the metric used at the decoder for the scheme in~\cite{WKW_isit19}  factors as the sum of two metrics, one which depends only on the source statistics, and the other which depends only on the channel statistics. Importantly, this hints that the  inner bound achieved by JHTCC scheme is not subsumed by that in~\cite{WKW_isit19}. That said, a direct  computational comparison appears difficult, as evaluating the latter requires optimization over several parameters as mentioned in the last paragraph of~\cite{WKW_isit19}.

 \subsection{Organization}
The remainder of the paper is organized as follows. Section \ref{Prelims} formulates the operational problem along with the required definitions. The main results are presented in Section \ref{mainresults}. The proofs are furnished in Section \ref{proofs-results}. Finally, concluding remarks are given in  Section \ref{conclu}.

\section{Preliminaries} \label{Prelims}
\subsection{Notation}
We use the following notation. All logarithms are with respect to the natural base $e$. $\mathbb{N}$, $\mathbb{R}$, $\mathbb{R}_{\geq 0}$, and  $\bar{\mathbb{R}}$ denotes the set of natural, real, non-negative real and  extended real numbers, respectively. For  $a, b \in \mathbb{R}_{\geq 0}$, $[a:b]:=\{n \in \mathbb{N}:~a \leq n \leq b\}$ and $[b]:=[1:b]$. Calligraphic letters, e.g., $\mathcal{X}$, denote sets, while $\X^c$ and  $|\mathcal{X}|$ stands for its complement and cardinality, respectively. For $n \in \mathbb{N}$, $\mathcal{X}^n$ denotes the $n$-fold Cartesian product of $\X$, and $x^n=(x_1, \cdots,x_n)$ denotes an element of $\mathcal{X}^n$. Bold-face letters denote vectors or sequences, e.g., $\mathbf{x}$ for $x^n$;  its length $n$ will be clear from the context. 
For $ i, j \in \mathbb{N}$ such that $i \leq j$, $x_i^j:=(x_i,x_{i+1},\cdots,x_j)$, the subscript is omitted when $i=1$. $\ind_\mathcal{A}$ denotes the indicator of set $\mathcal{A}$. For a real sequence $\{a_n\}_{n \in \mathbb{N}} $, $a_n \xrightarrow{(n)} b$ stands for $\lim_{n \rightarrow \infty} a_n=b$, while $a_n \gtrsim b$ denotes  $\lim_{n \rightarrow \infty}a_n \geq b$. Similar notations apply for other inequalities.    $O(\cdot)$, $\Omega(\cdot)$ and $o(\cdot)$ denote  standard asymptotic notations.

Random variables and their realizations are denoted by uppercase and lowercase letters, respectively, e.g., $X$ and $x$. Similar conventions apply for random vectors and their realizations.  The set of all probability mass functions (PMFs) on a finite set $\mathcal{X}$ 
is denoted by $\mathcal{P}(\mathcal{X})$. The joint PMF of two discrete random variables $X$ and $Y$ 
is denoted by $P_{XY}$; the corresponding marginals are $P_X$ and $P_Y$. The conditional PMF of $X$ given $Y$ is represented by $P_{X|Y}$. Expressions such as $P_{XY}=P_XP_{Y|X}$ are to be understood as pointwise equality, i.e.,  $P_{XY}(x,y)=P_X(x)P_{Y|X}(y|x)$, for all $(x,y)\in\mathcal{X}\times\mathcal{Y}$. When the joint distribution of a triple $(X,Y,Z)$ factors as $P_{XYZ}=P_{XY}P_{Z|X}$, these variables form a Markov chain $X-Y-Z$. 
If the entries of $X^n$ are drawn in an i.i.d.  manner, i.e., if $P_{X^n}(\mathbf{x})=\prod_{i=1}^n P_{X}(x_i)$, $\forall~ \mathbf{x} \in \mathcal{X}^n$, then the PMF $P_{X^n}$ is denoted by $P^{\otimes n}_X$. 
Similarly, if $P_{Y^n|X^n}(\mathbf{y}|\mathbf{x})=\prod_{i=1}^nP_{Y|X}(y_i|x_i)$ for all $(\mathbf{x},\mathbf{y}) \in \X^n \times \Y^n$, then we write $P^{\otimes n}_{Y|X}$ for $P_{Y^n|X^n}$. The conditional product PMF given a fixed $\mathbf{x} \in \mathcal{X}^n$ is designated by $P^{\otimes n}_{Y|X}(\cdot|\mathbf{x})$.  The probability measure induced by a PMF $P$ is denoted by $\mathbb{P}_P$.  
The corresponding expectation is designated by $\mathbb{E}_{P}$.

The \textit{type} or empirical PMF of a sequence $\mathbf{x}\in \mathcal{X}^n$ is designated by $P_{\mathbf{x}}$, i.e., $P_{\mathbf{x}}(x):=\frac{1}{n}\sum_{i=1}^n \ind_{\{x_i=x\}}$. 
The set of $n$-length sequences $\mathbf{x} \in \mathcal{X}^n$ of type $P_{X}$ is  $  \mathcal{T}_n(P_{X},\mathcal{X}^n):=\{\mathbf{x} \in \mathcal{X}^n: P_{\mathbf{x}}=P_{X} \}$. Whenever the underlying alphabet $\mathcal{X}^n$ is clear from the context,  $\mathcal{T}_n(P_{X},\mathcal{X}^n)$ is simplified to $\mathcal{T}_n(P_{X})$. The set of all possible types of $n$-length sequences $\mathbf{x} \in \mathcal{X}^n$  is $\mathcal{T}(\mathcal{\X}^n):=\big\{P_{X} \in \mathcal{P}(\X):  \big|\mathcal{T}_n(P_{X},\mathcal{X}^n)\big| \geq 1 \big\}$.  Similar notations are used for larger combinations, e.g., $P_{\mathbf{x}\mathbf{y}}$, $\mathcal{T}_n(P_{XY},\X \times \Y)$ and $\mathcal{T}(\mathcal{\X}^n \times \Y^n)$.  
 For a given  $\mathbf{x} \in \mathcal{T}_n(P_{X},\mathcal{X}^n)$ and a conditional PMF $P_{Y|X}$,
 $ \mathcal{T}_n(P_{ Y|X},\mathbf{x}):=\{\mathbf{y} \in \Y^n: (\mathbf{x},\mathbf{y}) \in \mathcal{T}_n(P_{XY}, \mathcal{X}^n \times \Y^n)\}$ stands for the $P_{Y|X}$-conditional type class of $\mathbf{x}$.

 For PMFs $P,Q\in\mathcal{P}(\X)$, the Kullback--Leibler (KL) divergence between $P$ and $Q$ is $\kl{P}{Q}:=\sum_{x\in \X}P(x)\log $ $\big(P(x)/Q(x)\big)$.  The conditional KL divergence between $P_{Y|X}$ and $Q_{Y| X}$  given $ P_X $  is 
$D\big(P_{Y|X}||Q_{Y|X}\big|P_X\big):= \sum_{x\in \X} P_{X}(x) D\big(P_{Y|X}(\cdot|x)||Q_{Y|X}(\cdot|x)\big)$.
  The mutual information and entropy terms are denoted by $I_P(\cdot)$ and $H_P(\cdot)$, respectively, where $P$ denotes the PMF of the relevant random variables. When the PMF is clear from the context, the subscript is omitted.
For $(\mathbf{x},\mathbf{y}) \in \mathcal{X}^n \times \Y^n$, the empirical conditional entropy of $\mathbf{y}$ given $\mathbf{x}$ is $ H_e(\mathbf{y}|\mathbf{x}):= H_P(\tilde Y|\tilde X)$, 
 where $P_{\tilde X\tilde Y}=P_{\mathbf{x}\mathbf{y}}$. For a given function $f: \Z \rightarrow \mathbb{R}$ and a random variable $Z\sim P_Z$, the log-moment generating function  of $Z$ with respect to  $f$ is $\psi_{P_Z,f}(\lambda):= \log \mathbb{E}_{P_Z} [  e^{\lambda f(Z)}]$
whenever the expectation exists. Finally, 
let 
\begin{align}
\psi_{P_Z,f}^*(\theta):= \sup_{\lambda \in \mathbb{R}} \theta \lambda- \psi_{P_Z,f}(\lambda),\label{Chernoffexp}
\end{align}
denote the rate function (see, e.g., Definition 15.5 in \cite{Polyanskiy-Wu-book}). 
\subsection{Problem Formulation}
\begin{figure}[t]
\centering
\includegraphics[trim=0cm 0cm 0cm 0cm, clip, width= 0.7\textwidth]{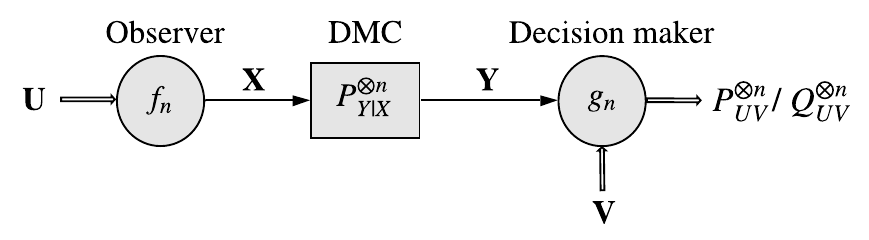}
\caption{DHT over a noisy channel. The observer observes an $n$-length i.i.d. sequence $\mathbf{U}$, and transmits $\mathbf{X}$ over the DMC $P_{Y|X}^{\otimes n}$. Based on the channel output $\mathbf{Y}$ and the $n$-length i.i.d. sequence $\mathbf{V}$, the decision maker performs a binary  HT to determine whether $(\mathbf{U},\mathbf{V})\sim P_{UV}^{\otimes n}$ or $(\mathbf{U},\mathbf{V})\sim Q_{UV}^{\otimes n}$. } \label{htnosiychnfig}
\end{figure}

Let $\Ucal$, $\V$, $\X$ and $\Y$ be finite sets, and $n \in \mathbb{N}$. The DHT over a noisy channel setting is depicted in Figure \ref{htnosiychnfig}. Herein, the  observer and the decision maker observe $n$ i.i.d. 
samples, denoted by $\mathbf{u}$ and $\mathbf{v}$, respectively. Based on its observations $\mathbf{u}$, the observer outputs a sequence $\mathbf{x} \in \mathcal{X}^n$ as the channel input sequence\footnote{In our problem formulation, we assume  that the ratio of the number of channel uses to the number of data samples, termed the bandwidth ratio,  is 1. However, the results easily generalize to arbitrary bandwidth ratios.}. The discrete memoryless channel (DMC) with transition kernel $P_{Y|X}$ produces a sequence $\mathbf{y} \in \Y^n$  according to the probability law $P_{Y|X}^{\otimes n}(\cdot|\mathbf{x})$  as its output. We will assume that $P_{Y|X}(\cdot|x)\ll P_{Y|X}(\cdot|x')$, $\forall ~(x,x') \in \X^2$, where $P \ll Q$ indicates absolute continuity of $P$ with respect to  $Q$. Based on its observations,  $\mathbf{y}$ and $\mathbf{v}$, the decision maker performs binary HT on the joint probability distribution of $(\mathbf{U},\mathbf{V})$ with the null ($H_0$) and alternative ($H_1$) hypotheses given by
\begin{subequations}\notag
\begin{align}
 & H_0:~ (\mathbf{U},\mathbf{V}) \sim P_{UV}^{\otimes n}, \\
 & H_1:~ (\mathbf{U},\mathbf{V}) \sim Q_{UV}^{\otimes n}.  
\end{align}
\end{subequations}
The decision maker outputs $\hat h \in \hat{\mathcal{H}}:=\{0,1\}$ as the decision of the hypothesis test, where $0$ and $1$ denote $H_0$ and $H_1$, respectively.  
 A length-$n$ DHT code $c_n$ is a pair of functions $(f_n,g_n)$, where
\begin{enumerate}[label=(\roman*)]
     \item $f_n: \Ucal^n \rightarrow \mathcal{P}(\mathcal{X}^n)$ denotes the encoding function, and
     \item $g_n: \V^n \times \Y^n \rightarrow \hat{\mathcal{H}}$ denotes a deterministic decision function\footnote{There is no loss in generality in restricting our attention  to a deterministic decision function for the objective of characterizing the error-exponents trade-off in HT (for example, see \cite[Lemma 3]{SCG-2019}).} specified by an acceptance region (for null hypothesis $H_0$) $\mathcal{A}_n \subseteq \V^n \times \Y^n$  as
$g_n(\mathbf{v},\mathbf{y})=1-\ind_{\{(\mathbf{v},\mathbf{y}) \in \mathcal{A}_n\}},~\forall (\mathbf{v},\mathbf{y}) \in \V^n \times \Y^n$. 
 \end{enumerate}
A code $c_n=(f_n,g_n)$ induces the joint PMFs $P^{(c_n)}_{\mathbf{U}\mathbf{V}\mathbf{X}\mathbf{Y}\hat H} $ and $Q^{(c_n)}_{\mathbf{U}\mathbf{V}\mathbf{X}\mathbf{Y}\hat H} $  under the null and alternative hypotheses, respectively, where 
\begin{align}
P^{(c_n)}_{\mathbf{U}\mathbf{V}\mathbf{X}\mathbf{Y}\hat H}(\mathbf{u},\mathbf{v},\mathbf{x},\mathbf{y},\hat h):=P_{UV}^{\otimes n}(\mathbf{u},\mathbf{v})~ f_n(\mathbf{x}|\mathbf{u})~P^{\otimes n}_{Y|X}(\mathbf{y}|\mathbf{x})~\ind_{\big\{g_n(\mathbf{v},\mathbf{y})=\hat h\big\}},
\end{align}
and 
\begin{align}
Q^{(c_n)}_{\mathbf{U}\mathbf{V}\mathbf{X}\mathbf{Y}\hat H}(\mathbf{u},\mathbf{v},\mathbf{x},\mathbf{y},\hat h):=Q_{UV}^{\otimes n}(\mathbf{u},\mathbf{v})~ f_n(\mathbf{x}|\mathbf{u})~P^{\otimes n}_{Y|X}(\mathbf{y}|\mathbf{x})~\ind_{\big\{g_n(\mathbf{v},\mathbf{y})=\hat h\big\}},
\end{align}
respectively. For a given code $c_n$, the type I and type II error probabilities are $\alpha_n(c_n):=\mathbb{P}_{P^{(c_n)}}(\hat H=1)$ and  
$\beta_n(c_n):=\mathbb{P}_{Q^{(c_n)}}(\hat H=0)$
respectively. The following  definition formally states the error-exponents trade-off we aim to characterize. 

\begin{definition}[\textbf{Error-exponent region}] \label{deft1t2expdistach} 
An error-exponent pair  $(\kappa_{\alpha}, \kappa_{\beta}) \in \mathbb{R}^2_{\geq 0}$ is  said to be achievable if there exists a  sequence of codes $\{c_n\}_{n \in \mathbb{N}}$ such that 
\begin{subequations} \label{seqkappa2}
\begin{equation}
 \liminf_{n \rightarrow \infty}  -\frac{1}{n}\log  \alpha_n \left(c_n \right)   \geq  \kappa_{\alpha}, \label{type1errprobconst} 
\end{equation}
\begin{equation}  
 \liminf_{n \rightarrow \infty}  -\frac{1}{n}\log  \beta_n \left(c_n \right)   \geq  \kappa_{\beta}.  \label{type2errprobconst} 
\end{equation}
\end{subequations} 
The error-exponent region $\bar {\mathcal R}$ is  the closure of the set of all achievable error-exponent pairs $(\kappa_{\alpha},\kappa_{\beta})$. Set $\mathcal{R}:=\{\big( \kappa_{\alpha}, \kappa(\kappa_{\alpha})\big):\kappa_{\alpha} \in (0,\kappa^{\star}_{\alpha})\}$,
 where $\kappa^{\star}_{\alpha}=\inf \{\kappa_{\alpha}:\kappa(\kappa_{\alpha})=0\}$ and $\kappa(\kappa_{\alpha}) :=  \sup \{ \kappa_{\beta}: (\kappa_{\alpha},\kappa_{\beta}) \in \bar {\mathcal R} \}$.
\end{definition}
We are interested in a computable characterization of $\mathcal{R}$, which pertains to the region of positive error-exponents (i.e., excluding the boundary points corresponding to Stein's exponent). To this end, we  present two inner bounds on $\mathcal{R}$ below.

\section{Main Results} \label{mainresults}
In this section, we obtain two inner bounds on $\mathcal{R}$, first using a separation-based scheme which performs independent HT and channel coding, termed the SHTCC scheme, and the second via a joint HT and channel coding scheme that uses hybrid coding for communication between the observer and the decision maker.
\subsection{Inner Bound on $\mathcal{R}$ via SHTCC Scheme}
Let $\mathcal{S}=\mathcal{X}$ and $P_{SXY}=P_{SX}P_{Y|X}$ be a PMF under which  $S-X-Y$ forms a Markov chain.  
For $x \in \X$, set $\Lambda_{x,P_{SXY}}(y):= \log \big(P_{Y|X=x}(y)/P_{Y|S=x}(y) \big)$
and 
\begin{align}
   E_{\mathrm{sp}}(P_{SX}, \theta):= \sum_{s \in \mathcal{S}} P_S(s) \psi^*_{P_{Y|S=s}, \Lambda_{s,P_{SXY}}}(\theta), \notag
\end{align}
where  the rate function $\psi^*$ is defined in \eqref{Chernoffexp}. 
For a fixed $P_{SX}$ and $R \geq 0$, let 
\begin{flalign} 
E_{\mathrm{ex}}(R,P_{SX}):= \max_{\rho \geq 1} -\rho R-\rho\log \Bigg(& \sum_{s,x,\tilde x}P_S(s)P_{X|S}(x|s)P_{X|S}(\tilde x|s) \Big( \sum_{y} \big(P_{Y|X}(y|x)P_{Y|X}(y|\tilde x)\big)^{\frac 12}\Big)^{\frac{1}{\rho}} \Bigg), \notag &&
\end{flalign}
denote the expurgated exponent \cite{Gallager-1965}\cite{Csiszar-Korner}.
Let $\mathcal{W}$ be a  finite set and $\mathfrak{F}$ denote the set of all continuous mappings from $\mathcal{P}(\Ucal)$ to   $\mathcal{P}(\W|\Ucal)$, where $\mathcal{P}(\W|\Ucal)$ is the set of all conditional distributions $P_{W|U}$.  Set  
$\theta_{\mathrm{l}}(P_{SX}):= \sum_{s \in \mathcal{S}} P_{S}(s) D \big(P_{Y|S=s}||P_{Y|X=s}\big)$, $\theta_{\mathrm{u}}(P_{SX}):= \sum_{s \in \mathcal{S}} P_{S}(s) D \big(P_{Y|X=s}||P_{Y|S=s}\big)$, and  $\Theta(P_{SX}):= \big(-\theta_{\mathrm{l}}(P_{SX}), $ $\theta_{\mathrm{u}}(P_{SX})\big)$. Denote an arbitrary element of $ \mathfrak{F}  \times \mathbb{R}_{\geq 0} \times  \mathcal{P}(\mathcal{S}\times \X)\times \Theta(P_{SX})$ by $(\omega,R, P_{SX}, \theta)$,  
and set 
\begin{subequations}
\begin{align}
&\mathcal{L}(\kappa_{\alpha}):=\left\{
\begin{aligned}
(\omega,R, P_{SX}, \theta) :~ &  \zeta(\kappa_{\alpha}, \omega)-  \rho(\kappa_{\alpha},\omega) \leq  R <  I_P(X;Y|S),~ P_{SXY}=P_{SX}P_{Y|X}      \\
& \min \left\lbrace E_{\mathrm{sp}}(P_{SX}, \theta), E_{\mathrm{ex}}\left(R,P_{SX}\right), E_{\mathrm{b}}(\kappa_{\alpha},\omega,R) \right\rbrace \geq \kappa_{\alpha} \end{aligned}
\right\}, \notag \\
 & \hat{\mathcal{L}}(\kappa_{\alpha},\omega):=\big\{
     P_{\hat U\hat V \hat W}:~ \kl{P_{\hat U\hat V \hat W}}{P_{UV \hat W}} \leq \kappa_{\alpha},  P_{\hat W|\hat U}= \omega(P_{\hat U}), P_{UV \hat W}=P_{UV}P_{\hat W|\hat U}  
\big\}, \label{lhatsetdef}\\
& E_{\mathrm{b}}(\kappa_{\alpha},\omega,R):=\begin{cases}
 R-  \zeta(\kappa_{\alpha}, \omega)+  \rho(\kappa_{\alpha},\omega), &\mbox{ if }0 \leq R <  \zeta(\kappa_{\alpha}, \omega), \notag  \\
~~~~~\infty, &\mbox{ otherwise},
\end{cases} \notag \\
&\zeta(\kappa_{\alpha}, \omega):=\underset{\substack{P_{\hat{U} \hat W}:~\exists~ P_{\hat V}, P_{\hat{U}\hat V \hat W} \in  \hat{\mathcal{L}}(\kappa_{\alpha},\omega)}}{\max} I_P(\hat U; \hat W), \label{totquantrate} \\
 &   \rho(\kappa_{\alpha},\omega):=\min_{\substack{P_{\hat V \hat W}:~\exists~ P_{\hat U},  P_{\hat U \hat V \hat W} \in \hat{\mathcal{L}}(\kappa_{\alpha}, \omega)}} I_P(\hat V; \hat W), \label{totsideinfrate} \\
  &E_1(\kappa_{\alpha},\omega):= \min_{\substack{(P_{\tilde U \tilde V \tilde W},Q_{\tilde U \tilde V \tilde W})  \in \mathcal{T}_1(\kappa_{\alpha},\omega)}} D(P_{\tilde U \tilde V \tilde W}||Q_{\tilde U \tilde V \tilde W}), \notag \\
  &E_2(\kappa_{\alpha},\omega,R):=
\begin{cases} \notag
\underset{\substack{(P_{\tilde U \tilde V \tilde W},Q_{\tilde U \tilde V \tilde W})  \in \mathcal{T}_2(\kappa_{\alpha},\omega)}}{\min} D(P_{\tilde U \tilde V \tilde W}||Q_{\tilde U \tilde V \tilde W})+E_{\mathrm{b}}(\kappa_{\alpha},\omega,R), &\mbox{ if } R < \zeta(\kappa_{\alpha}, \omega),\\
\qquad \qquad \qquad \qquad \qquad \qquad \infty, &\mbox{ otherwise},
\end{cases}  \\
&E_3(\kappa_{\alpha},\omega,R,P_{SX}):=
\begin{cases} \notag
\underset{\substack{(P_{\tilde U \tilde V \tilde W},Q_{\tilde U \tilde V \tilde W})  \in \mathcal{T}_3(\kappa_{\alpha},\omega)}}{\min} D(P_{\tilde U \tilde V \tilde W}||Q_{\tilde U \tilde V \tilde W})+E_{\mathrm{b}}(\kappa_{\alpha},\omega,R)+E_{\mathrm{ex}}\left(R,P_{SX}\right), &\\ \qquad \qquad \qquad \qquad \qquad \qquad\qquad \qquad \qquad ~~~\mbox{ if } R < \zeta(\kappa_{\alpha},\omega),&\\
\underset{\substack{(P_{\tilde U \tilde V \tilde W},Q_{\tilde U \tilde V \tilde W})  \in \mathcal{T}_3(\kappa_{\alpha},\omega)}}{\min} D(P_{\tilde U \tilde V \tilde W}||Q_{\tilde U \tilde V \tilde W}) +\rho(\kappa_{\alpha},\omega) +E_{\mathrm{ex}}\left(R,P_{SX}\right), &\\ \qquad \qquad \qquad \qquad \qquad \qquad\qquad \qquad \qquad ~~~ \mbox{ otherwise},&
\end{cases}\\
&E_4(\kappa_{\alpha},\omega,R,P_{SX}, \theta ):=
\begin{cases} \notag
\underset{P_{\hat V}: P_{\hat U \hat V \hat W} \in  \hat{\mathcal{L}}(\kappa_{\alpha},\omega)}{\min} D( P_{\hat V} ||Q_V)+ E_{\mathrm{b}}(\kappa_{\alpha},\omega,R)+E_{\mathrm{m}}(P_{SX}, \theta)-\theta, &\\ \qquad \qquad \qquad \qquad \qquad \qquad\qquad \qquad \qquad ~~~\mbox{ if } R < \zeta(\kappa_{\alpha}, \omega),&\\
\underset{P_{\hat V}: P_{\hat U \hat V \hat W} \in  \hat{\mathcal{L}}(\kappa_{\alpha},\omega)}{\min} D( P_{\hat V} ||Q_{ V})+ \rho(\kappa_{\alpha},\omega)+E_{\mathrm{m}}(P_{SX}, \theta)-\theta, &\\ \qquad \qquad \qquad \qquad \qquad \qquad\qquad \qquad \qquad ~~~ \mbox{ otherwise},&
\end{cases}\notag \\
&  \mathcal{T}_1(\kappa_{\alpha},\omega):=\left\lbrace
                \begin{array}{ll}
                  (P_{\tilde U \tilde V \tilde W},Q_{\tilde U \tilde V \tilde W}):  & P_{\tilde U \tilde W}= P_{\hat U  \hat W},~ P_{\tilde V \tilde W}= P_{\hat V  \hat W},~ Q_{\tilde U \tilde V \tilde W} \\&:=Q_{UV}P_{\tilde W|\tilde U}\mbox{ for some }   P_{\hat U \hat V \hat W} \in   \hat{\mathcal{L}}(\kappa_{\alpha},\omega)
                \end{array}
             \right\rbrace, \label{firstexpfactsep} \\
&\mathcal{T}_2(\kappa_{\alpha},\omega):=\left\lbrace
                \begin{array}{ll}
                  (P_{\tilde U \tilde V \tilde W},Q_{\tilde U \tilde V \tilde W}):  & P_{\tilde U \tilde W}= P_{\hat U  \hat W}, ~ P_{\tilde V}= P_{\hat V},~ H_P(\tilde W|\tilde V) \geq H_P(\hat W|\hat  V),\\& Q_{\tilde U \tilde V \tilde W}:=Q_{UV}P_{\tilde W|\tilde U}  \mbox{ for some }   P_{\hat U \hat V \hat W} \in  \hat{ \mathcal{L}}(\kappa_{\alpha},\omega)
                \end{array}
             \right\rbrace, \notag \\
& \mathcal{T}_3(\kappa_{\alpha},\omega):=\left\lbrace
                \begin{array}{ll}
                  (P_{\tilde U \tilde V \tilde W},Q_{\tilde U \tilde V \tilde W}):  & P_{\tilde U \tilde W}= P_{\hat U  \hat W}, ~ P_{\tilde V}= P_{\hat V},~Q_{\tilde U \tilde V \tilde W}:=Q_{UV}P_{\tilde W|\tilde U} \\& \mbox{ for some }   P_{\hat U \hat V \hat W} \in  \hat{ \mathcal{L}}(\kappa_{\alpha},\omega)
                \end{array}
             \right\rbrace. \notag
\end{align}
\end{subequations}
We have the following lower bound for $\kappa(\kappa_{\alpha})$ which translates to an inner bound for $\mathcal{R}$.
\begin{theorem}[Inner bound via SHTCC scheme]\label{lbbinningts}
$\kappa(\kappa_{\alpha}) \geq \kappa_{\mathrm{s}}^{\star}(\kappa_{\alpha}),$ where
\begin{align}
 \kappa_{\mathrm{s}}^{\star}(\kappa_{\alpha}) &:=\max_{\substack{(\omega,R,P_{SX}, \theta)  \in~    \mathcal{L}(\kappa_{\alpha})}}\min \big\{E_1(\kappa_{\alpha},\omega),E_2(\kappa_{\alpha},\omega,R),E_3(\kappa_{\alpha},\omega,R,P_{SX}),  E_4(\kappa_{\alpha},\omega,R,P_{SX},\theta)  \big\}. \label{innerbndwithbinning} 
\end{align}
\end{theorem}
The proof of Theorem \ref{lbbinningts} is presented in Section  \ref{lbbinningtsproof}. The SHTCC scheme, which achieves the error-exponent pair $(\kappa_{\alpha},\kappa_{\mathrm{s}}^{\star}(\kappa_{\alpha}))$, is a coding scheme analogous to separate source and channel coding for the lossy transmission of a source over a communication channel with correlated side-information at the receiver~\cite{MerhavShamai2003}, however, with the objective of reliable HT. In this scheme, the source samples are first compressed to an index, which acts as the message to be transmitted over the channel. But, in contrast to standard communication problems,  there is a need to protect certain messages more reliably than others; hence, an unequal error-protection scheme~\cite{Csiszar-1982,Borade-09} is used. To describe briefly, the SHTCC scheme involves $(i)$
 the quantization and binning of $\mathbf{u}$ sequences, whose type $P_{\mathbf{u}}$ is within a $\kappa_{\alpha}$-neighborhood (in terms of KL divergence) of $P_U$, using $\mathbf{V}$  as  side information  at the decision maker for decoding, and $(ii)$ unequal error-protection channel coding scheme  in~\cite{Borade-09} for protecting a special message which informs the decision maker that $P_{\mathbf{u}}$ lies outside the  $\kappa_{\alpha}$-neighborhood of $P_U$. The output of the channel decoder is processed by an empirical conditional entropy decoder which recovers the quantization codeword with the least conditional entropy with $\mathbf{V}$. Since this decoder depends only on the empirical distributions of the observations, it is universal and useful in the hypothesis testing context, where multiple distributions are involved (as was first noted in~\cite{Shimokawa}).   The various  factors $E_1$ to $E_4$ in \eqref{innerbndwithbinning} have natural interpretations in terms of events  that could possibly result in a hypothesis testing error.  Specifically, $E_1$ and $E_2$ correspond to the error events arising due to quantization and binning, respectively, while $E_3$ and $E_4$ correspond to the error events of wrongly decoding an ordinary channel codeword and special message codeword, respectively.
\begin{remark}[Generalization of Han-Kobayashi inner bound]
In \cite[Theorem 1] {HK-1989}, Han and Kobayashi obtained an inner bound on $\mathcal{R}$ for DHT over a noiseless channel.  At a high level, their coding scheme involves  type-based quantization of $\mathbf{u} \in \Ucal^n$ sequences, whose type $ P_{\mathbf{u}}$ lies  within a $\kappa_{\alpha}$-neighbourhood of $P_{U}$, where $\kappa_{\alpha}$ is the desired type I error-exponent. 
As a corollary, Theorem \ref{lbbinningts} recovers the lower bound for $\kappa( \kappa_{\alpha})$ obtained in \cite{HK-1989} by 
$(i)$ setting  $E_{\mathrm{ex}}\left(R,P_{SX}\right)$, $E_{\mathrm{m}}(P_{SX}, \theta)$ and $E_{\mathrm{m}}(P_{SX}, \theta)-\theta$ to $\infty$, which hold when the channel is noiseless; and
    $(ii)$  maximizing over the set $\big\{(\omega,R, P_{SX}, \theta) \in \mathfrak{F} \times \mathbb{R}_{\geq 0} \times \mathcal{P}(\mathcal{S}\times \X) \times \Theta(P_{SX}): \zeta(\kappa_{\alpha}, \omega) \leq R<I_P(X;Y|S),P_{SXY}:=P_{SX}P_{Y|X} \big\} \subseteq \mathcal{L}(\kappa_{\alpha})$ in \eqref{innerbndwithbinning}.  Then, note that the terms $ E_2(\kappa_{\alpha},\omega,R)$, $ E_3(\kappa_{\alpha},\omega,R,P_{SX})$ and $ E_4(\kappa_{\alpha},\omega,R,P_{SX}, \theta)$ all equal $\infty$, and thus the inner bound in Theorem \ref{lbbinningts} reduces to that given in \cite[Theorem 1]{HK-1989}.
 \end{remark}
\begin{remark}[Improvement via time-sharing]
Since the lower bound on $\kappa( \kappa_{\alpha})$ in Theorem \ref{lbbinningts} is  not necessarily concave, a tighter bound  can be obtained using the technique of time-sharing similar to \cite[Theorem 3]{HK-1989}. 
We omit its description as it is cumbersome, although straightforward. 
\end{remark}
 Theorem \ref{lbbinningts} also  recovers the lower bound for the optimal type II error-exponent  for a fixed type I error probability constraint established in \cite[Theorem 2]{SD_2020}  by letting $\kappa_{\alpha}\rightarrow0$. The details are  provided in Appendix  \ref{genHTsepbndproof}.
Further, specializing the lower bound in Theorem \ref{lbbinningts} to the case of TAI, i.e., when $Q_{UV}=P_UP_V$, we obtain the following corollary which   recovers the optimal type II error-exponent for TAI established in \cite[Proposition 7]{SD_2020}.
\begin{cor}[Inner bound for TAI] \label{boundwithoutbintai}
Let $P_{UV} \in \mathcal{P}(\Ucal \times \V)$ be an arbitrary distribution and $Q_{UV}=P_UP_V$. Then,
\begin{align}
    \kappa(\kappa_{\alpha}) \geq \kappa_{\mathrm{s}}^{\star}(\kappa_{\alpha}) \geq \kappa_{\mathrm{i}}^{\star}(\kappa_{\alpha}), \label{bndtaigenexptrd}
\end{align}
where 
\begin{align}
&\kappa_{\mathrm{i}}^{\star}(\kappa_{\alpha}):= \max_{\substack{(\omega,P_{SX}, \theta)\in  \mathcal{L}^{\star}(\kappa_{\alpha})}}\min \left\lbrace E_1^{\mathrm{i}}(\kappa_{\alpha},\omega),E_2^{\mathrm{i}}(\kappa_{\alpha},\omega,P_{SX}),E_3^{\mathrm{i}}(\kappa_{\alpha},\omega,P_{SX}, \theta)  \right\rbrace, \notag \\
&\mathcal{L}^{\star}(\kappa_{\alpha}):= \left\{
\begin{aligned} 
&(\omega, P_{SX}, \theta) \in \mathfrak{F} \times  \mathcal{P}(\mathcal{S}\times \X) \times \Theta(P_{SX}):  \zeta(\kappa_{\alpha}, \omega) <   I_P(X;Y|S), \\
& P_{SXY}:=P_{SX}P_{Y|X},~\min \left \lbrace E_{\mathrm{sp}}(P_{SX}, \theta), E_{\mathrm{ex}}\left(\zeta(\kappa_{\alpha}, \omega),P_{SX}\right) \right \rbrace \geq \kappa_{\alpha}
\end{aligned}
\right\},  \label{lstardeftai}
\\
&  E_1^{\mathrm{i}}(\kappa_{\alpha},\omega):=
  \underset{\substack{P_{\hat V \hat W}:\mspace{2 mu} \exists\mspace{2 mu} P_{\hat U \hat V \hat W} \in \hat{\mathcal{L}}(\kappa_{\alpha},\omega)}}{\min} I_P(\hat V; \hat W) + D(P_{\hat V}||P_V), \notag \\
  & E_2^{\mathrm{i}}(\kappa_{\alpha},\omega,P_{SX}):= 
  \rho(\kappa_{\alpha}, \omega)+E_{\mathrm{ex}}\big(\zeta(\kappa_{\alpha}, \omega),P_{SX}\big),\notag \\
 & E_3^{\mathrm{i}}(\kappa_{\alpha},\omega,P_{SX}, \theta):=
  \rho(\kappa_{\alpha}, \omega)+E_{\mathrm{sp}}(P_{SX}, \theta)-\theta,\notag 
\end{align}
and,  $\hat{\mathcal{L}}(\kappa_{\alpha},\omega)$, $\zeta(\kappa_{\alpha}, \omega)$ and $\rho(\kappa_{\alpha}, \omega)$ are  defined in \eqref{lhatsetdef}, \eqref{totquantrate} and \eqref{totsideinfrate}, respectively. In particular, \begin{align}
    \lim_{\kappa_{\alpha} \rightarrow 0} \kappa(\kappa_{\alpha}) =\kappa_{\mathrm{s}}^{\star}(0)=\kappa_{\mathrm{i}}^{\star}(0)= \underset{\substack{P_{W|U}:
    I_P(U;W) \leq C(P_{Y|X}),\\P_{UVW}=P_{UV}P_{W|U}}}{\max} I_P(V;W),\label{Steinregimetai}
\end{align}
where $|\W| \leq |\Ucal|+1$ and $C(P_{Y|X})$ denotes the capacity of the channel $P_{Y|X}$.
\end{cor}
The proof of Corollary \ref{boundwithoutbintai} is given in Section \ref{boundwithoutbintaiproof}. Its achievability follows from a special case of the SHTCC scheme without binning  at the encoder. 

Next, we consider \textit{testing against dependence} (TAD) for which $Q_{UV}$ is an arbitrary joint distribution and $P_{UV}=Q_U Q_V$.  Theorem \ref{lbbinningts} specialized to TAD gives the following corollary.
\begin{cor}[Inner bound for TAD]\label{innbndtad}
Let $Q_{UV} \in \mathcal{P}(\Ucal \times \V)$ be an arbitrary distribution and $P_{UV}=Q_UQ_V$. Then,
\begin{align}
    \kappa(\kappa_{\alpha}) \geq \kappa_{\mathrm{s}}^{\star}(\kappa_{\alpha}) = \kappa_{\mathrm{d}}^{\star}(\kappa_{\alpha}):= \max_{\substack{(\omega,P_{SX}, \theta)\\ \in  \mathcal{L}^{\star}(\kappa_{\alpha})}}\min \left\lbrace E_1^{\mathrm{d}}(\kappa_{\alpha},\omega),E_2^{\mathrm{d}}(\kappa_{\alpha},\omega,P_{SX}),E_3^{\mathrm{d}}(P_{SX}, \theta)  \right\rbrace, \label{bndtadexpgen}
\end{align}
where 
\begin{align}
  & E_1^{\mathrm{d}}(\kappa_{\alpha},\omega):=\min_{\substack{(P_{\tilde U \tilde V \tilde W},Q_{\tilde U \tilde V \tilde W}) \\ \in \mathcal{T}_1(\kappa_{\alpha},\omega)}} D(P_{\tilde U \tilde V \tilde W}||Q_{\tilde U \tilde V \tilde W}) \geq 
  \min_{\substack{(P_{\hat V \hat W},Q_{V \hat W}):~P_{\hat U \hat V \hat W} \in \hat{\mathcal{L}}(\kappa_{\alpha},\omega), \\Q_{UV \hat W}=Q_{UV}P_{\hat W|\hat U} }}  D(P_{\hat V \hat W}||Q_{ V \hat W}), \notag \\
 & E_2^{\mathrm{d}}(\kappa_{\alpha},\omega,P_{SX}):= E_{\mathrm{ex}}\left(\zeta(\kappa_{\alpha}, \omega),P_{SX}\right),\notag \\
  & E_3^{\mathrm{d}}(P_{SX}, \theta):=E_{\mathrm{sp}}(P_{SX}, \theta)-\theta,\notag 
\end{align}
and, $\hat{\mathcal{L}}(\kappa_{\alpha},\omega)$, $\mathcal{T}_1(\kappa_{\alpha},\omega)$ and $\mathcal{L}^{\star}(\kappa_{\alpha})$ are  given in \eqref{lhatsetdef}, \eqref{firstexpfactsep} and  \eqref{lstardeftai}, respectively.
In particular,
\begin{align}
  \lim_{\kappa_{\alpha} \rightarrow 0} \kappa(\kappa_{\alpha}) \geq \kappa_{\mathrm{s}}^{\star}(0)= \kappa_{\mathrm{d}}^{\star}(0)\geq   \kappa_{\mathrm{TAD}}^{\star}, \label{steinbndtadchar}
\end{align}
where 
\begin{align}
 \kappa_{\mathrm{TAD}}^{\star}= \max_{\substack{(P_{W|U},P_{SX}):\\
  I_Q(W;U)\leq I_{P}(X;Y|S),\\Q_{UVW}=Q_{UV}P_{W|U},\\P_{SXY}=P_{SX}P_{Y|X}}}  \min\big\{D(Q_{V}Q_W||Q_{VW}), E_{\mathrm{ex}}\left(I_Q(U;W),P_{SX}\right), \theta_{\mathrm{l}}(P_{SX})\big\}, \notag
\end{align}
and   $|\W| \leq |\Ucal|+1$.
\end{cor}
The proof of Corollary \ref{innbndtad} is given in Section \ref{innbndtadproof}. Note that the expression for $\kappa_{\mathrm{s}}^{\star}(\kappa_{\alpha}) $ given in \eqref{bndtadexpgen} is relatively simpler to compute compared to that in Theorem \ref{lbbinningts}. This will be handy in showing that the JHTCC scheme strictly outperforms the SHTCC scheme, which we highlight via an  example in Section \ref{compshtccjhtcc} below. 
\subsection{Inner Bound via JHTCC Scheme}
It is well known that joint source-channel coding schemes offer advantages over separation-based coding schemes in several information theoretic problems, such as the transmission of correlated sources over a multiple-access channel \cite{Cover-elgamal-salehi,Lim-minero-kim-2015} and the error-exponent in the lossless or lossy transmission of a source over a noisy channel \cite{Csiszar-1980,Csiszar-1982}.  Recently, it is shown via an example in \cite{SD_2020} that joint schemes also achieve a strictly larger type II error-exponent in DHT problems compared to a separation-based scheme in some scenarios.  Motivated by this, we present an inner bound on $\mathcal{R}$ using a generalization of the JHTCC scheme in \cite{SD_2020}.

Let $\W$ and $\mathcal{S}$ be arbitrary finite sets, and $\mathfrak{F}'$ denote the set of all continuous mappings from $\mathcal{P}({\Ucal} \times \mathcal{S})$ to   $\mathcal{P}(\W|~\Ucal \times \mathcal{S})$, where $\mathcal{P}(\W|~\Ucal \times \mathcal{S})$ is the set of all conditional distributions $P_{W|US}$. 
Let $\big(P_{S}, \omega'(\cdot,P_{S}), P_{X|USW},P_{X'|US}\big)$ denote an arbitrary element of $\mathcal{P}(\mathcal{S}) \times  \mathfrak{F}' \times \mathcal{P}(\X|\Ucal \times \mathcal{S} \times \W)\times \mathcal{P}(\X| \Ucal \times \mathcal{S})$, and define
\begin{flalign}
&\mathcal{L}_{\mathrm{h}}(\kappa_{\alpha}):=\big\{
\big(P_{S}, \omega'(\cdot,P_{S}), P_{X|USW},P_{X'|US}\big) :E_{\mathrm{b}}'(\kappa_{\alpha},\omega',P_S, P_{X|USW})\geq \kappa_{\alpha}
\big\}, \notag \\
 & \hat{\mathcal{L}}_{\mathrm{h}}(\kappa_{\alpha},\omega', P_S, P_{X|USW}) \notag \\[5 pt]
 &:=\left\{
\begin{aligned}
      (P_{\hat U\hat V \hat W \hat Y S} :&~ D(P_{\hat U \hat V \hat W \hat Y|S}||P_{UV\hat WY|S}|P_S) \leq \kappa_{\alpha},~ P_{SUV\hat WXY}:= P_S P_{UV}P_{\hat W|\hat US}P_{X|USW}P_{Y|X},\\ &P_{\hat W|\hat U S}= \omega'(P_{\hat U}, P_S)
 \end{aligned}
\right\}, \notag  \\[5 pt]
& E_{\mathrm{b}}'(\kappa_{\alpha},\omega',P_S, P_{X|USW}):= \rho'(\kappa_{\alpha}, \omega',P_S, P_{X|USW})-\zeta'_q(\kappa_{\alpha},\omega',P_S), \notag \\[5 pt]
&\zeta'(\kappa_{\alpha}, \omega',P_S):=\underset{\substack{P_{\hat{U} \hat WS}:~\exists ~P_{\hat V \hat Y} \mbox{ s.t. } \\ P_{\hat{U}\hat V \hat W \hat Y S} ~\in  \hat{\mathcal{L}}_{\mathrm{h}}(\kappa_{\alpha},\omega',P_S,P_{X|USW})}}{\max} I_P(\hat U; \hat W|S), \notag  \\
 &   \rho'(\kappa_{\alpha},\omega',P_S, P_{X|USW}):=\min_{\substack{P_{\hat V \hat W \hat Y S}:~\exists~ P_{\hat U} \mbox{ s.t. } \\P_{\hat{U}\hat V \hat W \hat Y S} ~\in  \hat{\mathcal{L}}_{\mathrm{h}}(\kappa_{\alpha},\omega',P_S,P_{X|USW})}} I_P(\hat Y,\hat V; \hat W|S), \notag  \\
    &E_1'(\kappa_{\alpha},\omega'):= \min_{\substack{(P_{\tilde U \tilde V \tilde W \tilde YS},Q_{\tilde U \tilde V \tilde W \tilde YS}) \in \mathcal{T}_1'(\kappa_{\alpha},\omega')}} D(P_{\tilde U \tilde V \tilde W \tilde Y|S}||Q_{ \tilde U \tilde V \tilde W \tilde  Y|S}|P_S), \notag \\[5 pt]
&E_2'(\kappa_{\alpha},\omega',P_{S},P_{X|USW}):=
\underset{\substack{(P_{\tilde U \tilde V \tilde W \tilde Y S},Q_{\tilde U \tilde V \tilde W \tilde Y S})  \in \mathcal{T}_2'(\kappa_{\alpha},\omega',P_S,P_{X|USW})}}{\min} D(P_{\tilde U \tilde V \tilde W \tilde Y|S}||Q_{ \tilde U \tilde V \tilde W \tilde  Y|S}|P_S) \notag \\
&\qquad \qquad \qquad \qquad\qquad \qquad\qquad \qquad \qquad \qquad \qquad \qquad\qquad+E_{\mathrm{b}}'(\kappa_{\alpha},\omega',P_{S},P_{X|USW}),\notag \\
&E_3'(\kappa_{\alpha},\omega',P_{S}, P_{X|USW},P_{X'|US}):=
\underset{\substack{P_{\hat V \hat YS}: P_{\hat U \hat V \hat W \hat Y S} \in   \hat{\mathcal{L}}_{\mathrm{h}}(\kappa_{\alpha},\omega', P_S, P_{X|USW})}}{\min} D( P_{\hat V \hat Y|S }||Q_{V Y'|S }|P_S) \notag \\
&\qquad \qquad \qquad \qquad\qquad \qquad\qquad \qquad \qquad \qquad \qquad\qquad\qquad+ 
 E_{\mathrm{b}}'(\kappa_{\alpha},\omega',P_S, P_{X|USW}),\notag \\
&Q_{S UV X'Y' }:=P_SQ_{UV}P_{X'|US} P_{Y'|X'},~ P_{Y'|X'}:=P_{Y|X}, \notag \\
&  \mathcal{T}_1'(\kappa_{\alpha},\omega',P_{S}, P_{X|USW})\mspace{-2 mu}:=\mspace{-2 mu}\left\lbrace
                \begin{array}{ll}
                 \mspace{-10 mu} (P_{\tilde U \tilde V \tilde W \tilde Y S},   & P_{\tilde U \tilde W S}= P_{\hat U  \hat W S},~ P_{\tilde V \tilde W \tilde Y S}= P_{\hat V  \hat W \hat Y S},\\
                 \mspace{-5 mu}~~~ Q_{\tilde U \tilde V \tilde W \tilde Y S}): &Q_{S \tilde U \tilde V \tilde W \tilde X \tilde Y}:= P_S Q_{ U V} P_{\tilde W|\tilde U S}~P_{X|U SW}P_{Y|X}\\
                  & \mbox{for some }   P_{\hat U \hat V \hat W \hat Y S }\in   \hat{\mathcal{L}}_{\mathrm{h}}(\kappa_{\alpha},\omega', P_S, P_{X|USW})
                \end{array}
             \mspace{-10 mu} \right\rbrace, \notag \\
&\mathcal{T}_2'(\kappa_{\alpha},\omega',P_{S}, P_{X|USW})\mspace{-2 mu}:=\mspace{-2 mu}\left\lbrace  
                \begin{array}{ll}
                  \mspace{-10 mu} (P_{\tilde U \tilde V \tilde W \tilde Y S},   & P_{\tilde U \tilde WS}= P_{\hat U  \hat WS}, ~ P_{\tilde V \tilde YS}= P_{\hat V \hat YS },\\  \mspace{-5 mu}~~~~ Q_{\tilde U \tilde V \tilde W \tilde Y S}):& H_P(\tilde W|\tilde V, \tilde Y, S) \geq  H_P(\hat W|\hat  V, \hat Y, S),\\
                  &Q_{S \tilde U \tilde V \tilde W \tilde X \tilde Y}:= P_S Q_{ U V} P_{\tilde W|\tilde U S}~P_{X|U SW}P_{Y|X}\\&
                  \mbox{for some }   P_{\hat U \hat V \hat W \hat Y S} \in  \hat{\mathcal{L}}_{\mathrm{h}}(\kappa_{\alpha},\omega', P_S, P_{X|USW})
                \end{array} \mspace{-10 mu}
             \right\rbrace. \notag &&
     \end{flalign}
     Then, we have the following result.
     \begin{theorem}[Inner bound via JHTCC scheme] \label{jhtccthm}
     \begin{flalign}
    &\kappa(\kappa_{\alpha}) \geq \max\left\lbrace\kappa_{\mathrm{h}}^{\star}(\kappa_{\alpha}),\kappa_{\mathrm{u}}^{\star}(\kappa_{\alpha})\right\rbrace,  \notag
    \end{flalign}
    where 
    \begin{flalign}
    & \kappa_{\mathrm{h}}^{\star}(\kappa_{\alpha}):= \max_{\substack{(P_S,\omega',P_{X|USW},P_{X'|US})  \in~  \mathcal{L}_{\mathrm{h}}(\kappa_{\alpha}) }} \min \Big \{ E_1'(\kappa_{\alpha}, \omega'),~ E_2'(\kappa_{\alpha}, \omega',P_S,P_{X|USW}), \notag \\
&\qquad \qquad \qquad \qquad\qquad \qquad\qquad \qquad \qquad E_3'(\kappa_{\alpha},\omega',P_{S}, P_{X|USW},P_{X'|US}) \Big\},\notag \\
    &\kappa_{\mathrm{u}}^{\star}(\kappa_{\alpha}):=\max_{\substack{(P_S,P_{X|US})  \in \mathcal{P}(\mathcal{S}) \times \mathcal{P}(\X|\mathcal{S} \times  \Ucal)}} \kappa_{\mathrm{u}}(\kappa_{\alpha},P_S,P_{X|US}),\notag \\ & \kappa_{\mathrm{u}}(\kappa_{\alpha},P_S,P_{X|US}):=\min_{\substack{P_S P_{\hat V\hat Y }:
    D\big(P_{\hat V\hat Y|S}||P_{VY|S}| P_S\big)\leq \kappa_{\alpha}}} D\big(P_{\hat V\hat Y|S}||Q_{VY|S}|P_S\big),\notag\\
    &P_{SUVXY}=P_S P_{UV}P_{X|US}P_{Y|X} \quad\mbox{ and } \quad Q_{SUVXY}=P_SQ_{UV}P_{X|US}P_{Y|X}. \notag
     \end{flalign}
     \end{theorem}
        The proof of Theorem \ref{jhtccthm} is given in Section \ref{jhtccthmproof}, and utilizes a generalization of hybrid coding scheme \cite{Lim-minero-kim-2015} to achieve the stated inner bound. Specifically, the error-exponent pair $\big(\kappa_{\alpha},\kappa_{\mathrm{h}}^{\star}(\kappa_{\alpha})\big)$ is achieved using type-based  hybrid coding, while $\big(\kappa_{\alpha}, \kappa_{\mathrm{u}}^{\star}(\kappa_{\alpha})\big)$ is realized by uncoded transmission, in which the channel input $\mathbf{X}$ is generated as the output of a DMC $P_{X|U}$ with input $\mathbf{U}$ (along with time sharing). In standard hybrid coding, the source sequence is first quantized  via  joint typicality and the channel input is then chosen as a function of both the original source sequence  and its quantization.   At the decoder, the quantized codeword is first recovered using the channel output  and side information via joint typicality decoding, and an estimate of the source sequence is output as a function of the channel output and recovered codeword. The quantization part forms the \emph{digital} part of the scheme, while the use of the source sequence for encoding and channel output for decoding comprises the \emph{analog} part. The scheme derives its name from these joint hybrid  digital-analog operations. In the HT context considered here, the aforementioned source quantization  is replaced by a type-based quantization at the encoder, and the joint typicality  decoder is replaced by a universal empirical conditional entropy decoder. We note that Theorem \ref{jhtccthm} recovers the lower bound on the optimal type II error-exponent  proved in~Theorem 5 in~\cite{SD_2020}. The details are provided in Appendix \ref{corjhtccsteinproof}.

Next, we provide a comparison between the SHTCC and JHTCC bounds via an example as  mentioned earlier. \subsection{Comparison of Inner Bounds} \label{compshtccjhtcc}
We compare the inner bounds established  in Theorem \ref{lbbinningts} and Theorem \ref{jhtccthm} for a simple setting of TAD over a BSC. For this purpose, we will use the inner bound $\kappa_{\mathrm{d}}^{\star}(\kappa_{\alpha})$ stated in Corollary \ref{innbndtad} and $\kappa_{\mathrm{u}}^{\star}(\kappa_{\alpha})$ that is achieved by uncoded transmission. Our objective is to illustrate that the JHTCC scheme achieves a strictly tighter bound on $\mathcal{R}$ compared to the SHTCC scheme, at least for some points of the trade-off.   
\begin{figure}     
  \subfloat[ $p=0.25$ and $q=0$ ]{%
      \label{fig025} \includegraphics[width=0.49\textwidth]{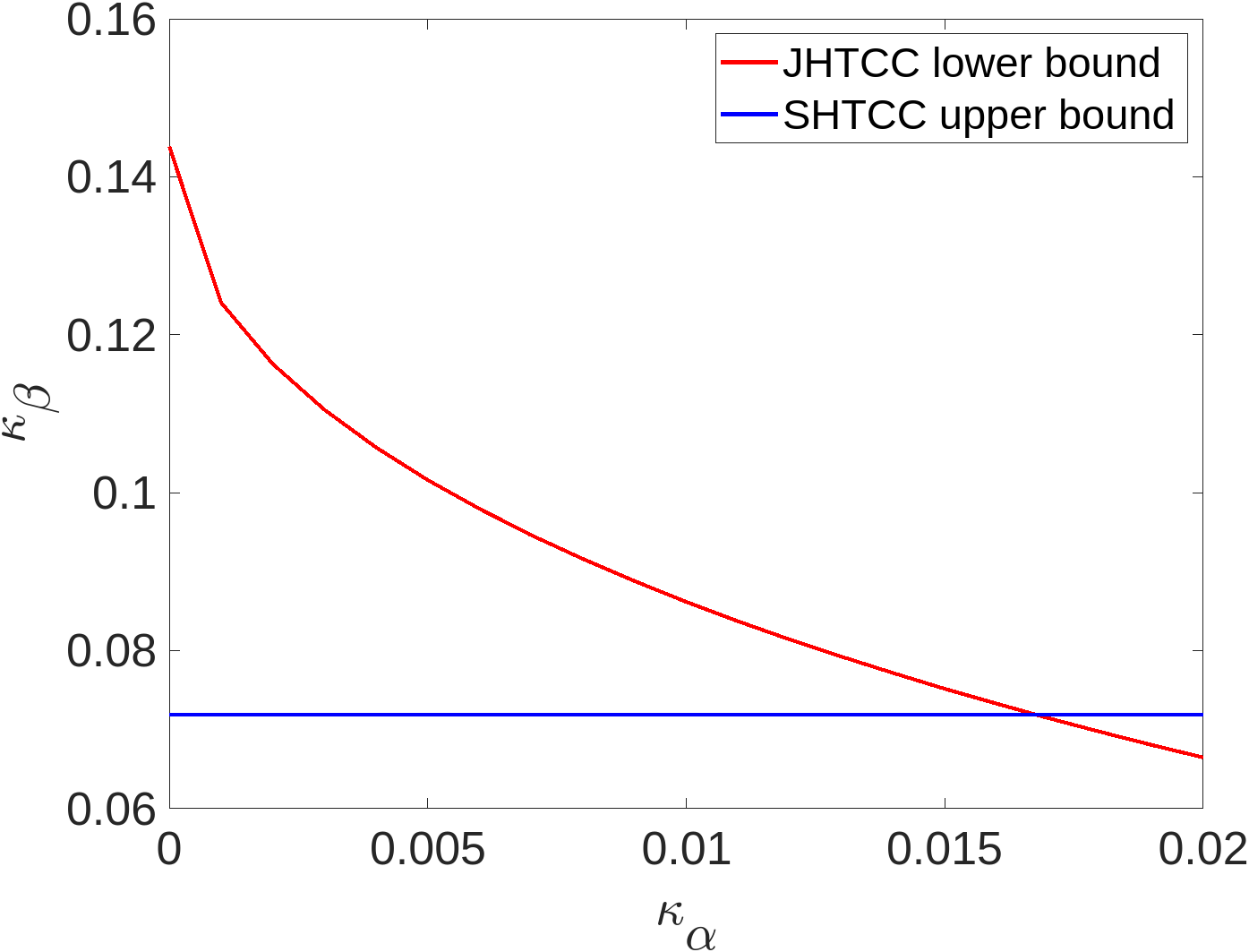}}  
          \subfloat[ $p=0.35$ and $q=0$ ]{\label{fig035}
        \hspace{3 pt}\includegraphics[width=0.485\textwidth]{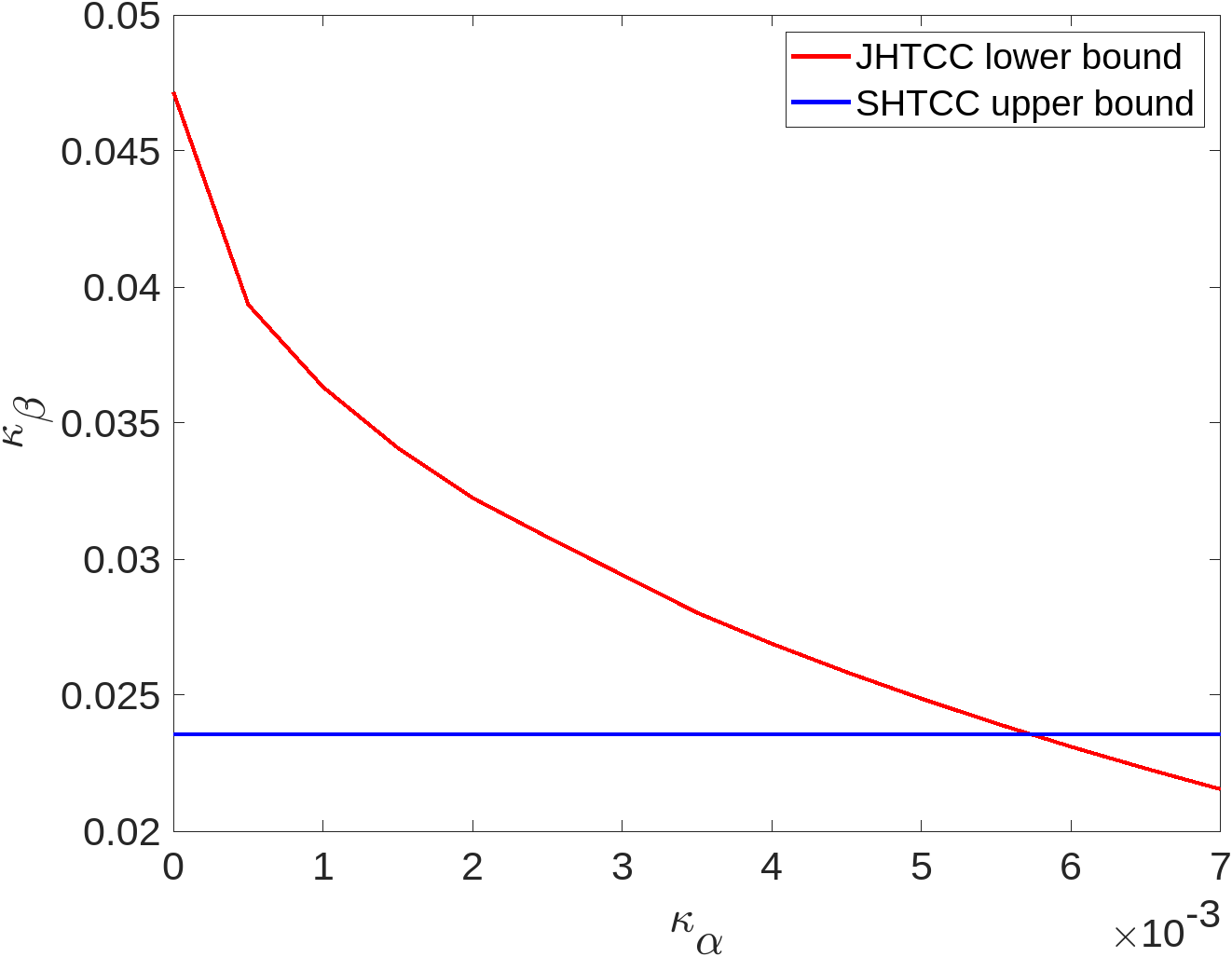}} 
           \caption{Comparison of the error-exponents trade-off achieved by the SHTCC and JHTCC schemes for TAD over a BSC in Example \ref{Excompsvsj} with parameters  $p=0.25, q=0$ for Figure \ref{fig025} and  $p=0.35,q=0$ for Figure \ref{fig035}. The red curve shows $\left(\kappa_{\alpha},\kappa_u^{\star}(\kappa_{\alpha})\right)$ pairs achieved by uncoded transmission while the blue line plots  $\left(\kappa_{\alpha},E_{\mathrm{ex}}(0)\right)$.  The joint scheme  clearly achieves a better error-exponent trade-off for values of $\kappa_{\alpha}$ below a threshold which depends on the  transition kernel of the channel. In particular, a more uniform channel results in a lesser threshold. } \label{fig:comparison1}
\end{figure}

\begin{figure}     
  \subfloat[  $p=0.25$ and $q=0.05$ ]{%
      \label{fig025045} \includegraphics[width=0.49\textwidth]{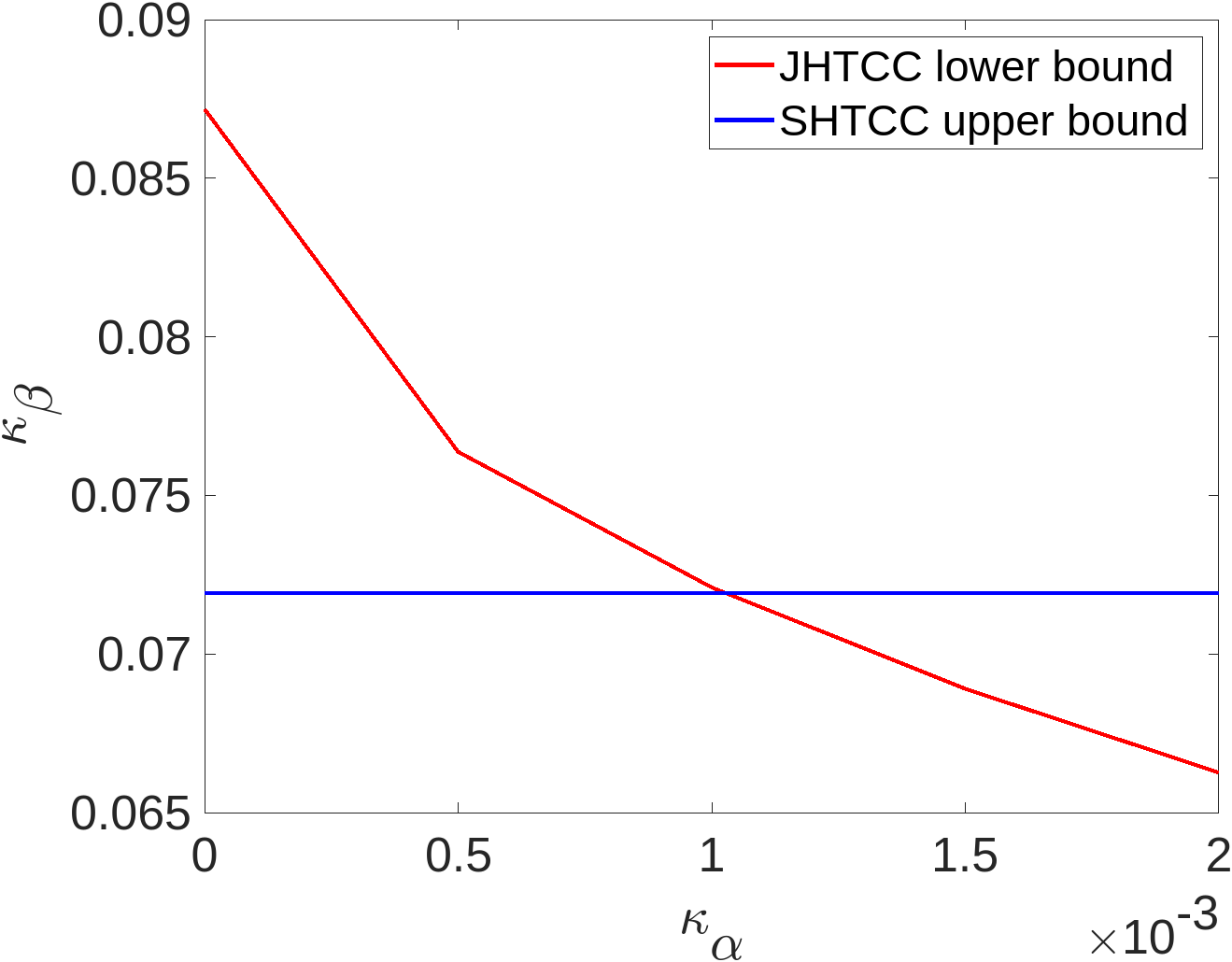}}  
          \subfloat[  $p=0.35$ and $q=0.05$  ]{\label{fig035045}
        \hspace{3 pt}\includegraphics[width=0.49\textwidth]{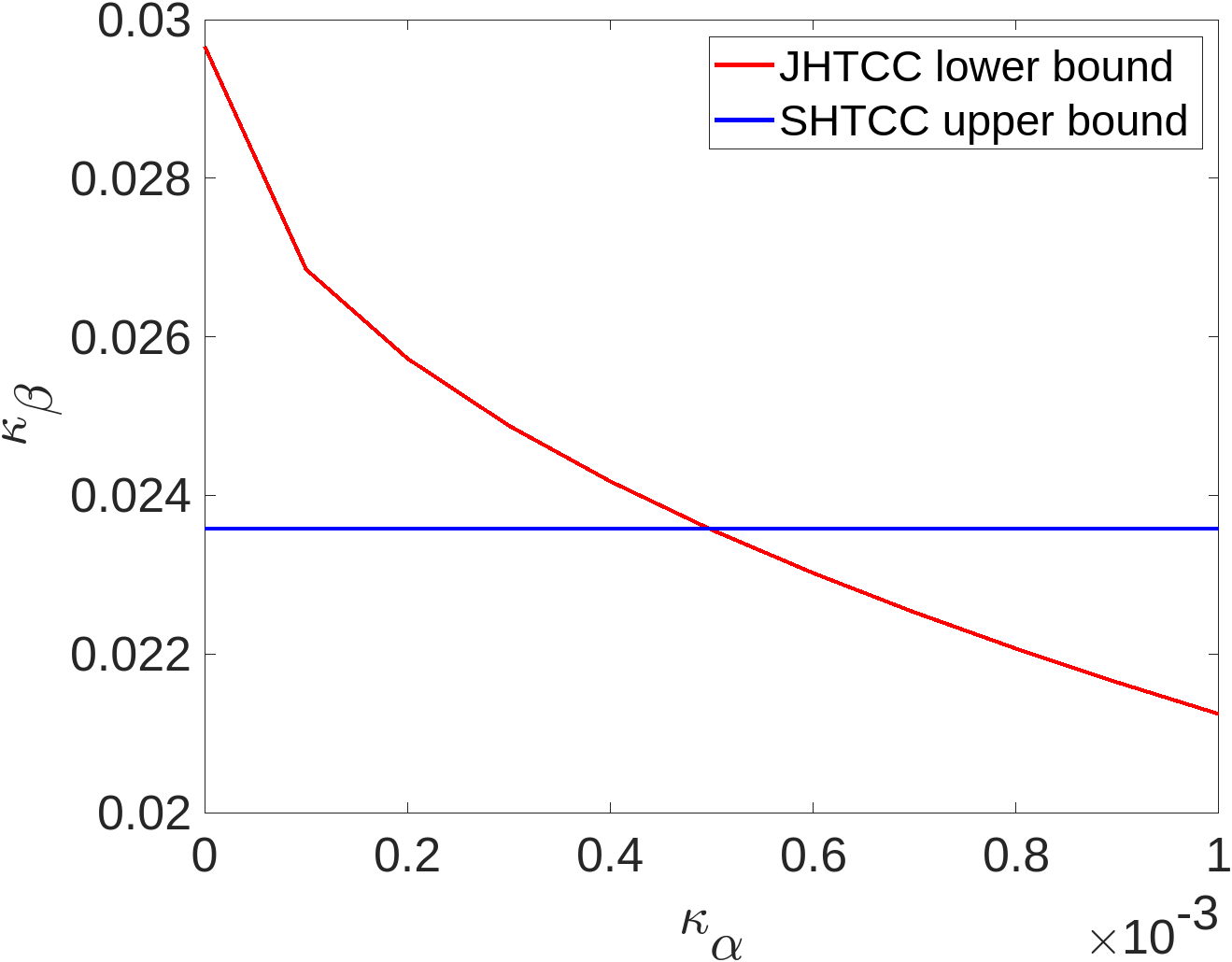}} 
           \caption{Comparison of the error-exponents trade-off achieved by the SHTCC and JHTCC schemes for Example \ref{Excompsvsj} with parameters  $p=0.25, q=0.05$ for Figure \ref{fig025045} and  $p=0.35,q=0.05$ for Figure \ref{fig035045}. The JHTCC scheme improves over the separation based scheme for small values of $\kappa_{\alpha}$; however, the region of improvement is reduced compared to Figure \ref{fig:comparison1} as the source is more uniformly distributed. } \label{fig:comparison2} 
\end{figure}

\begin{example}[Comparison of inner bounds] \label{Excompsvsj}
Let $p,q \in [0,0.5]$, $\Ucal=\V=\X=\Y=\mathcal{S}=\{0,1\}$,
\[Q_{UV}=
\begin{bmatrix}
q & 0.5-q\\ 0.5-q& q
\end{bmatrix},~~~~
P_{Y|X}=
\begin{bmatrix}
1-p & p \\p & 1-p
\end{bmatrix}, \mbox{ and }P_{UV}=Q_UQ_V.
\] 
\end{example}
A comparison of the inner bounds achieved by the SHTCC and JHTCC schemes for the above example are shown in Figures \ref{fig:comparison1} and \ref{fig:comparison2}, 
where we plot the error-exponents trade-off achieved by uncoded transmission (a lower bound for the JHTCC scheme), and the expurgated exponent at a zero rate:
\begin{align}
    E_{\mathrm{ex}}(0):=\max_{P_{SX} \in \mathcal{P}(\mathcal{S} \times \X)} E_{\mathrm{ex}}(P_{SX},0)=-0.25\log(4p(1-p)), \notag
\end{align}
which is  an upper bound on $\kappa_{\mathrm{d}}^{\star}(\kappa_{\alpha})$ for any $\kappa_{\alpha}\geq 0$. To compute   $E_{\mathrm{ex}}(0)$, we used the closed-form expression for $E_{\mathrm{ex}}(\cdot)$ given in~Problem 10.26(c) in \cite{Csiszar-Korner}. Clearly, it can be seen that the JHTCC scheme outperforms SHTCC scheme for $\kappa_{\alpha}$ below a threshold, which depends on the source  and channel distributions. In particular, the threshold below which improvement is seen is reduced when the channel or the source becomes more uniform. The former behavior can be seen directly by comparing the subplots in Figures \ref{fig:comparison1} and  \ref{fig:comparison2}, while the latter  can be noted by comparing Figure \ref{fig:comparison1}a with Figure  \ref{fig:comparison2}a, or Figure \ref{fig:comparison1}b with \mbox{Figure \ref{fig:comparison2}b.}

\section{Proofs} \label{proofs-results}

 \subsection{Proof of Theorem \ref{lbbinningts}}\label{lbbinningtsproof}
We will show the achievability of the error-exponent pair $(\kappa_{\alpha},\kappa_{\mathrm{s}}^{\star}(\kappa_{\alpha}))$ by constructing a suitable ensemble of HT codes, and showing that the expected  type I and type II error probabilities (over this ensemble)  satisfy \eqref{seqkappa2} for the pair $(\kappa_{\alpha},\kappa_{\mathrm{s}}^{\star}(\kappa_{\alpha}))$. Then, an expurgation argument\cite{Gallager-1965} will be used to show the existence of a HT code that satisfies \eqref{seqkappa2} for the same error-exponent pair, thus showing that $(\kappa_{\alpha},\kappa_{\mathrm{s}}^{\star}(\kappa_{\alpha})) \in \mathcal{R}$ as desired.

 Let $n \in \mathbb{N}$, $|\W| < \infty$, $\kappa_{\alpha}>0$, $(\omega,R, P_{SX}, \theta) \in \mathcal{L}(\kappa_{\alpha})$, $R':=\zeta(\kappa_{\alpha}, \omega)$, and $\eta>0$ be a small number. Also, suppose $R \geq 0$ satisfy
\begin{align}
&\zeta(\kappa_{\alpha}, \omega)-  \rho(\kappa_{\alpha},\omega) \leq  R <   I_P(X;Y|S), \label{chncodrateactval}
\end{align}
where $\zeta(\kappa_{\alpha}, \omega)$ and $\rho(\kappa_{\alpha},\omega) $ are defined in \eqref{totquantrate} and \eqref{totsideinfrate}, respectively. The SHTCC  scheme is as follows: \\
\textbf{Encoding:} The observer's encoder is composed of two stages, a \textit{source encoder} followed by a \textit{channel encoder}. \\
\textbf{Source encoder:} The source encoding comprises of a quantization scheme followed by binning to reduce the rate if necessary. \\
\textbf{Quantization codebook:} Let 
\begin{align}
    \mathcal{D}_n(P_U,\eta):=\big\{P_{\hat U} \in \mathcal{T}(\Ucal^n):  D(P_{\hat U}||P_U) \leq \kappa_{\alpha}+\eta\big\}. \label{quantsrcseqset}
\end{align}
Consider some ordering on the types in $ \mathcal{D}_n(P_U,\eta)$ and denote the elements as $P_{\hat U_i}$ for $i \in  \big[ | \mathcal{D}_n(P_U,\eta)|\big]$. For each type $P_{\hat U_i} \in \mathcal{D}_n(P_U,\eta)$, $i \in  \big[ | \mathcal{D}_n(P_U,\eta)|\big]$, 
choose a joint type variable $P_{\hat U_i \hat {W}_i} \in \mathcal{T}(\Ucal^n \times \W^n)$ such that
\begin{subequations}
\begin{align}
    &D\left(P_{\hat {W}_i|\hat U_i}|| P_{W_i| U }\big|  P_{\hat U_i}\right) \leq \frac{\eta}{3}, \label{condtypeclosediv}\\
    &I_P\Big(\hat U_i; \hat {W}_i\Big) \leq R'+\frac{\eta}{3}, \label{coverlemmaconst}
\end{align}
 \end{subequations}
 where  $P_{W_i|U}= \omega(P_{\hat U_i})$. Note that this is always possible for $n$ sufficiently large by definition of $R'$. 
 
 Let
 \begin{subequations}
\begin{align} \mathcal{D}_n(P_{UW},\eta)&:=\big\{P_{\hat U_i \hat {W}_i}:  i \in  \big[ | \mathcal{D}_n(P_U,\eta)|\big] \big\}, \label{deffuw}\\
R_i'&:=I_P\left(\hat U_i; \hat {W}_i\right)+ (\eta/3), i \in  \big[ | \mathcal{D}_n(P_U,\eta)|\big],    \label{quantrateeachtype}\\
\mathcal{M}_i'&:=\left[1+\sum_{k=1}^{i-1}e^{nR'_k}:\sum_{k=1}^{i}e^{nR'_k}\right],
\end{align}
 \end{subequations}
and $\mathbb{B}_{W,n}=\big\lbrace \mathbf{W}(j), 1 \leq j \leq  \sum_{i=1}^{| \mathcal{D}_n(P_U,\eta)|}|\mathcal{M}_i'| \big\rbrace$
denote a random quantization codebook such that the codeword  $\mathbf{W}(j)\sim \mathsf{Unif}\big[\mathcal{T}_n(P_{\hat {W}_i})\big]$, if  $j \in \mathcal{M}_i'$ for some $i \in  \big[ | \mathcal{D}_n(P_U,\eta)|\big]$. 
Denote a realization of $\mathbb{B}_{W,n}$ by $\mathcal{B}_{W,n}=\big\lbrace \mathbf{w}(j) \in \W^n, 1 \leq j \leq \sum_{i=1}^{| \mathcal{D}_n(P_U,\eta)|}|\mathcal{M}_i'|\big\rbrace$. \\ 
\textbf{Quantization scheme:} For a given codebook $\mathcal{B}_{W,n}$ and $\mathbf{u} \in \mathcal{T}_n\big(P_{\hat U_i}\big)$ such that  $P_{\hat U_i} \in \mathcal{D}_n(P_U,\eta)$ for some $i \in  \big[ | \mathcal{D}_n(P_U,\eta)|\big]$, let
 \begin{equation}
  \tilde M\big(\mathbf{u},\mathcal{B}_{W,n}\big) := \big\{j  \in \mathcal{M}_i':~\mathbf{w}(j) \in \mathcal{B}_{W,n},(\mathbf{u},\mathbf{w}(j)) \in \mathcal{T}_n\big(P_{\hat U_i \hat {W}_i}\big),~ P_{\hat U_i\hat {W}_i} \in  \mathcal{D}_n(P_{UW},\eta)\big\}. \notag
 \end{equation}
If $ | \tilde M(\mathbf{u},\mathcal{B}_{W,n})| \geq 1$, let $M'\left(\mathbf{u},\mathcal{B}_{W,n}\right)$ denote an index selected uniformly at random from the set $ \tilde M(\mathbf{u},\mathcal{B}_{W,n})$, otherwise, set $ M'\left(\mathbf{u},\mathcal{B}_{W,n}\right)=0$. Denoting the support of  $M'\left(\mathbf{u},\mathcal{B}_{W,n}\right)$ by $\mathcal{M}'$, 
we have for sufficiently large  $n$ that
\begin{flalign}
  |\mathcal{M}'| \leq 1+\sum_{i=1}^{| \mathcal{D}_n(P_U,\eta)|} e^{nR_i'} &\leq 1+| \mathcal{D}_n(P_U,\eta)|e^{ \underset{P_{\hat U \hat W} \in  \mathcal{D}_n(P_{UW},\eta)}{\max} n I(\hat U; \hat W)+ (n\eta/3)} \leq  e^{n (R'+\eta)},  \label{rateconstshow}  
\end{flalign}
where the last inequality uses \eqref{coverlemmaconst} and  $| \mathcal{D}_n(P_U,\eta)| \leq (n+1)^{|\Ucal|}$. \\
\textbf{Binning:} If $|\mathcal{M}'| > |\mathcal{M}|$, then the source encoder performs binning as given below.
Let $R_n:= \log \big(e^{nR}/| \mathcal{D}_n(P_U,\eta)|\big)$, $\mathcal{M}_i :=[1+(i-1)R_n: i R_n],~i \in  \big[ | \mathcal{D}_n(P_U,\eta)|\big]$, and $\mathcal{M}:=\{0\} \bigcup \big\{ \cup_{i \in [| \mathcal{D}_n(P_U,\eta)|]}\mathcal{M}_i\big\}.$ Note that 
 \begin{align}
 e^{n R_n} \geq e^{nR-|\Ucal|\log(n+1)}. \label{chncoderateact}
\end{align}
Let $f_{\mathbb{B}}$ denote the random binning function such that for each $j \in \mathcal{M}_i'$,  $f_{\mathbb{B}}(j)\sim \textsf{Unif}~[|\mathcal{M}_i|]$ for  $i \in  \big[ | \mathcal{D}_n(P_U,\eta)|\big]$, and $f_{\mathbb{B}}(0)=0$ with probability one.  
Denote a realization of $f_{\mathbb{B}}(j)$ by $f_b$, where $f_b:  \mathcal{M}' \rightarrow  \mathcal{M}$. 
 Given a codebook $\mathcal{B}_{W,n}$ and binning function $f_b$, the source encoder outputs $M=f_b\left(M'\left(\mathbf{u},\mathcal{B}_{W,n}\right)\right)$ for $\mathbf{u} \in \Ucal^n$.
 If $|\mathcal{M}'| \leq |\mathcal{M}|$, then $f_b$ is taken to be the identity map (no binning), and in this case, $M=M'\left(\mathbf{u},\mathcal{B}_{W,n}\right)$. \\
\textbf{Channel codebook:} Let  $\mathbb{B}_{X,n}:=\{\mathbf{X}(m) \in \mathcal{X}^n, m \in \mathcal{M}\}$ denote a random channel codebook generated as follows.  Without loss of generality (w.l.o.g.), denote the elements of the set $\mathcal{S}=\mathcal{X}$ as $1, \ldots, |\X|$. The codeword length $n$ is divided into $|\mathcal{S}|=|\mathcal{X}|$ blocks, where the length of the first block is $\lceil P_S(1)n \rceil$, the second block is $\lceil P_S(2)n \rceil$, so on so forth, and the length of the last block is chosen such that the total length is $n$. For $i \in [|\X|]$, let $ k_i:=\sum_{l=1}^{i-1}\lceil P_S(l)n \rceil+1$ and $\bar k_i:=\sum_{l=1}^{i} \lceil P_S(l)n \rceil$, 
 where the empty sum is defined to be zero. Let $\mathbf{s} \in \mathcal{X}^n$ be such that $s_{k_i}^{\bar k_i}=i$, i.e., the elements of $\mathbf{s}$ equal $i$ in the $i^{th}$ block for $ i \in [|\X|]$. Let $\mathbf{X}(0)=\mathbf{s}$ with probability one, and  
 the remaining 
 codewords $\mathbf{X}(m), ~m \in \mathcal{M}\backslash \{0\}$ be constant composition codewords \cite{Csiszar-Korner} selected such that 
 $X_{k_i}^{\bar k_i}(m) \sim \textsf{Unif}\big[\mathcal{T}_{\lceil P_S(i)n \rceil}(\hat P_{X| S}(\cdot|i))\big]$, where $\hat P_{X| S}$ is such that $\mathcal{T}_{\lceil P_S(i)n \rceil}\big(\hat P_{X| S}(\cdot|i)\big)$ is non-empty  and $D(\hat P_{X| S}||P_{X|S}|P_S)\leq \frac{\eta}{3}$. Denote a realization of $\mathbb{B}_{X,n}$ by $\mathcal{B}_{X,n}:=\{\mathbf{x}(m) \in \mathcal{X}^n, m \in \mathcal{M}\}$. Note that for $m \in  \mathcal{M}\backslash \{0\}$ and large $n$, the codeword pair $(\mathbf{x}(0),\mathbf{x}(m))$  has joint type (approx) $P_{\mathbf{x}(0)\mathbf{x}(j)}=\hat P_{SX}:=P_S\hat P_{X|S}$.
  \\
\textbf{Channel encoder:} For a given $\mathcal{B}_{X,n}$, the channel encoder outputs $\mathbf{x}=\mathbf{x}(m)$ for  output $m$ from the source encoder. 
 Denote this map by $f_{\mathcal{B}_{X,n}}:\mathcal{M} \rightarrow \mathcal{X}^n$. \\
\textbf{Encoder:}  Denote by $f_n:\Ucal^n \rightarrow \mathcal{P}(\mathcal{X}^n)$  the encoder  induced by all the above operations, i.e., $f_n(\cdot|\mathbf{u})=f_{\mathcal{B}_{X,n}} \circ f_b\big(M'(\mathbf{u},\mathcal{B}_{W,n})\big)$.\\
\textbf{Decision function:} The decision function  consists of three parts, a channel decoder, a source decoder and a tester. \\
\textbf{Channel decoder:} The channel decoder first performs a Neyman-Pearson test on the channel output $\mathbf{y}$ according to $\tilde{\Pi}_{\theta}:\Y^n \rightarrow \{0,1\}$, where
\begin{align}
\tilde{\Pi}_{\theta}(\mathbf{y})&:=  \ind \left(\sum_{i=1}^{n}  \log\left(\frac{P_{Y|X}(y_i|s_i)}{P_{Y|S}(y_i|s_i)}\right) \geq n \theta\right).   \label{NPtestchndec}
\end{align}
If $\tilde{\Pi}_{\theta}(\mathbf{y})=1$, then $\hat M=0$. Else,  for a given $\mathcal{B}_{X,n}$, maximum likelihood (ML) decoding is done on the remaining set of codewords $\{\mathbf{x}(m),m \in \mathcal{M}\backslash \{0\}\} $, and $\hat M$ is set equal to the ML estimate. Denote the channel decoder induced by the above  operations by $g_{\mathcal{B}_{X,n}}$, where $g_{\mathcal{B}_{X,n}}:\Y^n \rightarrow \mathcal{M}$. 

For a given codebook $\mathcal{B}_{X,n}$, the channel encoder-decoder pair described above induces a distribution
\begin{align} P^{(\mathcal{B}_{X,n})}_{\mathbf{X}\mathbf{Y}\hat M|M}(m,\mathbf{x},\mathbf{y},\hat m|m):=\ind_{\big\{f_{\mathcal{B}_{X,n}}(m)=\mathbf{x}\big\}}~P_{Y|X}^{\otimes n}(\mathbf{y}|\mathbf{x})\ind_{\big\{\hat m=g_{\mathcal{B}_{X,n}}\big\}}. \notag
\end{align}
Note that  $P_{\mathbf{x}(0)\mathbf{x}(m)}=\hat P_{SX}$, $\mathbf{Y} \sim \prod_{i=1}^{|\X|}P_{Y|X}^{\otimes \lceil P_S(i)n \rceil}(\cdot|i)$ for $M=0$ and $\mathbf{Y} \sim \prod_{i=1}^{|\X|} P_{Y|S}^{\otimes \lceil P_S(i)n \rceil}(\cdot|i)$ for $M=m \neq 0$. Then, it follows by an application of  Proposition \ref{genchernbnd} proved in Appendix \ref{App:auxres} that for any $\mathcal{B}_{X,n}$ and  $n$ sufficiently large, the Neyman-Pearson test in \eqref{NPtestchndec} yields
\begin{subequations} \label{chernbnddist}
\begin{align}
    \mathbb{P}_{P^{(\mathcal{B}_{X,n})}}\left( \hat M =0| M =m\right) &\leq e^{-n \left(   E_{\mathrm{sp}}(P_{SX}, \theta)-\eta \right)}, ~ m \in \mathcal{M}\backslash \{0\}, \label{chernbnddist1} \\
    \mathbb{P}_{P^{(\mathcal{B}_{X,n})}} \left( \hat M  \neq 0| M =0\right) &\leq e^{-n\left(E_{\mathrm{sp}}(P_{SX}, \theta) -\theta-\eta \right)}. \label{chernbnddist2}
\end{align}
\end{subequations}
Moreover, given $\hat M \neq 0$, a random coding argument over the ensemble of $\mathbb{B}_{X}^n$ (see \cite[Exercise 10.18, 10.24]{Csiszar-Korner} and \cite{Gallager-1965}) shows that there exists a deterministic codebook $\mathcal{B}_{X,n}$ such that \eqref{chernbnddist} holds, and the ML-decoding described above asymptotically achieves 
\begin{align}
    \mathbb{P}_{P^{(\mathcal{B}_{X,n})}} \left( \hat M \neq m| M =m \neq 0, \hat M \neq 0  \right) &\leq e^{-n \left(E_{\mathrm{ex}}\left(R,P_{SX}\right)-\eta\right)}. \label{chndecerrexp} 
\end{align}
This deterministic codebook $\mathcal{B}_{X,n}$ is used for channel coding.\\ 
\textbf{Source decoder:} For a given codebook $\mathcal{B}_{W,n}$ and inputs $\hat M=\hat m$ and $\mathbf{V}=\mathbf{v}$, the source decoder first decodes for the quantization codeword $\mathbf{w}(\hat m')$ (if required) using the empirical conditional entropy decoder, and then declares the output $\hat H$ of the hypothesis test  based on $\mathbf{w}(\hat m')$ and $\mathbf{v}$. 
More specifically, if binning is not performed, i.e., if $|\mathcal{M}| \geq |\mathcal{M}'|$,  $\hat M'=\hat m$. Otherwise,  $\hat{M}'=\hat m'$,  where $\hat m'= 0$ if $\hat m=0$ and  $\hat m'= \argmin_{j: f_b(j)=\hat m}~ H_e(\mathbf{w}(j)|\mathbf{v})$ otherwise.
Denote the source decoder induced by the above operations by $g_{\mathcal{B}_{W,n}}:\mathcal{M} \times \V^n \rightarrow \mathcal{M}'$.\\
 \textbf{Testing and Acceptance region:}  If $\hat m'=0$, $\hat H=1$ is declared.
 Otherwise, $\hat H=0$ or $\hat H=1$ is declared depending on whether $(\hat m', \mathbf{v}) \in \mathcal{A}_n$ or  $(\hat m', \mathbf{v}) \notin \mathcal{A}_n$, respectively, where   $\mathcal{A}_n$ denotes the acceptance region  for $H_0$ as specified next. For a given codebook $\mathcal{B}_{W,n}$,
 let $\mathcal{O}_{m'}$ denote the set of $\mathbf{u}$ such that the source encoder outputs $m'$, $m' \in \mathcal{M}'\backslash \{0\}$. For each $m' \in \mathcal{M}' \backslash \{0\}$ and $\mathbf{u} \in \mathcal{O}_{m'}$,
let  
\begin{align}
    \mathcal{Z}_{m'}(\mathbf{u})= \{\mathbf{v} \in \V^n: (\mathbf{w}(m'),\mathbf{u},\mathbf{v}) \in \mathcal{J}_n(\kappa_{\alpha}+\eta, P_{W_{m'}UV})\}, \notag
\end{align}
where $\mathcal{J}_n(r,P_X):= \{\mathbf{x} \in \mathcal{X}^n: \kl{P_{\mathbf{x}}}{P_X} \leq r \}$,  
\begin{align}
P_{UVW_{m'}}:=P_{UV}P_{W_{m'}|U} \mbox{ and } P_{W_{m'}|U}= \omega(P_{\mathbf{u}}).  \label{conddistdefn}
\end{align} 
For $m' \in  \mathcal{M}' \backslash \{0\}$, set $\mathcal{Z}_{m'}:= \{\mathbf{v}: \mathbf{v}  \in   \mathcal{Z}_{m'}(\mathbf{u}) \mbox{ for some } \mathbf{u} \in \mathcal{O}_{m'}\}$, and 
define the acceptance region for $H_0$ at the decision maker as $\mathcal{A}_n:= \cup_{m' \in \mathcal{M}' \backslash 0}~ m' \times \mathcal{Z}_{m'}$ or equivalently  as $\mathcal{A}_n^e:=  \cup_{m' \in \mathcal{M}' \backslash 0} \mathcal{O}_{m'} \times \mathcal{Z}_{m'}$. Note that $ \mathcal{A}_n$ is the same as  the acceptance region for $H_0$ in \cite[Theorem 1]{HK-1989}. Denote the decision function induced by $g_{\mathcal{B}_{X,n}} $, $g_{\mathcal{B}_{W,n}} $ and $\mathcal{A}_n$ by $g_n:\Y^n \times \V^n \rightarrow \hat {\mathcal{H}}$. \\
\textbf{Induced probability distribution:} The PMFs induced by a code  $c_n=(f_n,g_n)$ with respect to codebook $\mathcal{B}_n:=\left(\mathcal{B}_{W,n},f_b,\mathcal{B}_{X,n}\right)$ under $H_0$ and $H_1$ are
\begin{align}
&P^{(\mathcal{B}_n,c_n)}_{\mathbf{UV}M'M\mathbf{XY} \hat M \hat M'\hat H}(\mathbf{u},\mathbf{v},m',m,\mathbf{x},\mathbf{y},\hat m,\hat m',\hat h)\notag \\
 &\qquad :=P_{UV}^{\otimes n}(\mathbf{u},\mathbf{v})~\ind_{\left\{M'\left(\mathbf{u},\mathcal{B}_{W,n}\right)=m',~ f_b(m')=m\right\}} P^{(\mathcal{B}_{X,n})}_{\mathbf{X}\mathbf{Y}\hat M|M}(\mathbf{x},\mathbf{y},\hat m|m)~\ind_{\big\{g_{\mathcal{B}_{W,n}}(m,\mathbf{v})=\hat m', \mspace{1 mu}\hat h=\ind_{\left\{(\hat m',\mathbf{v}) \in \mathcal{A}_n^c \right\}}\big\}}, \notag \\[5 pt] &Q^{(\mathcal{B}_n,c_n)}_{\mathbf{UV}M'M\mathbf{XY}\hat M \hat M'\hat H}(\mathbf{u},\mathbf{v},m',m,\mathbf{x},\mathbf{y},\hat m,\hat m',\hat h)\notag \\
 &\qquad :=Q_{UV}^{\otimes n}(\mathbf{u},\mathbf{v})~\ind_{\left\{M'\left(\mathbf{u},\mathcal{B}_{W,n}\right)=m', \mspace{1 mu}f_b(m')=m\right\}} P^{(\mathcal{B}_{X,n})}_{\mathbf{X}\mathbf{Y}\hat M|M}(\mathbf{x},\mathbf{y},\hat m|m)~\ind_{\left\{g_{\mathcal{B}_{W,n}}(m,\mathbf{v})=\hat m',\mspace{1 mu}\hat h=\ind_{\left\{(\hat m',\mathbf{v}) \in \mathcal{A}_n^c \right\}}\right\}}, \notag 
\end{align}
respectively. For simplicity, we will denote the above distributions by $ P^{(\mathcal{B}_n)}$ and  $Q^{(\mathcal{B}_n)}$. Let $\mathbb{B}_n:=\left(\mathbb{B}_{W,n},f_{\mathbb{B}},\mathcal{B}_{X,n}\right)$, $\mathfrak{B}_n$, and $\mu_n$ denote the random codebook, its support, and the probability measure induced by its random construction, respectively. Also, define $\bar{\mathbb{P}}_{P^{(\mathbb{B}_n)}}:=  \mathbb{E}_{\mu_n}\big[\mathbb{P}_{P^{(\mathbb{B}_n)}}\big]$ and  $\bar{\mathbb{P}}_{Q^{(\mathbb{B}_n)}}:=  \mathbb{E}_{\mu_n}\big[\mathbb{P}_{Q^{(\mathbb{B}_n)}}\big]$.
\\
 \textbf{Analysis of the type I and type II error probabilities:} We analyze the  type I and type II error probabilities averaged over the random ensemble of quantization and binning codebooks $(\mathbb{B}_W, f_{\mathbb{B}})$. Then, an  expurgation technique \cite{Gallager-1965} guarantees the existence of a sequence of deterministic codebooks $\{\mathcal{B}_n\}_{n \in \mathbb{N}}$ and a code $\{c_n=(f_n,g_n)\}_{n \in \mathbb{N}}$  that achieves the lower bound given in Theorem \ref{lbbinningts}. \\
\textbf{Type I error probability:} In the following, random sets where the randomness is induced due to $\mathbb{B}_n$ will be written using blackboard bold letters, e.g., $\mathbb{A}_n$ for the random acceptance region for $H_0$. Note that a type I error can occur only under the following events:
\begin{enumerate}[label=(\roman*)]
\item $ \mathcal{E}_{\textrm{EE}}:= \underset{P_{\hat U} \in \mathcal{D}_n(P_U,\eta)}{\bigcup}~ \underset{\mathbf{u} \in \mathcal{T}_n(P_{\hat U})}{\bigcup}\mathcal{E}_{\textrm{EE}}(\mathbf{u})$, where
\begin{flalign}   \mathcal{E}_{\textrm{EE}}(\mathbf{u})&:= \big\{ \nexists~j \in \mathcal{M}'\backslash \{0\} \mbox{ s.t. } (\mathbf{u},\mathbf{W}(j))  \in \mathcal{T}_n\big({P_{\hat U_i\hat {W}_i}}\big), P_{\hat U_i}=P_{\mathbf{u}}, P_{\hat U_i\hat {W}_i} \in  \mathcal{D}_n(P_{UW},\eta)\big\}, \notag &&
\end{flalign}
    \item  $\mathcal{E}_{\textrm{NE}}:=\{\hat M'=M' \mbox{ and } (\hat M',\mathbf{V}) \notin \mathbb{A}_n\},$
    \item  $\mathcal{E}_{\textrm{OCE}}:=\{M' \neq 0,\hat M \neq M \mbox{ and }(\hat M',\mathbf{V}) \notin \mathbb{A}_n\},$
    \item $\mathcal{E}_{\textrm{SCE}}:=\{M'=M=0, \hat M \neq M \mbox{ and }(\hat M',\mathbf{V}) \notin \mathbb{A}_n\},$
        \item $\mathcal{E}_{\textrm{BE}}:=\{M' \neq 0,~\hat M = M, \hat M' \neq M' \mbox{ and } (\hat M',\mathbf{V}) \notin \mathbb{A}_n\}$.
\end{enumerate}
Here, $ \mathcal{E}_{\textrm{EE}}$ corresponds to the event that there does not exist a quantization codeword corresponding to at least one sequence $\mathbf{u}$ of type $P_{\mathbf{u}}  \in \mathcal{D}_n(P_U,\eta)$; $ \mathcal{E}_{\textrm{NE}}$ corresponds to the event, in which, there is neither an error at the channel decoder nor at the empirical conditional entropy decoder;  $ \mathcal{E}_{\textrm{OCE}}$ and $ \mathcal{E}_{\textrm{SCE}}$ corresponds to the case, in which, there is an error at the channel decoder (hence also at the empirical conditional entropy decoder); and, $ \mathcal{E}_{\textrm{BE}}$  corresponds to the case  that there is an error (due to binning) only at the empirical conditional entropy decoder. 
For the event $\mathcal{E}_{\textrm{EE}}$, it follows from a slight generalization of the type-covering lemma \cite[Lemma 9.1]{Csiszar-Korner} that
\begin{align}
 \bar{\mathbb{P}}_{P^{(\mathbb{B}_{n})}}(\mathcal{E}_{\textrm{EE}}) \leq e^{-e^{n\Omega(\eta)}}. \label{doubleexpdecaycov}
\end{align} 
Since $e^{n\Omega(\eta)}/n \xrightarrow{(n)} \infty$ for $\eta>0$, the event $\mathcal{E}_{\textrm{EE}}$ may be safely ignored from the analysis of the error-exponents.  
Given $\mathcal{E}_{\textrm{EE}}^c$ holds for some $\mathcal{B}_{W,n}$, it follows from  \cite[Equation 4.22]{HK-1989} that 
\begin{align}
    \bar{\mathbb{P}}_{P^{(\mathbb{B}_{n})}} \left( \mathcal{E}_{\textrm{NE}} |\mathcal{E}_{\textrm{EE}}^c\right)\leq e^{-n \kappa_{\alpha}}, \label{noerreventHK}
\end{align}
 for sufficiently large $n$ since the acceptance region is the same as that in \cite[Theorem 1]{HK-1989}.

Next, consider the event $\mathcal{E}_{\textrm{OCE}}$.  We have for sufficiently large $n$ that
\begin{flalign}
   \bar{\mathbb{P}}_{P^{(\mathbb{B}_{n})}} \left( \mathcal{E}_{\textrm{OCE}} \right)  & \leq  \bar{\mathbb{P}}_{P^{(\mathbb{B}_{n})}} \left( M' \neq 0 \right)\bar{\mathbb{P}}_{P^{(\mathbb{B}_{n})}} \left( \hat M \neq M|M' \neq 0 \right) \notag \\
   & \stackrel{(a)}{\leq}  \bar{\mathbb{P}}_{P^{(\mathbb{B}_{n})}} \left( \hat M \neq M|M \neq 0 \right) \notag \\
   &\leq  \bar{\mathbb{P}}_{P^{(\mathbb{B}_{n})}} \left( \hat M= 0|M \neq 0 \right)+ \bar{\mathbb{P}}_{P^{(\mathbb{B}_{n})}} \left( \hat M \neq  M|M \neq 0, \hat M \neq 0 \right) \notag \\
    & \stackrel{(b)}{\leq} e^{-n \left(E_{\mathrm{m}}(P_{SX}, \theta)-\eta\right)}+ e^{-n \left(E_{\mathrm{ex}}\left(R,P_{SX}\right)-\eta\right)} \notag \\
   &= e^{-n \left(  \min \left\{E_{\mathrm{m}}(P_{SX}, \theta), E_{\mathrm{ex}}\left(R,P_{SX}\right)\right\}-\eta\right)}, \label{chncoderrexptp1} &&
\end{flalign}
where 
\begin{enumerate}[label=\emph{(\alph*)}]
    \item holds since the event $\{ M' \neq 0\}$ is equivalent to $ \{M \neq 0\}$;
    \item holds due to \eqref{chernbnddist1} and \eqref{chndecerrexp}, which holds for $\mathcal{B}_{X,n}$.
\end{enumerate}
Also,  the probability of $\mathcal{E}_{\textrm{SCE}}$ can be upper bounded as
\begin{flalign}
    \bar{\mathbb{P}}_{P^{(\mathbb{B}_{n})}} \left(\mathcal{E}_{\textrm{SCE}}\right) &\leq   \bar{\mathbb{P}}_{P^{(\mathbb{B}_{n})}} \left( M' = 0 \right)\notag \\
    &\leq \bar{\mathbb{P}}_{P^{(\mathbb{B}_{n})}} \left(M'=0|\mathbf{U} \in  \mathcal{D}_n(P_U,\eta)\right)+\bar{\mathbb{P}}_{P^{(\mathbb{B}_{n})}} \big(\mathbf{U} \notin  \mathcal{D}_n(P_U,\eta)\big)\notag \\
    &= \bar{\mathbb{P}}_{P^{(\mathbb{B}_{n})}} \left(\mathcal{E}_{\textrm{EE}}\right)+\bar{\mathbb{P}}_{P^{(\mathbb{B}_{n})}} \left(\mathbf{U} \notin  \mathcal{D}_n(P_U,\eta)\right) \notag \\
    &\leq e^{-n \kappa_{\alpha}}, \label{spclmsgprobt1} &&
\end{flalign}
where \eqref{spclmsgprobt1} is due to \eqref{doubleexpdecaycov}, the definition of $\mathcal{D}_n(P_U,\eta)$ in \eqref{quantsrcseqset} and \cite[Lemma 2.2, Lemma 2.6]{Csiszar-Korner}. 

Finally, consider the event $\mathcal{E}_{\textrm{BE}}$. Note that this event occurs only when $|\mathcal{M}| \leq |\mathcal{M}'|$. Also,  $M=0$ iff $M'=0$, and hence $M'\neq 0$ and $\hat M=M$ implies that $\hat M \neq 0$. 
Let 
\begin{align}
    \mathcal{D}_n(P_{VW},\eta) 
  &:= \left\lbrace P_{\hat V \hat  W}: \begin{aligned}  &\exists~(\mathbf{w},\mathbf{u},\mathbf{v}) \in \underset{m' \in \mathcal{M}' \backslash \{0\}}{\cup}\mathcal{J}_n(\kappa_{\alpha}+\eta, P_{W_{m'}UV}), P_{W_{m'}UV} \mbox{ satisfies } \\
  &\eqref{conddistdefn} \mbox{ and }P_{\mathbf{w}\mathbf{u}\mathbf{v}}=P_{\hat  W\hat U \hat V}\end{aligned}\right\rbrace. \notag
\end{align}
We have
\begin{flalign}
    \bar{\mathbb{P}}_{P^{(\mathbb{B}_{n})}} \left(\mathcal{E}_{\textrm{BE}}\right)&= \bar{\mathbb{P}}_{P^{(\mathbb{B}_{n})}} \left( \mathcal{E}_{\textrm{BE}}, (M',\mathbf{V}) \in \mathbb{A}_n\right) +\bar{\mathbb{P}}_{P^{(\mathbb{B}_{n})}} \left( \mathcal{E}_{\textrm{BE}},(M',\mathbf{V}) \notin \mathbb{A}_n\right). \label{splittermevnt5} 
      \end{flalign} 
      The second term in \eqref{splittermevnt5} can be upper-bounded as 
      \begin{flalign}
     \bar{\mathbb{P}}_{P^{(\mathbb{B}_{n})}} \big( \mathcal{E}_{\textrm{BE}},(M',\mathbf{V}) \notin \mathbb{A}_n\big) &\leq \bar{\mathbb{P}}_{P^{(\mathbb{B}_{n})}} \big(  (M',\mathbf{V}) \notin \mathbb{A}_n, \mathcal{E}_{\textrm{EE}}\big) +\bar{\mathbb{P}}_{P^{(\mathbb{B}_{n})}} \big(  (M',\mathbf{V}) \notin \mathbb{A}_n, \mathcal{E}_{\textrm{EE}}^c\big)\notag \\
   & \leq    e^{-e^{n\Omega(\eta)}} +\bar{\mathbb{P}}_{P^{(\mathbb{B}_{n})}} \big(  (M',\mathbf{V}) \notin \mathbb{A}_n| \mathcal{E}_{\textrm{EE}}^c\big)\notag \\
      &  \leq   e^{-e^{n\Omega(\eta)}} +\bar{\mathbb{P}}_{P^{(\mathbb{B}_{n})}} \big( (\mathbf{U},\mathbf{V}) \notin \mathbb{A}_n^e\big) \notag\\
       &\leq  e^{-e^{n\Omega(\eta)}} + e^{-n\kappa_{\alpha}}, \label{hkschemet1err} &&
    \end{flalign}
    where the inequality in \eqref{hkschemet1err} follows from \cite[Equation 4.22]{HK-1989} for sufficiently large $n$, since the acceptance region $\mathbb{A}_n^e$ is the same as that in \cite{HK-1989}. To bound the first term in \eqref{splittermevnt5}, define  
$\mathcal{D}_n(P_V,\eta):= \{P_{\hat V}: \exists~ P_{\hat V \hat W} \in  \mathcal{D}_n(P_{VW},\eta)\}$, and observe that since $(M',\mathbf{V}) \in \mathbb{A}_n$ implies  $M' \neq 0$, we have
    \begin{flalign}
  & \bar{\mathbb{P}}_{P^{(\mathbb{B}_{n})}} \big( \mathcal{E}_{\textrm{BE}}, (M',\mathbf{V}) \in \mathbb{A}_n\big) \notag \\
  &=\sum_{(m',m) \in \mathcal{M}' \times  \mathcal{M}}\bar{\mathbb{P}}_{P^{(\mathbb{B}_{n})}} \big( \mathcal{E}_{\textrm{BE}}, (M',\mathbf{V}) \in \mathbb{A}_n,M=m,M'=m'\big) \notag \\
  &= \sum_{(m',m) \in \mathcal{M}' \times  \mathcal{M}}\bar{\mathbb{P}}_{P^{(\mathbb{B}_{n})}} \big(M=m,M'=m',\hat M=M\big)~ \notag \\
  & \qquad \qquad  \qquad \qquad   \bar{\mathbb{P}}_{P^{(\mathbb{B}_{n})}} \left(  \hat M' \neq M' , (\hat M',\mathbf{V}) \notin \mathbb{A}_n, (M',\mathbf{V}) \in \mathbb{A}_n\big|M'=m',M=m,\hat M=M\right) \notag \\
  &\leq   \sum_{(m',m) \in \mathcal{M}' \times  \mathcal{M}}\bar{\mathbb{P}}_{P^{(\mathbb{B}_{n})}} \big(M=m,M'=m',\hat M=M\big)~ \notag \\& \qquad \qquad  \qquad \qquad   \bar{\mathbb{P}}_{P^{(\mathbb{B}_{n})}} \left(  \hat M' \neq M' ,  (M',\mathbf{V}) \in \mathbb{A}_n\big|M'=m',M=m,\hat M=M\right) \label{termindpind} \\
  &\stackrel{(a)}{=}   \bar{\mathbb{P}}_{P^{(\mathbb{B}_{n})}} \left(  \hat M' \neq M' ,  (M',\mathbf{V}) \in \mathbb{A}_n\big|M'=1,M=1,\hat M=M\right)
\notag \\
  & \stackrel{(b)}{\leq }  \sum_{\substack{P_{\mathbf{v} } \in   \mathcal{D}_n(P_V,\eta)}} \sum_{\mathbf{v}  \in P_{\mathbf{v} }} \bar{\mathbb{P}}_{P^{(\mathbb{B}_{n})}} (\mathbf{V}=\mathbf{v}|M'=1)\notag \\
  &\qquad  \qquad    \qquad \qquad \qquad \bar{\mathbb{P}}_{P^{(\mathbb{B}_{n})}}\Big(\exists ~j  \in f_{\mathbb{B}}^{-1}(1),~ j \neq 1, H_e(\mathbf{W}(j)|\mathbf{v}) \leq H_e(\mathbf{W}(1)|\mathbf{v})\big|M'=1,\mathbf{V}=\mathbf{v}\Big), \label{secondtermbnd} &&
    \end{flalign}
where $(a)$ follows since by the symmetry of the source encoder, binning function and random codebook construction, the term in \eqref{termindpind} is independent of $(m,m')$; and $(b)$ holds since $(M',\mathbf{V}) \in \mathbb{A}_n$ implies that $P_{\mathbf{v}} \in  \mathcal{D}_n(P_V,\eta)$ and $(\mathbf{V},\mathbb{B}_W)-M'-(M,\hat M)$ form a Markov chain. 
 Defining  $P_{\hat V}=P_{\mathbf{v}}$, and the event
   $\mathcal{E}_1':=\{M'=1,\mathbf{V}=\mathbf{v}\}$,
we obtain
\begin{flalign}
& \bar{\mathbb{P}}_{P^{(\mathbb{B}_{n})}} \left(\exists ~j  \in f_{\mathbb{B}}^{-1}(1),~ j \neq 1, H_e(\mathbf{W}(j)|\mathbf{v}) \leq H_e(\mathbf{W}(1)|\mathbf{v}) ~\big| ~  \mathcal{E}_1' \right) \notag \\
& = \sum_{j \in \mathcal{M}' \backslash \{0,1\}} \bar{\mathbb{P}}_{P^{(\mathbb{B}_{n})}} \left(f_{\mathbb{B}}(j)=1, H_e(\mathbf{W}(j)|\mathbf{v}) \leq H_e(\mathbf{W}(1)|\mathbf{v}) ~\big| ~  \mathcal{E}_1' \right) \notag \\
& \stackrel{(a)}{\leq } \frac{1}{e^{nR_n}} \sum_{j \in \mathcal{M}' \backslash \{0,1\}} \bar{\mathbb{P}}_{P^{(\mathbb{B}_{n})}} \left( H_e(\mathbf{W}(j)|\mathbf{v}) \leq H_e(\mathbf{W}(1)|\mathbf{v}) ~\big| ~ \mathcal{E}_1' \right) \notag\\
& \stackrel{(b)}{\leq } \frac{1}{e^{nR_n}} \sum_{j \in \mathcal{M}' \backslash \{0,1\}} ~\sum_{\substack{P_{\hat W}:P_{\hat V \hat W} \in  \mathcal{D}_n(P_{VW},\eta)}} \sum_{\substack{\mathbf{w}:  (\mathbf{v},\mathbf{w}) \in \mathcal{T}_n(P_{\hat V \hat W})}} \bar{\mathbb{P}}_{P^{(\mathbb{B}_{n})}} \left(\mathbf{W}(1)=\mathbf{w}~\big| ~  \mathcal{E}_1' \right)~\notag \\
& \qquad \qquad \qquad \qquad \qquad \sum_{\substack{\tilde{\mathbf{w}} \in \mathcal{T}_n(P_{\hat W}):H_e(\tilde{\mathbf{w}}|\mathbf{v}) \leq   H(\hat W|\hat V) }} \bar{\mathbb{P}}_{P^{(\mathbb{B}_{n})}} \left( \mathbf{W}(j)=\tilde{\mathbf{w}}~\big| ~  \mathcal{E}_1'  \cup \{ \mathbf{W}(1)=\mathbf{w}\}\right) \notag \\
& \stackrel{(c)}{\leq } \frac{1}{e^{nR_n}} \sum_{j \in \mathcal{M}' \backslash \{0,1\}} ~\sum_{\substack{P_{\hat W}:P_{\hat V \hat W} \in  \mathcal{D}_n(P_{VW},\eta)}} \sum_{\substack{\mathbf{w}:  (\mathbf{v},\mathbf{w}) \in \mathcal{T}_n(P_{\hat V \hat W})}}\bar{\mathbb{P}}_{P^{(\mathbb{B}_{n})}} \left( \mathbf{W}(1)=\mathbf{w}~\big| ~  \mathcal{E}_1' \right) \notag \\
& \qquad \qquad \qquad \qquad \qquad \qquad \qquad \qquad \sum_{\substack{\tilde{\mathbf{w}} \in \mathcal{T}_n(P_{\hat W}):H_e(\tilde{\mathbf{w}}|\mathbf{v}) \leq   H(\hat W|\hat V) }} 2~ \bar{\mathbb{P}}_{P^{(\mathbb{B}_{n})}} \left( \mathbf{W}(j)=\tilde{\mathbf{w}}\right), \label{bndtwicrpbsideinf} &&
\end{flalign}
where 
\begin{enumerate}[label=\emph{(\alph*)}]
    \item follows since $f_{\mathbb{B}}(\cdot)$ is the uniform binning function independent of $\mathbb{B}_{W,n}$;
    \item holds due to the fact that if  $P_{\mathbf{v}} \in  \mathcal{D}_n(P_V,\eta) $, then $M'=1$ implies that $(\mathbf{W}(1),\mathbf{v}) \in \mathcal{T}_n(P_{\hat V \hat W})$ with probability one  for some $P_{\hat V \hat W} \in  \mathcal{D}_n(P_{VW},\eta)$;
    \item holds since $ \bar{\mathbb{P}}_{P^{(\mathbb{B}_{n})}} \left( \mathbf{W}(j)=\tilde{\mathbf{w}}~\big| ~  \mathcal{E}_1'\cup \{ \mathbf{W}(1)=\mathbf{w}\}\right) \leq 2~  \bar{\mathbb{P}}_{P^{(\mathbb{B}_{n})}}  \left( \mathbf{W}(j)=\tilde{\mathbf{w}}\right)$, 
which follows similarly to \cite[Equation (101)]{SD_2020}.
\end{enumerate}
 Continuing, we can write for sufficiently large $n$,
\begin{flalign}
 & \bar{\mathbb{P}}_{P^{(\mathbb{B}_{n})}} \left(\exists ~j  \in f_{\mathbb{B}}^{-1}(1),~ j \neq 1, H_e(\mathbf{W}(j)|\mathbf{v}) \leq H_e(\mathbf{W}(1)| \mathbf{v})~\big| ~  \mathcal{E}_1' \right) \notag \\
  & \stackrel{(a)}{\leq} \frac{1}{e^{nR_n}} \sum_{j \in \mathcal{M}' \backslash \{0,1\}} ~\sum_{\substack{P_{\hat W}:P_{\hat V \hat W} \in  \mathcal{D}_n(P_{VW},\eta)}} \sum_{\substack{\mathbf{w}:  (\mathbf{v},\mathbf{w}) \in \mathcal{T}_n(P_{\hat V \hat W})}} \bar{\mathbb{P}}_{P^{(\mathbb{B}_{n})}} \left( \mathbf{W}(1)=\mathbf{w}~\big| ~  \mathcal{E}_1' \right)\notag \\
  & \qquad \qquad \qquad \qquad \qquad \qquad \sum_{\substack{\tilde{\mathbf{w}} \in \mathcal{T}_n(P_{\hat W}):H_e(\tilde{\mathbf{w}}|\mathbf{v})  \leq   H(\hat W|\hat V) }} 2 ~e^{-n(H(\hat W)-\eta)} \notag \\
  & \stackrel{(b)}{\leq} \frac{1}{e^{nR_n}} \sum_{j \in \mathcal{M}' \backslash \{0,1\}} ~\sum_{\substack{P_{\hat W}:P_{\hat V \hat W} \in  \mathcal{D}_n(P_{VW},\eta)}} \sum_{\substack{\mathbf{w}:  (\mathbf{v},\mathbf{w}) \in \mathcal{T}_n(P_{\hat V \hat W})}} \bar{\mathbb{P}}_{P^{(\mathbb{B}_{n})}} \left( \mathbf{W}(1)=\mathbf{w}~\big| ~ \mathcal{E}_1' \right)~\notag \\
  & \qquad \qquad \qquad \qquad  \qquad \qquad   \qquad \qquad  \qquad \qquad \qquad \qquad \qquad  (n+1)^{|\V||\W|} e^{nH(\hat W|\hat V)} 2~ e^{-n(H(\hat W)-\eta)} \notag \\
    & \leq \frac{1}{e^{nR_n}} \sum_{j \in \mathcal{M}' \backslash \{0,1\}} ~\sum_{\substack{P_{\hat W}:P_{\hat V \hat W} \in  \mathcal{D}_n(P_{VW},\eta)}}  2~(n+1)^{|\V||\W|}~e^{-n\left(I(\hat W;\hat V)-\eta\right)}  \notag \\
 & \stackrel{(c)}{\leq} \frac{1}{e^{nR_n}} \sum_{j \in \mathcal{M}' \backslash \{0,1\}}  2~(n+1)^{|\W|}~(n+1)^{|\V||\W|}~e^{-n\left(\underset{P_{\hat V \hat W} \in \mathcal{D}_n(P_{VW},\eta) }{\min} I(\hat W;\hat V)-\eta \right)}  \notag \\
 &\stackrel{(d)}{\leq} e^{-n( R-R'+\rho_n-\eta_n')}, \label{finbndsideinfcdwrd} 
\end{flalign}
where $\rho_n:=\min_{P_{\hat V \hat W} \in   \mathcal{D}_n(P_{VW},\eta)} I(\hat V;\hat W)$ and $\eta_n':= 3\eta+ o(1)$.
In the above 
\begin{enumerate}[label=\emph{(\alph*)}]
    \item used \cite[Lemma 2.3]{Csiszar-Korner} and the fact that the codewords are chosen uniformly at random from $\mathcal{T}_n(P_{\hat W})$;
    \item follows since the total number of sequences $\tilde{\mathbf{w}} \in \mathcal{T}_n(P_{\hat  W}) $ such that $P_{\tilde{\mathbf{w}}\mathbf{v}}=P_{\tilde W \tilde V} $ and $H(\tilde W|\tilde V) \leq H(\hat W|\hat V)$ is upper bounded by $e^{n H(\hat W|\hat V)}$, and  $|\mathcal{T}(\W^n \times \V^n)| \leq (n+1)^{|\V||\W|}$;
    \item holds due to \cite[Lemma 2.2]{Csiszar-Korner};
    \item follows from  $R':=\zeta(\kappa_{\alpha}$, \eqref{chncodrateactval}, \eqref{rateconstshow} and  \eqref{chncoderateact}.
\end{enumerate}
Thus, since $\rho_n \rightarrow \rho(\kappa_{\alpha},\omega)+O(\eta) $, we have from \eqref{splittermevnt5}, \eqref{hkschemet1err}, \eqref{secondtermbnd}, \eqref{finbndsideinfcdwrd}  for  large enough $n$ that   
    \begin{align}
     \bar{\mathbb{P}}_{P^{(\mathbb{B}_{n})}} \left( \mathcal{E}_{\textrm{BE}}\right) \leq e^{-n \left( \min \left\{ \kappa_{\alpha}, R-\zeta(\kappa_{\alpha}, \omega)+\rho(\kappa_{\alpha},\omega)-O(\eta)\right\}\right)}. \label{t1errfifthevnt}
    \end{align}
By choice of $(\omega, P_{SX}, \theta) \in \mathcal{L}(\kappa_{\alpha})$, it follows from \eqref{doubleexpdecaycov},  \eqref{noerreventHK}, \eqref{chncoderrexptp1}, \eqref{spclmsgprobt1} and \eqref{t1errfifthevnt} that the type I error probability is upper bounded by $e^{-n\left(\kappa_{\alpha}-O(\eta)\right)}$ for large $n$.\\
\textbf{Type II error probability:} 
    We  analyze the type II error probability averaged over $\mathbb{B}_n$. A type II error can occur only under the following events:
  \begin{enumerate}[label=(\roman*)]
      \item
    $
      \mathcal{E}_a:=
      \left\lbrace
  \begin{array}{ll}
    & \hat M=M,\hat M'=M'\neq 0, (\mathbf{U},\mathbf{V},\mathbf{W}(M')) \in \mathcal{T}_n\left(P_{\hat U \hat V \hat W}\right) \notag \\
      &\mbox{ s.t. } P_{\hat U \hat W} \in  \mathcal{D}_n(P_{UW},\eta) \mbox{ and } P_{\hat V \hat W} \in  \mathcal{D}_n(P_{VW},\eta)\end{array}
             \right\rbrace,$\vspace{5pt}
      \item $
         \mathcal{E}_b:= \left\lbrace
                \begin{array}{ll}
                  &M' \neq 0, \hat M = M, \hat M' \neq  M', f_{\mathbb{B}}(\hat M')=f_{\mathbb{B}}( M'), (\mathbf{U},\mathbf{V},\mathbf{W}(M'), \\ &\mathbf{W}(\hat M')) \in\mathcal{T}_n\left(P_{\hat U \hat V \hat W \hat W_d}\right)\mbox{ s.t. }   P_{\hat U \hat W} \in  \mathcal{D}_n(P_{UW},\eta) ,\\&P_{\hat V \hat W_d} \in  \mathcal{D}_n(P_{VW},\eta) \mbox{  and }H_e\left(\mathbf{W}(\hat M')|\mathbf{V} \right)\leq H_e\left(\mathbf{W}(M')|\mathbf{V}\right)
                \end{array}
             \right\rbrace,$\vspace{5pt}
 \item $\mathcal{E}_c:= \left\lbrace
                \begin{array}{ll}
                  & M' \neq 0, \hat M \neq M \mbox{ or }0, (\mathbf{U}, \mathbf{V},\mathbf{W}(M'),\mathbf{W}(\hat M')) \in \mathcal{T}_n\left(P_{\hat U \hat V \hat W \hat W_d}\right) \mbox{ s.t. }  \\  & P_{\hat U \hat W} \in  \mathcal{D}_n(P_{UW},\eta) \mbox{ and } P_{\hat V \hat W_d} \in  \mathcal{D}_n(P_{VW},\eta)
                \end{array}
             \right\rbrace,$\vspace{5pt}
 \item $\mathcal{E}_d:=\left\{ M=M'=0, \hat M \neq M,  (\mathbf{V},\mathbf{W}(\hat M')) \in \mathcal{T}_n\left(P_{\hat V \hat W_d}\right) \mbox{ s.t. }P_{\hat V \hat W_d} \in  \mathcal{D}_n(P_{VW},\eta)\right\}$.
  \end{enumerate}
Similar to \eqref{doubleexpdecaycov}, it follows that
$\bar{ \mathbb{P}}_{Q^{(\mathbb{B}_{n})}}(\mathcal{E}_{\textrm{EE}}) \leq e^{-e^{n\Omega(\eta)}}$. Hence, we may assume that $\mathcal{E}_{\textrm{EE}}^c$ holds for the type II error-exponent analysis. It then follows from the analysis in \cite[Eq. 4.23-4.27]{HK-1989} that for sufficiently large $n$,  
   \begin{align}
      \bar{ \mathbb{P}}_{Q^{(\mathbb{B}_{n})}} \left(\mathcal{E}_a|\mathcal{E}_{\textrm{EE}}^c \right) \leq e^{-n \left(E_1(\kappa_{\alpha}, \omega)-O(\eta) \right)}. \notag
   \end{align}
  The analysis of the error events $\mathcal{E}_b$, $\mathcal{E}_c$ and $\mathcal{E}_d$ follows similarly to that in the proof of \cite[Theorem 2]{SD_2020}, and results in
  \begin{flalign}
    & -\frac {1}{n} \log \left(\bar{ \mathbb{P}}_{Q^{(\mathbb{B}_{n})}} \left(\mathcal{E}_b \right)\right)  \notag \\
    &\gtrsim
\begin{cases} \notag
\underset{\substack{(P_{\tilde U \tilde V \tilde W},Q_{\tilde U \tilde V \tilde W})  \in \mathcal{T}_2(\kappa_{\alpha},\omega)}}{\min} D(P_{\tilde U \tilde V \tilde W}||Q_{\tilde U \tilde V \tilde W})+E_{\mathrm{b}}(\kappa_{\alpha},\omega,R)-O(\eta), &\mbox{ if } R< \zeta(\kappa_{\alpha}, \omega)+\eta,\\
\qquad \qquad \qquad \qquad \qquad \qquad \infty, &\mbox{ otherwise},
\end{cases} \\
& = E_2(\kappa_{\alpha},\omega,R)-O(\eta).  \notag\\
    & \frac {-1}{n} \log \left(\bar{\mathbb{P}}_{Q^{(\mathbb{B}_{n})}} \left(\mathcal{E}_c \right)\right)  \notag \\
    &\gtrsim
\begin{cases} \notag
\underset{\substack{(P_{\tilde U \tilde V \tilde W},Q_{\tilde U \tilde V \tilde W})  \in \mathcal{T}_3(\kappa_{\alpha},\omega)}}{\min} D(P_{\tilde U \tilde V \tilde W}||Q_{\tilde U \tilde V \tilde W})+E_{\mathrm{b}}(\kappa_{\alpha},\omega,R) +E_{\mathrm{ex}}\left(R,P_{SX}\right)-O(\eta)  &  \mbox{ if } R< \zeta(\kappa_{\alpha}, \omega)+\eta  \\[-5 pt]
\underset{\substack{(P_{\tilde U \tilde V \tilde W},Q_{\tilde U \tilde V \tilde W})  \in \mathcal{T}_3(\kappa_{\alpha},\omega)}}{\min} D(P_{\tilde U \tilde V \tilde W}||Q_{\tilde U \tilde V \tilde W})+\rho(\kappa_{\alpha},\omega)+E_{\mathrm{ex}}\left(R,P_{SX}\right)-O(\eta) &\mbox{ otherwise}, 
\end{cases} \\
& = E_3(\kappa_{\alpha},\omega,R,P_{SX})-O(\eta).  \notag\\
    & \frac {-1}{n} \log \left(\bar{ \mathbb{P}}_{Q^{(\mathbb{B}_{n})}} \left(\mathcal{E}_d \right)\right)  \notag \\
    &\gtrsim
\begin{cases}
\underset{P_{\tilde V}: P_{\tilde V \tilde W }\in  \mathcal{D}_n(P_{VW},\eta)}{\min}
D( P_{\tilde V}||Q_V)+ E_{\mathrm{b}}(\kappa_{\alpha},\omega,R)+ E_{\mathrm{sp}}(P_{SX}, \theta) -\theta  -O(\eta) &\mbox{ if } R< \zeta(\kappa_{\alpha}, \omega)+\eta,\\[-5 pt]
\min_{\substack{P_{\tilde V}: P_{\tilde V \tilde W }\in  \mathcal{D}_n(P_{VW},\eta)}}D( P_{\tilde V}||Q_V)+\rho(\kappa_{\alpha},\omega) +E_{\mathrm{sp}}(P_{SX}, \theta) -\theta -O(\eta) &\mbox{otherwise}, 
\end{cases} \notag \\
& = E_4(\kappa_{\alpha},\omega,R,P_{SX}, \theta )-O(\eta).  \notag &&
\end{flalign} 
Since the exponent of the type II error probability is lower bounded by the minimum of the exponent of the type II error causing events, we have shown above that for a fixed $(\omega,R,P_{SX},\theta) \in    \mathcal{L}(\kappa_{\alpha})$ and sufficiently large $n$,
\begin{subequations}
\begin{align}
\bar{\mathbb{P}}_{P^{(\mathbb{B}_{n})}} \left(\hat H=1 \right)&\leq e^{-n(\kappa_{\alpha}-O(\eta))}, \label{t1errconstfin}\\
    \bar{ \mathbb{P}}_{Q^{(\mathbb{B}_{n})}} \left(\hat H=0 \right)&\leq e^{-n(\bar{\kappa}_s(\kappa_{\alpha},\omega,R,P_{SX},\theta) -O(\eta))},\label{t2errconstfin}
\end{align}
\end{subequations}
where 
\begin{align}  \bar{\kappa}_s(\kappa_{\alpha},\omega,R,P_{SX},\theta):= \min \big\{E_1(\kappa_{\alpha},\omega),E_2(\kappa_{\alpha},\omega,R),E_3(\kappa_{\alpha},\omega,R,P_{SX}),E_4(\kappa_{\alpha},\omega,R,P_{SX},\theta)  \big\}. \notag
\end{align}
\textbf{Expurgation:}
To complete the proof, we extract a deterministic codebook $\mathcal{B}_n^{\star}$ that satisfies 
\begin{align}
\mathbb{P}_{P^{(\mathcal{B}^{\star}_{n})}} \left(\hat H=1 \right)&\leq e^{-n(\kappa_{\alpha}-O(\eta))}, \notag\\  \mathbb{P}_{Q^{(\mathcal{B}^{\star}_{n})}} \left(\hat H=0 \right)&\leq e^{-n(\bar{\kappa}_s(\kappa_{\alpha},\omega,R,P_{SX},\theta) -O(\eta))}.\notag
\end{align}
For this purpose, remove a set $\mathfrak{B}_n'\subset \mathfrak{B}_n$ of highest type I error probability codebooks such that the remaining set  $\mathfrak{B}_n \backslash \mathfrak{B}_n'$ has a probability of $\tau \in (0.25,0.5)$, i.e., $\mu_n\left(\mathfrak{B}_n \backslash \mathfrak{B}_n'\right)=\tau$. Then, it follows from \eqref{t1errconstfin} and \eqref{t2errconstfin}  that for all $\mathcal{B}_n \in \mathfrak{B}_n \backslash \mathfrak{B}_n'$,
\begin{align*}
    \mathbb{P}_{P^{(\mathcal{B}_{n})}} \left(\hat H=1 \right)&\leq 2 e^{-n(\kappa_{\alpha}-O(\eta))}, \\
     \tilde{\mathbb{P}}_{ Q^{(\mathbb{B}_{n})}} \left(\hat H=0 \right)&\leq 4 e^{-n(\bar{\kappa}_s(\kappa_{\alpha},\omega,R,P_{SX},\theta) -O(\eta))},
\end{align*}
where $\tilde{\mathbb{P}}_{Q^{(\mathbb{B}_{n})}} =\frac{1}{\tau}\mathbb{E}_{\mu_n}\Big[\mathbb{P}_{Q^{\mathbb{B}_n}}\ind_{\left\{\mathbb{B}_n \in \mathfrak{B}_n \backslash \mathfrak{B}_n'\right\}}\Big]$ is a PMF. Perform one more similar expurgation step to obtain  $\mathcal{B}_n^{\star}=\big(\mathcal{B}^{\star}_{W,n},f_b^{\star},\mathcal{B}^{\star}_{X,n}\big)\in \mathfrak{B}_n \backslash \mathfrak{B}_n'$ such that for all sufficiently large $n$
\begin{align*}
\mathbb{P}_{P^{(\mathcal{B}^{\star}_{n})}} \left(\hat H=1 \right)&\leq 2 e^{-n(\kappa_{\alpha}-O(\eta))} \leq  e^{-n\big(\kappa_{\alpha}-O(\eta)-(\log 2/n)\big)}, \\
  \mathbb{P}_{ Q^{(\mathcal{B}^{\star}_{n})}} \left(\hat H=0 \right)&\leq  4 e^{-n\big(\bar{\kappa}_s(\kappa_{\alpha},\omega,R,P_{SX},\theta) -O(\eta)\big)}\leq e^{-n\big(\bar{\kappa}_s(\kappa_{\alpha},\omega,R,P_{SX},\theta) -O(\eta)-(\log 4/n)\big)}.
\end{align*}
Maximizing over $(\omega,R,P_{SX}, \theta) \in    \mathcal{L}(\kappa_{\alpha})$  and noting that $\eta>0$ is arbitrary completes the proof.

\subsection{Proof of Corollary \ref{boundwithoutbintai}} \label{boundwithoutbintaiproof}
Consider $(\omega, P_{SX}, \theta) \in \mathcal{L}^{\star}(\kappa_{\alpha})$ and $R=\zeta(\kappa_{\alpha}, \omega)$.  Then,  $(\omega, R, P_{SX}, \theta) \in \mathcal{L}(\kappa_{\alpha})$. Also, for any $(P_{\tilde U \tilde V \tilde W},Q_{\tilde U \tilde V \tilde W}) \in \mathcal{T}_1(\kappa_{\alpha}, \omega)$, we have
\begin{flalign}
   D(P_{\tilde U \tilde V \tilde W}||Q_{\tilde U \tilde V \tilde W})& =   D(P_{\tilde  U \tilde W}||Q_{\tilde U \tilde W})+ D\left(P_{\tilde V| \tilde U \tilde W}||Q_{\tilde V|\tilde U \tilde W}|P_{\tilde  U \tilde W}\right) \notag \\
   &\stackrel{(a)}{\geq}   D\left(P_{\tilde V| \tilde U \tilde W}||P_{ V}|P_{\tilde  U \tilde W}\right) \notag \\
   &=D\left(P_{\tilde V\tilde U \tilde W}||P_{ V}P_{\tilde  U \tilde W}\right) \notag \\
   & \stackrel{(b)}{\geq}  D\left(P_{\tilde V\tilde W}||P_{ V}P_{ \tilde W}\right) \notag \\
   &\stackrel{(c)}{=}    D\left(P_{\hat V\hat W}||P_{ V}P_{ \hat W}\right)  \notag\\
   &=  I_P(\hat V; \hat W) + D(P_{\hat V}||P_V),\label{firstexptaisimp} &&
\end{flalign}
where $(a)$ is due to the non-negativity of KL divergence and since $Q_{\tilde V|\tilde U \tilde W}=P_V$;
$(b)$ is because of the monotonicity of KL divergence \cite[Theorem 2.14]{Polyanskiy-Wu-book}; and
$(c)$ follows since for $(P_{\tilde U \tilde V \tilde W},Q_{\tilde U \tilde V \tilde W}) \in \mathcal{T}_1(\kappa_{\alpha}, \omega)$,  $P_{\tilde V \tilde W}= P_{\hat V \hat W}$ for some $P_{\hat U \hat V \hat W} \in   \hat{\mathcal{L}}(\kappa_{\alpha},\omega) $. Minimizing over all $P_{\hat U \hat V \hat W} \in \hat{\mathcal{L}}(\kappa_{\alpha},\omega) $ yields that 
\begin{flalign}  E_1(\kappa_{\alpha},\omega)&=\min_{\substack{(P_{\tilde U \tilde V \tilde W},Q_{\tilde U \tilde V \tilde W})  \in \mathcal{T}_1(\kappa_{\alpha},\omega)}} D(P_{\tilde U \tilde V \tilde W}||Q_{\tilde U \tilde V \tilde W}) \notag \\
   &\geq \min_{P_{\hat  U \hat V \hat W} \in \hat{\mathcal{L}}(\kappa_{\alpha},\omega)} \left[I_P(\hat V; \hat W) + D(P_{\hat V}||P_V) \right]\notag \\
   &= \min_{\substack{P_{\hat V \hat W}:P_{\hat  U \hat V \hat W} \in \hat{\mathcal{L}}(\kappa_{\alpha},\omega)}} \left[I_P(\hat V; \hat W) + D(P_{\hat V}||P_V) \right]\notag
   :=  E_1^{\mathrm{i}}(\kappa_{\alpha},\omega),\notag &&
\end{flalign}
where the inequality above follows from \eqref{firstexptaisimp}.
Next, since $\zeta(\kappa_{\alpha}, \omega) =  R$, we have that $E_2(\kappa_{\alpha},\omega,R)=\infty$. Also, by  non-negativity of KL divergence
\begin{flalign}
E_3(\kappa_{\alpha},\omega,R,P_{SX}) &= \underset{\substack{(P_{\tilde U \tilde V \tilde W},Q_{\tilde U \tilde V \tilde W}) \\ \in \mathcal{T}_3(\kappa_{\alpha},\omega)}}{\min} D(P_{\tilde U \tilde V \tilde W}||Q_{\tilde U \tilde V \tilde W}) +\rho(\kappa_{\alpha},\omega) +E_{\mathrm{ex}}\left(R,P_{SX}\right) \notag \\
&\geq  \rho(\kappa_{\alpha}, \omega)+E_{\mathrm{ex}}\left(\zeta(\kappa_{\alpha}, \omega),P_{SX}\right):=  E_2^{\mathrm{i}}(\kappa_{\alpha},\omega,P_{SX}), \notag \\
E_4(\kappa_{\alpha},\omega,P_{SX}, \theta)&=
\underset{P_{\hat V}: P_{\hat U \hat V \hat W} \in  \hat{\mathcal{L}}(\kappa_{\alpha},\omega)}{\min} D( P_{\hat V} ||P_V)+ \rho(\kappa_{\alpha},\omega)+E_{\mathrm{m}}(P_{SX}, \theta)-\theta \notag \\
& =\rho(\kappa_{\alpha},\omega)+E_{\mathrm{m}}(P_{SX}, \theta)-\theta := E_3^{\mathrm{i}}(\kappa_{\alpha},\omega,P_{SX}, \theta), \notag  &&
\end{flalign}
 where  the final equality is since $P_{UV}P_{W|U} \in \hat{\mathcal{L}}(\kappa_{\alpha},\omega)$ for $P_{W|U}:=\omega(P_U)$. The claim in \eqref{bndtaigenexptrd} now follows from Theorem \ref{lbbinningts}.

Next, we prove \eqref{Steinregimetai}. Note that $    \hat{\mathcal{L}}(0,\omega)=\{ P_{UVW}=P_{UV}P_{W|U}: P_{W|U}=\omega(P_U)\}$
  and $\mathcal{L}^{\star}(0)=   \{   (\omega, P_{SX}, \theta) \in \mathfrak{F} \times  \mathcal{P}(\mathcal{S}\times \X) \times \Theta(P_{SX}):~   I_P( U;W) <  I_P(X;Y|S),  P_{W|U}=\omega(P_U),~ P_{SXY}:=P_{SX}P_{Y|X}\}$
since $E_{\mathrm{sp}}(P_{SX}, \theta) \geq 0$ and  $E_{\mathrm{ex}}\left(I_P( U;W),P_{SX}\right) \geq 0$. Hence,  we have 
\begin{align}
   E_1^{\mathrm{i}}(0,\omega) &\geq   \min_{P_{\hat  U \hat V \hat W} \in \hat{\mathcal{L}}(0,\omega)} I_P(\hat V; \hat W) = I_P( V; W). \notag
\end{align}
Also, $\rho(0,\omega) = I_P(V;W)$, $E_2^{\mathrm{i}}(0,\omega,P_{SX}) \geq   \rho(0,\omega)$ and $E_3^{\mathrm{i}}(0,\omega,P_{SX},\theta) \geq   \rho(0,\omega)$.
By choosing $P_{XS}=P^{\star}_XP_S$ where $P^{\star}_X$ is the capacity achieving input distribution, we have $I_P(X;Y|S)=C$. Then,  it follows from \eqref{bndtaigenexptrd} and the continuity of $ E_1^{\mathrm{i}}(\kappa_{\alpha},\omega)$, $E_2^{\mathrm{i}}(\kappa_{\alpha},\omega,P_{SX})$ and $E_3^{\mathrm{i}}(\kappa_{\alpha},\omega,P_{SX},\theta)$ in $\kappa_{\alpha}$ that 
$ \lim_{\kappa_{\alpha} \rightarrow 0} \kappa(\kappa_{\alpha}) \geq \kappa_{\mathrm{i}}^{\star}(0)$. On the other hand, $\lim_{\kappa_{\alpha} \rightarrow 0} \kappa(\kappa_{\alpha}) \leq \kappa_{\mathrm{i}}^{\star}(0)$
follows from the converse proof in \cite[Proposition 7 ]{SD_2020}. The proof of the cardinality bound $|\W| \leq |\Ucal|+1$ follows from a standard application of the Eggleston-Fenchel-Carath{\'e}odory Theorem~\cite[Theorem 18]{Eggleston_Convexity1958}, thus completing the proof.

\subsection{Proof of Corollary \ref{innbndtad}}\label{innbndtadproof}
 Specializing Theorem \ref{lbbinningts} to TAD, note that $\rho(\kappa_{\alpha},\omega)=0$ since $P_{\hat U \hat V \hat W}=Q_U Q_V P_{\hat W|\hat U} \in \hat{\mathcal{L}}(\kappa_{\alpha},\omega)$ and $I_{P}(\hat V;\hat W)=0$. Also, for  $R \geq \zeta(\kappa_{\alpha}, \omega)$, $E_{\mathrm{b}}(\kappa_{\alpha},\omega,R)=\infty$.  Hence, 
\begin{align}
  &  \mathcal{L}(\kappa_{\alpha})\mspace{-3 mu}=\mspace{-3 mu}\left\{
(\omega,R, P_{SX}, \theta) : \begin{aligned}
& \zeta(\kappa_{\alpha}, \omega)\leq   R <  I_P(X;Y|S),P_{SXY}=P_{SX}P_{Y|X},       \\
& \min \left\lbrace E_{\mathrm{sp}}(P_{SX}, \theta), E_{\mathrm{ex}}\left(R,P_{SX}\right) \right\rbrace \geq \kappa_{\alpha}  
\end{aligned}
\right\}, \notag \\
 & \hat{\mathcal{L}}(\kappa_{\alpha},\omega):=\left\{
     P_{\hat U\hat V \hat W}:~ \kl{P_{\hat U\hat V \hat W}}{P_{UV\hat W}} \leq \kappa_{\alpha},  P_{\hat W|\hat U}= \omega(P_{\hat U}), P_{UV\hat W}=Q_UQ_VP_{\hat W|\hat U}
\right\}. \notag
\end{align}
Then, we have 
\begin{flalign} E_1(\kappa_{\alpha},\omega):=E_1^{\mathrm{d}}(\kappa_{\alpha},\omega)&:=\min_{\substack{(P_{\tilde U \tilde V \tilde W},Q_{\tilde U \tilde V \tilde W})  \in \mathcal{T}_1(\kappa_{\alpha},\omega)}} D(P_{\tilde U \tilde V \tilde W}||Q_{\tilde U \tilde V \tilde W}) \notag \\
   &\stackrel{(a)}{\geq} \min_{\substack{(P_{\tilde U \tilde V \tilde W},Q_{\tilde U \tilde V \tilde W})  \in \mathcal{T}_1(\kappa_{\alpha},\omega)}} D(P_{\tilde V \tilde W}||Q_{\tilde V \tilde W}) \notag \\
   & \stackrel{(b)}{=} \min_{\substack{(P_{\hat V \hat W},Q_{V \hat W}):P_{\hat U \hat V \hat W} \in \hat{\mathcal{L}}(\kappa_{\alpha},\omega), \\Q_{UV \hat W}=Q_{UV}P_{\hat W|\hat U} }} D(P_{\hat V \hat W}||Q_{V \hat W}),  \label{firstexpfacttad} &&
\end{flalign}
where $(a)$  follows due to the data processing inequality for KL divergence \cite[Theorem 2.15]{Polyanskiy-Wu-book}; and $(b)$ is since $(P_{\tilde U \tilde V \tilde W},Q_{\tilde U \tilde V \tilde W})  \in \mathcal{T}_1(\kappa_{\alpha},\omega)$ implies that $P_{ \tilde V \tilde W}=P_{\hat V \hat W}$ and $Q_{\tilde U \tilde V \tilde W}=Q_{UV}P_{\hat W|\hat U} $ for some $P_{\hat U \hat V \hat W} \in  \hat{\mathcal{L}}(\kappa_{\alpha},\omega)$. 
Next, note that  since $R \geq \zeta(\kappa_{\alpha}, \omega)$, $E_2(\kappa_{\alpha},\omega,R)=\infty$. Also,
\begin{subequations}
\begin{flalign}
E_3(\kappa_{\alpha},\omega,R,P_{SX}) &= \underset{\substack{(P_{\tilde U \tilde V \tilde W},Q_{\tilde U \tilde V \tilde W})  \in \mathcal{T}_3(\kappa_{\alpha},\omega)}}{\min} D(P_{\tilde U \tilde V \tilde W}||Q_{\tilde U \tilde V \tilde W})  +E_{\mathrm{ex}}\left(R,P_{SX}\right) \label{thirdfactsimpfurth}\\
     & \stackrel{(a)}{=}  E_{\mathrm{ex}}\left(R,P_{SX}\right), \label{expurgpostad} \\
E_4(\kappa_{\alpha},\omega,P_{SX}, \theta)&=
\underset{P_{\hat V}: P_{\hat U \hat V \hat W} \in  \hat{\mathcal{L}}(\kappa_{\alpha},\omega)}{\min} D( P_{\hat V} ||Q_V)+E_{\mathrm{m}}(P_{SX}, \theta)-\theta \notag \\
&\stackrel{(b)}{=}E_{\mathrm{m}}(P_{SX}, \theta)-\theta=:E_3^{\mathrm{d}}(P_{SX},\theta), \label{specpostad}&&
\end{flalign}
\end{subequations}
where 
\begin{enumerate}[label=\emph{(\alph*)}]
    \item is obtained by taking $P_{\hat U \hat V \hat W}=Q_U Q_VP_{W|U} \in \hat{\mathcal{L}}(\kappa_{\alpha},\omega)$ and $P_{W|U}=\omega(Q_U)$ in the definition of $\mathcal{T}_3(\kappa_{\alpha},\omega)$. This  implies that $(P_{\tilde U \tilde V \tilde W},Q_{\tilde U \tilde V \tilde W})=(Q_{UV}P_{W|U},Q_{UV}P_{W|U}) \in \mathcal{T}_3(\kappa_{\alpha},\omega)$, and hence that the first term in the right hand side (RHS) of \eqref{thirdfactsimpfurth} is zero;
    \item is due to $Q_{U}Q_VP_{W|U} \in  \hat{\mathcal{L}}(\kappa_{\alpha},\omega)$ for $P_{W|U}=\omega(Q_U)$.
\end{enumerate}
Since $E_{\mathrm{ex}}\left(R,P_{SX}\right)$ is a non-increasing function of $R$ and $R \geq \zeta(\kappa_{\alpha}, \omega)$, selecting $R=\zeta(\kappa_{\alpha}, \omega)$ maximizes $ E_3(\kappa_{\alpha},\omega,R,P_{SX})$. Then, \eqref{bndtadexpgen} follows from \eqref{firstexpfacttad}, \eqref{expurgpostad} and \eqref{specpostad}.

Next, we  prove \eqref{steinbndtadchar}. Note that $\zeta(0, \omega)=I_Q(U;W)$, where $Q_{UW}=Q_UP_{W|U}$, $P_{W|U}=\omega(Q_U)$, and  since $E_{\mathrm{sp}}(P_{SX},\theta)\geq 0$ and  $E_{\mathrm{ex}}\left(I_Q(U;W),P_{SX}\right) \geq 0$,
\begin{flalign}
\mathcal{L}^{\star}(0)&= \left\{
\begin{aligned}
&(\omega, P_{SX}, \theta) \in \mathfrak{F} \times  \mathcal{P}(\mathcal{S}\times \X) \times \Theta(P_{SX}):~  I_Q(U;W) <   I_P(X;Y|S),\\&Q_{UVW}=Q_{UV}P_{W|U}, ~P_{W|U}=\omega(Q_U),  P_{SXY}:=P_{SX}P_{Y|X}
\end{aligned}
\right\}. \notag 
\end{flalign}
 Also, $\hat{\mathcal{L}}(0,\omega)=\big\{ Q_UQ_VP_{W|U}:  P_{W|U}= \omega(Q_U)\big\}$.
By choosing  $\theta=-\theta_{\mathrm{l}}(P_{SX})$ (defined above \eqref{lhatsetdef}) that maximizes $E_3^{\mathrm{d}}(P_{SX},\theta)$, we have
\begin{subequations} \label{expfactall}
\begin{flalign}
 & E_1^{\mathrm{d}}(0,\omega) \geq \min_{\substack{(P_{\hat V \hat W},Q_{V \hat W}):~P_{\hat U \hat V \hat W} \in \hat{\mathcal{L}}(0,\omega), \\Q_{UV \hat W}=Q_{UV}P_{\hat W|\hat U} }} D(P_{\hat V \hat W}||Q_{V \hat W})= \min_{\substack{(P_{W|U},P_{SX}):~I_Q(U;W)\leq I_P(X;Y|S), \\Q_{ U V W}=Q_{UV}P_{ W| U},P_{SXY}=P_{SX}P_{Y|X} }} D(Q_V Q_W||Q_{ V W}),\label{firexpsingtad} \\
 &  E_2^{\mathrm{d}}(0,\omega,P_{SX}) =E_{\mathrm{ex}}\left(I_Q(U;W),P_{SX}\right),\label{secexpsingtad}\\
&E_3^{\mathrm{d}}(P_{SX},-\theta_{\mathrm{l}}(P_{SX}))=E_{\mathrm{m}}(P_{SX}, -\theta_{\mathrm{l}}(P_{SX}))+\theta_{\mathrm{l}}(P_{SX})=\theta_{\mathrm{l}}(P_{SX}),\label{thirdexpsingtad} 
\end{flalign}
\end{subequations}
where  \eqref{thirdexpsingtad} is due to $E_{\mathrm{m}}(P_{SX}, -\theta_{\mathrm{l}}(P_{SX}))=0$. The latter  in turn follows  similar to \eqref{equshowdiv} and \eqref{equshowdiv2} from the definition of $E_{\mathrm{m}}(\cdot,\cdot)$. From \eqref{bndtadexpgen}, \eqref{expfactall}, and the continuity of $E_1^{\mathrm{d}}(\kappa_{\alpha},\omega)$, $E_2^{\mathrm{d}}(\kappa_{\alpha},\omega,P_{SX})$  in $\kappa_{\alpha}$, \eqref{steinbndtadchar} follows.  The proof of the cardinality bound $|\W| \leq |\Ucal|+1$ in the RHS of \eqref{firexpsingtad} follows via a standard application of the Eggleston-Fenchel-Carath{\'e}odory Theorem~\cite[Theorem 18]{Eggleston_Convexity1958}. To see this, note that it is sufficient to preserve $\{Q_U(u), u \in \Ucal\}$, $D(Q_V Q_W||Q_{ V W})$ and $H_Q(U|W)$,  all of which can be written as a linear combination of functionals of $Q_{U|W}(\cdot|w)$ with weights $Q_W(w)$. Thus, it requires $|\Ucal|-1$ points to preserve $\{Q_U(u), u \in \Ucal\}$ and one each for $D(Q_V Q_W||Q_{ V W})$ and $H_Q(U|W)$. This completes the proof.
\subsection{Proof of Theorem \ref{jhtccthm}} \label{jhtccthmproof}
We will show that the error-exponent pairs $\big(\kappa_{\alpha}, \kappa_{\mathrm{h}}^{\star}(\kappa_{\alpha})\big)$ and $\big(\kappa_{\alpha}, \kappa_{\mathrm{u}}^{\star}(\kappa_{\alpha})\big)$ are  achieved by a hybrid coding scheme and uncoded transmission scheme, 
respectively. First we describe the hybrid coding scheme.

 Let $n \in \mathbb{N}$, $|\W| < \infty$, $\kappa_{\alpha}>0$, and $\big(P_{S}, \omega'(\cdot,P_{S}), P_{X|USW},$ $P_{X'|US}\big) \in \mathcal{L}_{\mathrm{h}}(\kappa_{\alpha})$. Further, let $\eta>0$ be a small number, and choose a sequence $\mathbf{s} \in \mathcal{T}_n\left(P_{\hat S}\right)$, 
 where $P_{\hat S}$ satisfies $D\left(P_{\hat S}||P_S\right) \leq \eta$. Set $R':=\zeta'(\kappa_{\alpha}, \omega',P_{\hat S})$. \\
\textbf{Encoding:} The encoder performs type-based quantization followed by  hybrid coding \cite{Lim-minero-kim-2015}. The details are as follows:\\
\textbf{Quantization codebook:} Let $\mathcal{D}_n(P_U,\eta)$ be as defined in \eqref{quantsrcseqset}.
Consider some ordering on the types in $  \mathcal{D}_n(P_U,\eta)$ and denote the elements as $P_{\hat U_i}$, $ i \in  \big[|\mathcal{D}_n(P_U,\eta)|\big]$.
For each  joint type $P_{\hat S\hat U_i}$ such that $P_{\hat U_i} \in   \mathcal{D}_n(P_U,\eta)$ and  $\hat S$ is independent of $\hat U_i$,
choose a joint type variable $P_{\hat S \hat U_i\hat {W}_i}$, $P_{\hat {W}_i} \in \mathcal{T}(\W^n)$, such that $D\big(P_{\hat {W}_i|\hat U_i\hat S}|| P_{W_i| U\hat S}\big| P_{\hat U_i\hat S} \big) \leq \eta/3$ and $I(\hat S, \hat U_i; \hat {W}_i) \leq R'+(\eta/3)$, 
where $P_{W_i|U,S}= \omega'(P_{\hat U_i},P_{\hat S})$. 
Define $\mathcal{D}_n(P_{SUW}, \eta):=\big\{P_{\hat S\hat U_i\hat {W}_i}:  i \in  \big[|\mathcal{D}_n(P_U,\eta)|\big] \big\}$, $R_i':=I_P(\hat S,\hat U_i; \hat {W}_i)+ (\eta/3)$ for $  i \in\big[|\mathcal{D}_n(P_U,\eta)|\big]$ and $\mathcal{M}_i':=\big[1+\sum_{m=1}^{i-1}e^{nR'_m}:\sum_{m=1}^{i}e^{nR'_m}\big], $ $i \in  \big[|\mathcal{D}_n(P_U,\eta)|\big]$. 
 Let $\mathbb{B}_{W,n}=\big\lbrace \mathbf{W}(j) \in \W^n, 1 \leq j \leq \sum_{i=1}^{| \mathcal{D}_n(P_U,\eta)|}e^{nR_i'} \big\rbrace$ denote a random  quantization codebook such that for $ i \in  \big[|\mathcal{D}_n(P_U,\eta)|\big]$, each codeword $\mathbf{W}(j)$, $j \in \mathcal{M}_i'$, is independently selected from $\mathcal{T}_n(P_{\hat {W}_i})$ according to uniform distribution, i.e., $\mathbf{W}(j) \sim~ \textsf{Unif}\big[\mathcal{T}_n(P_{\hat {W}_i})\big]$.
 Let $\mathcal{B}_{W,n}$ denote a realization of $\mathbb{B}_{W,n}$.\\
\textbf{Type-based hybrid coding:} For $\mathbf{u} \in \mathcal{T}_n\big(P_{\hat U_i}\big)$ such that  $P_{\hat U_i} \in \mathcal{D}_n(P_U,\eta)$ for some $i \in  \big[ | \mathcal{D}_n(P_U,\eta)|\big]$, let
 \begin{equation}
  \bar M\left(\mathbf{u},\mathcal{B}_{W,n}\right) \mspace{-2 mu}:= \mspace{-2 mu}\Big\{j  \in \mathcal{M}_i':~\mathbf{w}(j) \in \mathcal{B}_{W,n},~(\mathbf{s},\mathbf{u},\mathbf{w}(j)) \in \mathcal{T}_n(P_{\hat S\hat U_i\hat {W}_i}),~ P_{\hat S\hat U_i\hat {W}_i} \in  \mathcal{D}_n(P_{SUW}, \eta)\Big\}. \notag
 \end{equation}
If $ | \bar M\left(\mathbf{u},\mathcal{B}_{W,n}\right)| \geq 1$, let $ M'\left(\mathbf{u},\mathcal{B}_{W,n}\right)$ denote an index selected uniformly at random from the set $  \bar M\left(\mathbf{u},\mathcal{B}_{W,n}\right)$, otherwise, set $ M'\left(\mathbf{u},\mathcal{B}_{W,n}\right)=0$.  
Given $\mathcal{B}_{W,n}$ and $\mathbf{u} \in \Ucal^n$, the quantizer outputs $M'= M'\left(\mathbf{u},\mathcal{B}_{W,n}\right)$, where the support of $M'$ is $\mathcal{M}':=\{0\} \bigcup_{i=1}^{| \mathcal{D}_n(P_U,\eta)|}\mathcal{M}'_i$.
Note that for sufficiently large  $n$, it follows similarly to \eqref{rateconstshow} that $|\mathcal{M}'|\leq  e^{n (R'+\eta)}$. For a given $\mathcal{B}_{W,n}$ and $\mathbf{u} \in \Ucal^n$,  the encoder transmits $\mathbf{X}\sim  P_{X|USW}^{\otimes n}(\cdot|\mathbf{u},\mathbf{s},\mathbf{w}(m'))$ if $M'=m' \neq 0$, and   $\mathbf{X}' \sim  P_{X'|US}^{\otimes n}(\cdot|\mathbf{u},\mathbf{s})$ if $M'=0$.\\
\textbf{Acceptance region:} For a given codebook $\mathcal{B}_{W,n}$ and $m' \in \mathcal{M}'\backslash \{0\}$,
 let $\mathcal{O}_{m'}$ denote the set of $\mathbf{u}$ such that $M'\left(\mathbf{u},\mathcal{B}_{W,n}\right)=m'$. For each $m' \in \mathcal{M}' \backslash \{0\}$ and $\mathbf{u} \in \mathcal{O}_{m'}$,
set 
\begin{align}
   \mathcal{Z}'_{m'}(\mathbf{u})= \Big\{(\mathbf{v},\mathbf{y}) \in \V^n \times \Y^n: (\mathbf{s},\mathbf{u},\bar{\mathbf{w}}_{m'},\mathbf{v},\mathbf{y}) \in \mathcal{J}_n\big(\kappa_{\alpha}+\eta, P_{\hat SUW_{m'}VY}\big)\Big\}, \notag
\end{align}
where recall that $\mathcal{J}_n(r,P_X):= \{\mathbf{x} \in \mathcal{X}^n: \kl{P_{\mathbf{x}}}{P_X} \leq r \}$, and 
\begin{subequations}\label{conddistdefnh1}
   \begin{equation}
       P_{\hat SUW_{m'}VXY}=P_{\hat S}P_{UV}P_{W_{m'}|U \hat S}P_{X|U \hat S W_{m'}}P_{Y|X},~
   \end{equation}
   \begin{equation}
       P_{W_{m'}|U \hat S}= \omega'(P_{\mathbf{u}},P_{\hat S}) \mbox{ and }P_{X|U \hat S W_{m'}}=P_{X|U S W}.
   \end{equation}
\end{subequations}
For $m' \in  \mathcal{M}' \backslash \{0\}$, define
$ \mathcal{Z}'_{m'}:= \{(\mathbf{v},\mathbf{y}): (\mathbf{v},\mathbf{y})  \in  \mathcal{Z}'_{m'}(\mathbf{u}) \mbox{ for some } \mathbf{u} \in \mathcal{O}_{m'}\}.$
The acceptance region for $H_0$ is given by $  \mathcal{A}_n:= \cup_{m' \in \mathcal{M}' \backslash 0}~ \mathbf{s} \times m' \times\mathcal{Z}'_{m'}$
or equivalently  as $\mathcal{A}_n^e:= \cup_{m' \in \mathcal{M}' \backslash 0} ~\mathbf{s} \times  \mathcal{O}_{m'} \times\mathcal{Z}'_{m'}.$ \\
\textbf{Decoding:} Given codebook $\mathcal{B}_{W,n}$, $\mathbf{Y}=\mathbf{y}$, and $\mathbf{V}=\mathbf{v}$,  if $( \mathbf{v},\mathbf{y}) \in  \bigcup_{m' \in \M' \backslash \{0\} }\mathcal{Z}'_{m'}$, then $\hat M'=\hat m'$, where $\hat m':=\argmin_{j \in \mathcal{M}'\backslash 0} H_e(\mathbf{w}(j)|\mathbf{v},\mathbf{y},\mathbf{s})$. Otherwise, $\hat M'=0$. Denote the decoder induced by the above operations by $g_{\mathcal{B}_{W,n}}:\mathcal{S}^n \times \V^n \times \Y^n \rightarrow \mathcal{M}'$.\\
\textbf{Testing:} If $\hat M'=0$, $\hat H=1$ is declared.
 Otherwise, $\hat H=0$ or $\hat H=1$ is declared depending on whether $(\mathbf{s}, \hat m', \mathbf{v},\mathbf{y}) \in \mathcal{A}_n$ or  $(\mathbf{s}, \hat m', \mathbf{v},\mathbf{y}) \notin \mathcal{A}_n$, respectively. Denote the decision function induced by $g_{\mathcal{B}_{W,n}}$ and $\mathcal{A}_n$ by $g_n:\mathcal{S}^n \times  \V^n \times \Y^n \rightarrow \hat{\mathcal{H}}$.\\
\textbf{Induced probability distribution:}
The PMFs induced by a code  $c_n=(f_n,g_n)$ with respect to codebook $\mathcal{B}_{W,n}$ under $H_0$ and $H_1$ are
\begin{flalign}
&P^{(\mathcal{B}_{W,n},c_n)}_{\mathbf{UV}M'\mathbf{XY}  \hat M'\hat H}(\mathbf{u},\mathbf{v},m',\mathbf{x},\mathbf{y},\hat m',\hat h) \notag \\
 &:=
 \begin{cases}
   P_{UV}^{\otimes n}(\mathbf{u},\mathbf{v})~\ind_{\left\{M'\left(\mathbf{u},\mathcal{B}_{W,n}\right)=m'\right\}} ~ P_{X|USW}^{\otimes n}(\mathbf{x}|\mathbf{s},\mathbf{u},\mathbf{w}(m'))~P_{Y|X}^{\otimes n}(\mathbf{y}|\mathbf{x})~\\
   \qquad \qquad  \qquad \qquad \qquad \qquad\qquad \qquad \ind_{\big\{g_{\mathcal{B}_{W,n}}(\mathbf{v},\mathbf{y},\mathbf{s})=\hat m'\big\}}~\ind_{\Big\{\hat h=\ind_{\left\{(\mathbf{s}, \hat m',\mathbf{v},\mathbf{y}) \in \mathcal{A}_n^c \right\}}\Big\}},  & \mbox{ if } m' \neq 0, \notag \\
    P_{UV}^{\otimes n}(\mathbf{u},\mathbf{v})~\ind_{\left\{M'\left(\mathbf{u},\mathcal{B}_{W,n}\right)=m'\right\}} ~ P_{X'|US}^{\otimes n}(\mathbf{x}|\mathbf{s},\mathbf{u})~P_{Y|X}^{\otimes n}(\mathbf{y}|\mathbf{x}) ~\\
   \qquad \qquad  \qquad \qquad \qquad \qquad\qquad \qquad \ind_{\big\{g_{\mathcal{B}_{W,n}}(\mathbf{v},\mathbf{y},\mathbf{s})=\hat m'\big\}}~\ind_{\Big\{\hat h=\ind_{\left\{(\mathbf{s}, \hat m',\mathbf{v},\mathbf{y}) \in \mathcal{A}_n^c \right\}}\Big\}},  & \mbox{ otherwise},\notag 
 \end{cases} &&
 \end{flalign}
 and 
 \begin{flalign}
&Q^{(\mathcal{B}_{W,n},c_n)}_{\mathbf{UV}M'\mathbf{XY}  \hat M'\hat H}(\mathbf{u},\mathbf{v},m',\mathbf{x},\mathbf{y},\hat m',\hat h) \notag \\
 &:=
 \begin{cases}
   Q_{UV}^{\otimes n}(\mathbf{u},\mathbf{v})~\ind_{\left\{M'\left(\mathbf{u},\mathcal{B}_{W,n}\right)=m'\right\}} ~ P_{X|USW}^{\otimes n}(\mathbf{x}|\mathbf{s},\mathbf{u},\mathbf{w}(m'))~P_{Y|X}^{\otimes n}(\mathbf{y}|\mathbf{x})~\\
   \qquad \qquad  \qquad \qquad \qquad \qquad\qquad \qquad\ind_{\left\{g_{\mathcal{B}_{W,n}}(\mathbf{v},\mathbf{y},\mathbf{s})=\hat m'\right\}}~\ind_{\left\{\hat h=\ind_{\left\{(\mathbf{s}, \hat m',\mathbf{v},\mathbf{y}) \in \mathcal{A}_n^c \right\}}\right\}},  & \mbox{ if } m' \neq 0, \notag \\
    Q_{UV}^{\otimes n}(\mathbf{u},\mathbf{v})~\ind_{\left\{M'\left(\mathbf{u},\mathcal{B}_{W,n}\right)=m'\right\}} ~ P_{X'|US}^{\otimes n}(\mathbf{x}|\mathbf{s},\mathbf{u})~P_{Y|X}^{\otimes n}(\mathbf{y}|\mathbf{x}) ~\\
   \qquad \qquad  \qquad \qquad \qquad \qquad\qquad \qquad \ind_{\left\{g_{\mathcal{B}_{W,n}}(\mathbf{v},\mathbf{y},\mathbf{s})=\hat m'\right\}}~\ind_{\left\{\hat h=\ind_{\left\{(\mathbf{s}, \hat m',\mathbf{v},\mathbf{y}) \in \mathcal{A}_n^c \right\}}\right\}},  & \mbox{ otherwise},\notag 
 \end{cases} &&
 \end{flalign}
respectively. For brevity, we will denote $\mathcal{B}_{W,n}$ by $\mathcal{B}_n$,  $\mathbb{B}_{W,n}$ by $\mathbb{B}_n$, and the above probability distributions by  $ P^{(\mathcal{B}_n)}$ and $Q^{(\mathcal{B}_n)}$.  Let $\mathfrak{B}_n$ and $\mu_n$  stand for  the support and  probability measure of $\mathbb{B}_n$, respectively, and set $\bar{\mathbb{P}}_{P^{(\mathbb{B}_n)}}:=  \mathbb{E}_{\mu_n}\big[\mathbb{P}_{P^{(\mathbb{B}_n)}}\big]$,   $\bar{\mathbb{P}}_{Q^{(\mathbb{B}_n)}}:=  \mathbb{E}_{\mu_n}\big[\mathbb{P}_{Q^{(\mathbb{B}_n)}}\big]$.\\
\textbf{Analysis of the type I and type II error probabilities:}
We  analyze the expected type I and type II error probabilities, where the expectation is with respect to the randomness of $\mathbb{B}_n$, followed  by the expurgation technique to extract a sequence of deterministic codebooks $\{\mathcal{B}_n\}_{n \in \mathbb{N}}$ and a code $\{c_n=(f_n,g_n)\}_{n \in \mathbb{N}}$ that achieves the lower bound  in Theorem \ref{jhtccthm}.\\
\textbf{Type I error probability:} Denoting by $\mathbb{A}_n$  the random acceptance region for $H_0$, note that a type I error can occur only under the following events:
\begin{enumerate}[label=(\roman*)]
\item $ \mathcal{E}_{\textrm{EE}}':= \underset{P_{\hat U} \in \mathcal{D}_n(P_{U},\eta)}{\bigcup}~ \underset{\mathbf{u} \in \mathcal{T}_n(P_{\hat U})}{\bigcup}\mathcal{E}'_{\textrm{EE}}(\mathbf{u})$, where
\begin{flalign}
  \mathcal{E}'_{\textrm{EE}}(\mathbf{u}):= \left\lbrace
               \nexists~j \in \mathcal{M}'\backslash\{0\} \mbox{ s.t. } (\mathbf{s},\mathbf{u},\mathbf{W}(j)) \in \mathcal{T}_n(P_{\hat S\hat U_i \hat W_i}),~ P_{\hat S\hat U_i}=P_{\mathbf{s}\mathbf{u}},  ~P_{\hat S\hat U_i \hat W_i} \in  \mathcal{D}_n(P_{SUW}, \eta) 
             \right\rbrace, \notag
\end{flalign}

    \item  $  \mathcal{E}'_{\textrm{NE}}:=\{\hat M'=M'$ and  $(\mathbf{s},\hat M',\mathbf{V},\mathbf{Y}) \notin \mathbb{A}_n\}$,
    \item  $  \mathcal{E}'_{\textrm{ODE}}:=\{M' \neq 0$, $\hat M' \neq M'$ and  $(\mathbf{s},\hat M',\mathbf{V},\mathbf{Y}) \notin \mathbb{A}_n\}$,
    \item $  \mathcal{E}'_{\textrm{SDE}}:=\{M'=0$, $\hat M' \neq M'$ and  $(\mathbf{s},\hat M',\mathbf{V},\mathbf{Y}) \notin \mathbb{A}_n\}$.
\end{enumerate}
By definition of $R_i'$, we have similar to \eqref{doubleexpdecaycov} that
\begin{align}
\bar{\mathbb{P}}_{P^{\mathbb{B}_n}}(\mathcal{E}'_{\textrm{EE}}) \leq e^{-e^{n\Omega(\eta)}}. \label{doubleexpdecaycovhyb}
\end{align}
Next, the event $\mathcal{E}'_{\textrm{NE}}$ can be upper bounded as
\begin{flalign}
   \bar{\mathbb{P}}_{P^{\mathbb{B}_n}} \left(\mathcal{E}'_{\textrm{NE}}|\mathcal{E}'^c_{\textrm{EE}} \right) &\leq    \bar{\mathbb{P}}_{P^{\mathbb{B}_n}} \left((\mathbf{s},\hat M',\mathbf{V},\mathbf{Y}) \notin \mathbb{A}_n| \hat M'=M',\mathcal{E}'^c_{EE} \right)=1-\bar{\mathbb{P}}_{P^{\mathbb{B}_n}} \left((\mathbf{s},\mathbf{U},\mathbf{V},\mathbf{Y}) \in \mathbb{A}^e_n|\mathcal{E}'^c_{EE}\right). \label{finerrevntt1} 
\end{flalign}
For $\mathbf{u} \in \mathbb{O}_{m'}$, note that  similar to \cite[Equation 4.17]{HK-1989}, we have
\begin{flalign}
\bar{\mathbb{P}}_{P^{\mathbb{B}_n}}\big((\mathbf{V},\mathbf{Y}) \in  \mathbb{Z}'_{m'}(\mathbf{u})|\mathbf{U}=\mathbf{u},\mathbf{W}(m')=\bar{\mathbf{w}}_{m'},\mathcal{E}'^c_{EE}\big) \geq 1-e^{-n\left(\kappa_{\alpha}+\frac{\eta}{3}-D(P_{\mathbf{u}}||P_U)\right)}. \notag
\end{flalign}
From this and \eqref{quantsrcseqset}, we obtain  similar to \cite[Equation 4.22]{HK-1989} that
\begin{align}
    \bar{\mathbb{P}}_{P^{\mathbb{B}_n}}((\mathbf{s},\mathbf{U}, \mathbf{V},\mathbf{Y}) \in  \mathbb{A}_e^n|\mathcal{E}'^c_{EE})  \geq 1-e^{-n\kappa_{\alpha}}. \label{cordectype1errp}
\end{align}
Substituting \eqref{cordectype1errp} in \eqref{finerrevntt1} yields
\begin{align}
   &    \bar{\mathbb{P}}_{P^{\mathbb{B}_n}} \left(\mathcal{E}'_{\textrm{NE}}|\mathcal{E}'^c_{\textrm{EE}} \right) \leq  e^{-n\kappa_{\alpha}}. \label{frthevtt1hyb}
\end{align}
Next, we bound the probability of the event $ \mathcal{E}'_{\textrm{ODE}}$ as follows:
\begin{flalign}
    \bar{\mathbb{P}}_{P^{\mathbb{B}_n}} \left( \mathcal{E}'_{\textrm{ODE}}\right)
    &=\bar{\mathbb{P}}_{P^{\mathbb{B}_n}} \left( M' \neq 0, \hat M' \neq M',(\mathbf{s},M',\mathbf{V},\mathbf{Y}) \in \mathbb{A}_n,(\mathbf{s},\hat M',\mathbf{V},\mathbf{Y}) \notin \mathbb{A}_n\right) \notag \\
    &   \qquad \qquad \qquad \qquad + \bar{\mathbb{P}}_{P^{\mathbb{B}_n}} \left( M' \neq 0, \hat M' \neq M',(\mathbf{s},M',\mathbf{V},\mathbf{Y}) \notin \mathbb{A}_n,(\mathbf{s},\hat M',\mathbf{V},\mathbf{Y}) \notin \mathbb{A}_n\right) \notag \\
     &\leq \bar{\mathbb{P}}_{P^{\mathbb{B}_n}} \left( M' \neq 0, \hat M' \neq M',(\mathbf{s},M',\mathbf{V},\mathbf{Y}) \in \mathbb{A}_n,(\mathbf{s},\hat M',\mathbf{V},\mathbf{Y}) \notin \mathbb{A}_n\right) \notag \\
    &   \qquad \qquad \qquad \qquad + \bar{\mathbb{P}}_{P^{\mathbb{B}_n}} \left( M' \neq 0, \hat M' \neq M',(\mathbf{s},M',\mathbf{V},\mathbf{Y}) \notin \mathbb{A}_n\right) \notag \\
    &    \stackrel{(a)}{\leq} \bar{\mathbb{P}}_{P^{\mathbb{B}_n}} \left( M' \neq 0, \hat M' \neq M', (\mathbf{s},M',\mathbf{V},\mathbf{Y}) \in \mathbb{A}_n,(\mathbf{s},\hat M',\mathbf{V},\mathbf{Y}) \notin \mathbb{A}_n\right)+  e^{-e^{n\Omega(\eta)}} + e^{-n\kappa_{\alpha}}  \label{hkschemet1errhyb} \\
    & \leq \bar{\mathbb{P}}_{P^{\mathbb{B}_n}} \left( \hat M' \neq M'|M' \neq 0, (\mathbf{s},M',\mathbf{V},\mathbf{Y}) \in \mathbb{A}_n\right)+  e^{-e^{n\Omega(\eta)}} + e^{-n\kappa_{\alpha}} \notag\\
  &  \stackrel{(b)}{=} \bar{\mathbb{P}}_{P^{\mathbb{B}_n}} \left( \hat M' \neq M'|M' \neq 0, \hat M' \neq 0, (\mathbf{s},M',\mathbf{V},\mathbf{Y}) \in \mathbb{A}_n\right) \label{secondstageimplfirst} \\
   & \stackrel{(c)}{\leq} e^{-n \left(\rho'(\kappa_{\alpha},\omega',P_S, P_{X|USW})-\zeta'(\kappa_{\alpha}, \omega',P_{\hat S})-O(\eta) \right)}, \label{applyhybcdbnd}  
\end{flalign}
where $(a)$ follows similar to \eqref{hkschemet1err} using \eqref{doubleexpdecaycovhyb} and \eqref{cordectype1errp}; $(b)$ is since  $(\mathbf{s},M',\mathbf{V},\mathbf{Y}) \in \mathbb{A}_n$ implies that $\hat M' \neq 0$; and $(c)$ follows similar to \eqref{finbndsideinfcdwrd}.
Further, 
\begin{flalign}
    \bar{\mathbb{P}}_{P^{\mathbb{B}_n}} \left(\mathcal{E}'_{\textrm{SDE}} \right)  &\leq \bar{\mathbb{P}}_{P^{\mathbb{B}_n}} \left(M'= 0 \right) \notag \\
   &\leq \bar{\mathbb{P}}_{P^{\mathbb{B}_n}} \left(M'= 0|\mathcal{E}'^c_{EE} \right)+ \bar{\mathbb{P}}_{P^{\mathbb{B}_n}} \left(\mathcal{E}'_{EE}\right) \notag \\
   & = \sum_{\substack{\mathbf{u}:P_{\mathbf{u}} \notin \mathcal{D}_n(P_U,\eta)}}P_U^{\otimes n}(\mathbf{u})+ \bar{\mathbb{P}}_{P^{\mathbb{B}_n}} \left(\mathcal{E}'_{EE}\right)\notag \\
   & \leq   e^{-n \kappa_{\alpha}}+ e^{-e^{n\Omega(\eta)}}, && \label{thirdevnthybt1}
\end{flalign}
where the penultimate equality is since given $\mathcal{E}'^c_{EE} $,  $M'=0$ occurs only for $\mathbf{U}=\mathbf{u}$ such that $P_{\mathbf{u}} \notin \mathcal{D}_n(P_U,\eta)$, and the final inequality follows from \eqref{doubleexpdecaycovhyb}, the definition of $\mathcal{D}_n(P_U,\eta)$ and \cite[Lemma 1.6]{Csiszar-Korner}.
From \eqref{doubleexpdecaycovhyb}, \eqref{frthevtt1hyb}, \eqref{applyhybcdbnd} and \eqref{thirdevnthybt1}, the expected  type I error probability satisfies $e^{-n(\kappa_{\alpha}-O(\eta))}$ for sufficiently large $n$ via the union bound.\\
\textbf{Type II error probability:}
Next, we analyze the expected type II error probability over $\mathbb{B}_n$. Let
\begin{align}
   & \mathcal{D}_n(P_{SVWY},\eta) 
  := \left\lbrace
                \begin{array}{ll} P_{\hat S \hat V \hat  W \hat Y}:& ~\exists~(\mathbf{s},\mathbf{u},\mathbf{v},\bar{\mathbf{w}},\mathbf{y}) \in \underset{m' \in \mathcal{M}' \backslash \{0\}}{\cup}\mathcal{J}_n\left(\kappa_{\alpha}+\eta, P_{\hat SUVW_{m'}Y}\right), \\& P_{\hat SUVW_{m'}Y} \mbox{ satisfies } \eqref{conddistdefnh1} \mbox{ and }P_{\mathbf{s}\mathbf{u}\mathbf{v}\bar{\mathbf{w}}\mathbf{y}}=P_{\hat S \hat U \hat V \hat W \hat Y} \end{array}
             \right\rbrace, \notag \\
             &  \mathcal{F}'_{1,n}(\eta):= \left\lbrace
                \begin{array}{ll}  P_{\hat S\tilde U \tilde V \tilde W \tilde Y}  \in \mathcal{T}\left(\mathcal{S}^n \times \Ucal^n \times \mathcal{V}^n \times \W^n \times \Y^n\right)&:P_{\hat S \tilde U \tilde W} \in  \mathcal{D}_n(P_{SUW}, \eta),  \\
    & P_{\hat S\tilde V \tilde W \tilde Y} \in\mathcal{D}_n(P_{SVWY},\eta) \end{array}
             \right\rbrace.   \notag
\end{align}
  A type II error can occur only under the following events:
  \begin{enumerate}[label=\emph{(\alph*)}]
      \item$
      \mathcal{E}'_a:= \left\lbrace
                \begin{array}{ll} & \hat M'=M'\neq 0, (\mathbf{s},\mathbf{U},\mathbf{V},\mathbf{W}(M'),\mathbf{Y}) \in \mathcal{T}_n\left(P_{\hat S\hat U \hat V \hat W \hat Y}\right) \\ & \mbox{ s.t. } P_{\hat U \hat W} \in \mathcal{D}_n(P_{SUW}, \eta)\mbox{ and } P_{\hat S \hat V \hat W \hat Y} \in  \mathcal{D}_n(P_{SVWY},\eta) \end{array}
             \right\rbrace$, \vspace{5 pt}
      \item $\mathcal{E}'_b:= \left\lbrace
                \begin{array}{ll}
                  &M' \neq 0, \hat M' \neq  M', (\mathbf{s},\mathbf{U},\mathbf{V},\mathbf{W}(M'), \mathbf{Y},\mathbf{W}(\hat M')) \in \mathcal{T}_n\left(P_{\hat S\hat U \hat V \hat W \hat Y \hat W_d}\right)  \\  &  \mbox{ s.t. } P_{\hat S \hat U \hat W} \in  \mathcal{D}_n(P_{SUW}, \eta), P_{\hat S \hat V  \hat W_d \hat Y} \in  \mathcal{D}_n(P_{SVWY},\eta),\\ &\mbox{ and }H_e\left(\mathbf{W}(\hat M')|\mathbf{s}, \mathbf{V}, \mathbf{Y} \right) \leq H_e\left(\mathbf{W}(M')|\mathbf{s}, \mathbf{V}, \mathbf{Y}\right)
                \end{array}
             \right\rbrace$, \vspace{5 pt}
 \item $\mathcal{E}'_c:=\left\lbrace
                 M'=0, \hat M' \neq M',  (\mathbf{s}, \mathbf{V},\mathbf{Y}, \mathbf{W}(\hat M')) \in \mathcal{T}_n\left(P_{\hat S\hat V  \hat Y \hat W_d }\right)\mbox{ s.t. }P_{\hat S \hat V  \hat W_d \hat Y} \in  \mathcal{D}_n(P_{SVWY},\eta)
             \right\rbrace.$
  \end{enumerate}
  Considering the event $\mathcal{E}_a' $, we have
   \begin{flalign}
      & \bar{\mathbb{P}}_{Q^{\mathbb{B}_n}}\left(\mathcal{E}_a' \right) \notag \\
      &\leq \sum_{\substack{ P_{\hat S \tilde U \tilde V \tilde W \tilde Y}  \\
       \in  \mathcal{F}'_{1,n}(\eta) }} \sum_{\substack{(\mathbf{u},\mathbf{v},\bar{\mathbf{w}},\mathbf{y}): \\(\mathbf{s}, \mathbf{u},\mathbf{v},\bar{\mathbf{w}},\mathbf{y}) \\ \in \mathcal{T}_n( P_{\hat S \tilde U \tilde V \tilde W \tilde Y}   )}} \sum_{\substack{m' \in \mathcal{M'} \backslash \{0\}}}   \bar{\mathbb{P}}_{Q^{\mathbb{B}_n}} \left(\mathbf{U}=\mathbf{u},\mathbf{V}=\mathbf{v},M'=m', \mathbf{W}(m')=\bar{\mathbf{w}},\mathbf{Y}=\mathbf{y}| \mathbf{S}=\mathbf{s}\right) \notag \\
       & \leq \sum_{\substack{ P_{\hat S \tilde U \tilde V \tilde W \tilde Y}  
       \in  \mathcal{F}'_{1,n}(\eta) }} \sum_{\substack{(\mathbf{u},\mathbf{v},\bar{\mathbf{w}},\mathbf{y}): \\(\mathbf{s}, \mathbf{u},\mathbf{v},\bar{\mathbf{w}},\mathbf{y})  \in \mathcal{T}_n( P_{\hat S \tilde U \tilde V \tilde W \tilde Y}   )}} \sum_{m' \in \mathcal{M'} \backslash \{0\} } \bar{\mathbb{P}}_{Q^{\mathbb{B}_n}} \left(\mathbf{U}=\mathbf{u},\mathbf{V}=\mathbf{v},M'=m'|\mathbf{S}=\mathbf{s}\right) \notag  \\
       &\qquad  \qquad \qquad \qquad  \qquad \qquad   \qquad \qquad  \bar{\mathbb{P}}_{Q^{\mathbb{B}_n}} \left( \mathbf{W}(m')=\bar{\mathbf{w}}|\mathbf{U}=\mathbf{u},\mathbf{V}=\mathbf{v},M'=m', \mathbf{S}=\mathbf{s}\right) \notag \\
       & \qquad \qquad \qquad \qquad  \qquad \qquad   \qquad \qquad  \bar{\mathbb{P}}_{Q^{\mathbb{B}_n}} \left(\mathbf{Y}=\mathbf{y}|\mathbf{U}=\mathbf{u},\mathbf{V}=\mathbf{v},M'=m',\mathbf{W}(m')=\bar{\mathbf{w}}, \mathbf{S}=\mathbf{s}\right) \notag \\
      & \stackrel{(a)}{\leq}\sum_{\substack{ P_{\hat S \tilde U \tilde V \tilde W \tilde Y}  
       \in  \mathcal{F}'_{1,n}(\eta) }} \sum_{\substack{(\mathbf{u},\mathbf{v},\bar{\mathbf{w}},\mathbf{y}): \\(\mathbf{s}, \mathbf{u},\mathbf{v},\bar{\mathbf{w}},\mathbf{y})  \in \mathcal{T}_n( P_{\hat S \tilde U \tilde V \tilde W \tilde Y}   )}}  e^{-n \left(H(\tilde U ,\tilde V)+D\left(P_{\tilde U \tilde V}|| Q_{ U V}\right) \right)}~ e^{-n \left(H(\tilde W|\hat S,\tilde U)-\eta \right)}\notag \\
       & \qquad \qquad \qquad 
 \qquad\qquad\qquad \qquad  \qquad \qquad   \qquad   e^{-n\left(H( \tilde Y|\tilde U, \hat S ,\tilde W)+ D\left(P_{\tilde Y|\tilde U \hat S \tilde W}|| P_{Y|USW}|P_{\tilde U \hat S \tilde W}\right)\right)} \notag \\
      & \leq  \sum_{\substack{ P_{\hat S \tilde U \tilde V \tilde W \tilde Y}  
       \in  \mathcal{F}'_{1,n}(\eta) }}  e^{n H(\tilde U ,\tilde V ,\tilde W,\tilde Y|\hat S)} e^{-n \left(H(\tilde U ,\tilde V)+D\left(P_{\tilde U \tilde V}|| Q_{ U V}\right) \right)}~ 
      e^{-n \left(H(\tilde W|\hat S,\tilde U)-\eta \right)} \notag \\ & \qquad \qquad \qquad \qquad \qquad  \qquad \qquad  \qquad   \qquad \qquad  e^{-n\left(H( \tilde Y|\tilde U, \hat S ,\tilde W)+ D\left(P_{\tilde Y|\tilde U \hat S \tilde W}|| P_{Y|USW}|P_{\tilde U \hat S \tilde W}\right)\right)} \notag \\
       & \leq e^{-n E'_{1,n}}, \label{t2er1hyb} &&
   \end{flalign}
   where
   \begin{align}
    E'_{1,n}&:= \min_{ P_{\hat S \tilde U \tilde V \tilde W \tilde Y}  
       \in  \mathcal{F}'_{1,n}(\eta)} H(\tilde U ,\tilde V)+D\left(P_{\tilde U \tilde V}|| Q_{ U V}\right)+ H(\tilde W|\hat S,\tilde U)-\eta+H( \tilde Y|\tilde U, \hat S ,\tilde W)\notag \\
       &\qquad  + D\left(P_{\tilde Y|\tilde U \hat S \tilde W}|| P_{Y|USW}|P_{\tilde U \hat S \tilde W}\right)-H(\tilde U ,\tilde V ,\tilde W,\tilde Y|\hat S)- \frac{1}{n}||\Ucal||\V||\W||\Y|\log(n+1)\notag \\
        & \gtrsim\min_{\substack{ (P_{\tilde U \tilde V \tilde W\tilde Y S},Q_{\tilde U \tilde V \tilde W\tilde Y S}) \\\in  \mathcal{T}_1'(\kappa_{\alpha},\omega',P_{S}, P_{X|USW})}}  D(P_{\tilde U \tilde V \tilde W \tilde Y|S} ||Q_{ U V W Y|S}|P_S)-O(\eta) \notag \\
     & =E_1'(\kappa_{\alpha},\omega')-O(\eta). \notag
   \end{align}
  For the inequality in $(a)$ above, we used $\sum_{}\bar{\mathbb{P}}_{Q^{\mathbb{B}_n}}\big(M'=m'|\mathbf{U}=\mathbf{u},\mathbf{V}=\mathbf{v},\mathbf{S}=\mathbf{s}\big) \leq 1$  and  
    \begin{align}
   \bar{\mathbb{P}}_{Q^{\mathbb{B}_n}} \left( \mathbf{W}(m')=\bar{\mathbf{w}}|\mathbf{U}=\mathbf{u},\mathbf{V}=\mathbf{v},\mathbf{S}=\mathbf{s},M'=m'\right) \leq
   \begin{cases}
    e^{-n \big(H(\tilde W|\hat S,\tilde U )-\eta \big)}, & \mbox{ if } \bar{\mathbf{w}} \in \mathcal{T}_n(\tilde W),\\
    0,& \mbox{ otherwise},
   \end{cases}  \notag
    \end{align}
which in turn follows from the fact that  given $M'=m'$ and  $\mathbf{U}=\mathbf{u}$, $\mathbf{W}(m')$ is uniformly distributed in the set $\mathcal{T}_n\big(P_{\tilde W|\hat S \tilde U},\mathbf{s}, \mathbf{u}\big)$ and that for sufficiently large $n$ $\big|\mathcal{T}_n\big(P_{\tilde W|\hat S \tilde U},\mathbf{s},\mathbf{u}\big)\big| \geq e^{n \left(H(\tilde W|\hat S,\tilde U)-\eta \right)}$.

 Next, we analyze the probability of the event $\mathcal{E}'_b$. 
 Let 
   \begin{align}
    \mathcal{F}'_{2,n}(\eta):= \left\lbrace
                \begin{array}{ll}
                   P_{\hat S \tilde U \tilde V \tilde W \tilde Y  \tilde W_d}:&  P_{\hat S \tilde U \tilde W} \in  \mathcal{D}_n(P_{SUW}, \eta), P_{\hat S\tilde V \tilde W_d \tilde Y} \in  \mathcal{D}_n(P_{SVWY},\eta) \\& H\left(\tilde W_d| \hat S,\tilde V,\tilde Y\right) \leq H\left(\tilde W| \hat S,\tilde V,\tilde Y\right)
                \end{array}
             \right\rbrace.&& \notag  
   \end{align}    
   Then,
   \begin{flalign}
      & \bar{\mathbb{P}}_{Q^{\mathbb{B}_n}}\left(\mathcal{E}_b' \right) \notag \\
      &\leq \sum_{\substack{ P_{\hat S \tilde U \tilde V \tilde W \tilde Y  \tilde W_d} \\
       \in  \mathcal{F}'_{2,n}(\eta) }} \sum_{\substack{(\mathbf{u},\mathbf{v},\bar{\mathbf{w}},\mathbf{y},\mathbf{w}'): \\(\mathbf{s}, \mathbf{u},\mathbf{v},\bar{\mathbf{w}},\mathbf{y}, \mathbf{w}') \\ \in \mathcal{T}_n( P_{\hat S \tilde U \tilde V \tilde W \tilde Y  \tilde W_d} )}} \sum_{\substack{m' \in \\\mathcal{M'} \backslash \{0\} }}   \bar{\mathbb{P}}_{Q^{\mathbb{B}_n}} \left(\mathbf{U}=\mathbf{u},\mathbf{V}=\mathbf{v}, M'=m', \mathbf{W}(m')=\bar{\mathbf{w}}, \mathbf{Y}=\mathbf{y}|\mathbf{S}=\mathbf{s}\right) \notag \\
       &\qquad \qquad \qquad \qquad\qquad \qquad \qquad\sum_{\substack{\hat m' \in \mathcal{M'} \backslash \{0,m'\} }} \bar{\mathbb{P}}_{Q^{\mathbb{B}_n}} \left( \bar{\mathbf{W}}(\hat m')=\mathbf{w}'| \mathbf{U}=\mathbf{u},M'=m', \mathbf{W}(m')=\bar{\mathbf{w}}, \mathbf{S}=\mathbf{s}\right) \notag \\
   &\leq \sum_{\substack{ P_{\hat S \tilde U \tilde V \tilde W \tilde Y  \tilde W_d} \\
       \in  \mathcal{F}'_{2,n}(\eta) }} \sum_{\substack{(\mathbf{u},\mathbf{v},\bar{\mathbf{w}},\mathbf{y}): \\(\mathbf{s}, \mathbf{u},\mathbf{v},\bar{\mathbf{w}},\mathbf{y})  \in \mathcal{T}_n( P_{\hat S \tilde U \tilde V \tilde W \tilde Y}   )}}  2 e^{-n \left(H(\tilde U ,\tilde V)+D\left(P_{\tilde U \tilde V}|| Q_{ U V}\right) \right)}~
       e^{-n \big(H(\tilde W|\hat S,\tilde U)-\eta \big)}~  \notag \\
       & \qquad \qquad \qquad\qquad e^{-n\left(H( \tilde Y|\tilde U, \hat S ,\tilde W)+ D\left(P_{\tilde Y|\tilde U \hat S \tilde W}|| P_{Y|USW}|P_{\tilde U \hat S \tilde W}\right)\right)}~e^{n\left(\zeta'(\kappa_{\alpha}, \omega',P_{\hat S})+\eta\right)} e^{nH(\tilde W_d|\hat S,\tilde V, \tilde Y)}e^{-n\left(H(\tilde W_d)-\eta\right)} \notag \\
       &\leq e^{-n E'_{2,n}}, \label{t2er2hyb} &&
    \end{flalign}
    where 
    \begin{flalign}
       E'_{2,n}
       & \gtrsim \underset{\substack{(P_{\tilde U \tilde V \tilde W\tilde Y S},Q_{\tilde U \tilde V \tilde W\tilde Y S})  \in \mathcal{T}_2'(\kappa_{\alpha},\omega',P_S,P_{X|USW})}}{\min} D(P_{\tilde U \tilde V \tilde W \tilde Y|S} ||Q_{ U V W Y| S}|P_S)+ \rho'(\kappa_{\alpha},\omega',P_S, P_{X|USW}) \notag \\&\qquad \qquad \qquad \qquad \qquad \qquad \qquad \qquad \qquad \qquad -\zeta'(\kappa_{\alpha}, \omega',P_{S})-O(\eta) \notag \\
       &=E_2'(\kappa_{\alpha},\omega',P_{S},P_{X|USW})-O(\eta). \notag &&
    \end{flalign}
Finally, considering the event $\mathcal{E}_c'$, we have
\begin{flalign}
    & \bar{\mathbb{P}}_{Q^{\mathbb{B}_n}}\left(\mathcal{E}_c' \right)=\sum_{\substack{\mathbf{u} \in \mathcal{T}_n(P_{\tilde U}): \\ P_{\tilde U} \in  \mathcal{D}_n(P_U,\eta)}} \bar{\mathbb{P}}_{Q^{\mathbb{B}_n}} \left(\mathbf{U}=\mathbf{u}, \mathcal{E}'_{\textrm{EE}},\mathcal{E}'_c|\mathbf{S}=\mathbf{s}\right)+ \sum_{\substack{\mathbf{u} \in \mathcal{T}_n(P_{\tilde U}): \\ P_{\tilde U} \notin  \mathcal{D}_n(P_U,\eta)}} \bar{\mathbb{P}}_{Q^{\mathbb{B}_n}} \left(\mathbf{U}=\mathbf{u},\mathcal{E}'_c|\mathbf{S}=\mathbf{s}\right). \notag
\end{flalign}
 The first term in the RHS decays double exponentially as $ e^{-e^{n\Omega(\eta)}}$, while the second term  can be handled as follows: 
 \begin{flalign}
 &\sum_{\substack{\mathbf{u} \in \mathcal{T}_n(P_{\tilde U}):  P_{\tilde U} \notin  \mathcal{D}_n(P_U,\eta)}} \bar{\mathbb{P}}_{Q^{\mathbb{B}_n}} \left(\mathbf{U}=\mathbf{u},\mathcal{E}'_c|\mathbf{S}=\mathbf{s}\right) \notag \\
     & \leq\sum_{\substack{\mathbf{u} \in \mathcal{T}_n(P_{\tilde U}): \\ P_{\tilde U} \notin  \mathcal{D}_n(P_U,\eta)}} \sum_{\substack{(\mathbf{v},\mathbf{y}, \mathbf{w}'):\\(\mathbf{s}, \mathbf{v},\mathbf{y}, \mathbf{w}') \in \mathcal{T}_n\left(P_{\hat S \tilde V \tilde Y \tilde W_d}\right), \\ P_{\hat S \tilde V \tilde W_d \tilde Y}  \in  \mathcal{D}_n(P_{SVWY},\eta)}}~ \sum_{\hat m \in \mathcal{M} \backslash \{0\}}  \bar{\mathbb{P}}_{Q^{\mathbb{B}_n}} \left(\mathbf{U}=\mathbf{u},\mathbf{V}=\mathbf{v},M'=0,\mathbf{Y}=\mathbf{y} |\mathbf{S}=\mathbf{s}\right) \notag \\
 &  \qquad \qquad \qquad \qquad  \qquad \qquad \qquad  \qquad  \qquad  \qquad \qquad  \qquad\sum_{\hat m' \in \mathcal{M}'\backslash \{0\}} \bar{\mathbb{P}}_{Q^{\mathbb{B}_n}} \left( \mathbf{W}(\hat m')=\bar{\mathbf{w}} \right) \notag \\   & \leq \sum_{\substack{ P_{\tilde U \hat S \tilde V \tilde W_d \tilde Y} \mspace{2 mu} \in  \mspace{2 mu} \mathcal{D}_n(P_U,\eta)^c \\ \times  \mathcal{D}_n(P_{SVWY},\eta)}}   e^{n H(\tilde U, \tilde V, \tilde Y|\hat S )}   e^{-n \left(H(\tilde U, \tilde V, \tilde Y|\hat S)+D\left(P_{\tilde U \tilde V \tilde Y|\hat S}|| Q_{ U V Y'|\hat S}|P_{\hat S}\right) \right)}~\frac{e^{nH(\tilde W_d|\hat S,\tilde V, \tilde Y)}e^{n(R'+\eta)}}{e^{n\left(H(\tilde W_d)-\eta \right)}}\notag \\[5 pt]
    & \leq e^{-nE'_{3,n}}, \label{t2er3hyb} &&
\end{flalign}
where 
\begin{align}
  E'_{3,n}
  &\gtrsim\underset{\substack{P_{\hat V \hat YS}: P_{\hat U \hat V \hat W \hat Y S}~ \in \\  \hat{\mathcal{L}}_{\mathrm{h}}(\kappa_{\alpha},\omega', P_S, P_{X|USW})}}{\min} \mspace{-10 mu}D\big(P_{ \hat V \hat Y|S}||Q_{ V Y'|S}|P_S\big)+\rho'(\kappa_{\alpha},\omega',P_S, P_{X|USW})-\zeta'(\kappa_{\alpha}, \omega',P_{S}) -O(\eta) \notag \\
  &=E_3'(\kappa_{\alpha},\omega',P_{S}, P_{X|USW},P_{X'|US})-O(\eta).\notag
\end{align}
Since the exponent of the type II error probability is lower bounded by the minimum of the exponent of the type II error causing events, it follows from \eqref{t2er1hyb}, \eqref{t2er2hyb} and \eqref{t2er3hyb} that for a fixed $\big(P_{S}, \omega'(\cdot,P_{S}), P_{X|USW},P_{X'|US}\big) \in \mathcal{L}_{\mathrm{h}}(\kappa_{\alpha})$
\vspace{-15 pt}
\begin{subequations} \label{finavgerrexphyb}
   \begin{align}   \bar{\mathbb{P}}_{P^{(\mathbb{B}_{n})}} \left(\hat H=1 \right)&\leq e^{-n(\kappa_{\alpha}-O(\eta))}, \\
   \bar{ \mathbb{P}}_{Q^{(\mathbb{B}_{n})}} \left(\hat H=0 \right) &\leq e^{-n\big(\bar{\kappa}_{\mathrm{h}}(\kappa_{\alpha},\omega',P_{S}, P_{X|USW},P_{X'|US}) -O(\eta)\big)},
   \end{align}
\end{subequations}
where $\bar{\kappa}_{\mathrm{h}} = \min \big\{E_1'(\kappa_{\alpha},\omega'),E_2'(\kappa_{\alpha},\omega',P_{S},P_{X|USW}),E_3'(\kappa_{\alpha},\omega',P_{S}, P_{X|USW},P_{X'|US}) \big\}$. 
Performing expurgation as in the proof of Theorem \ref{lbbinningts} to obtain a deterministic codebook $\mathcal{B}_n$ satisfying \eqref{finavgerrexphyb}, maximizing over $\big(P_{S}, \omega'(\cdot,P_{S}), P_{X|USW},P_{X'|US}\big) \in \mathcal{L}_{\mathrm{h}}(\kappa_{\alpha})$ and noting that $\eta>0$ is arbitrary  yields that $\kappa(\kappa_{\alpha}) \geq \kappa_{\mathrm{h}}^{\star}(\kappa_{\alpha})$.

Finally, we show that $\kappa(\kappa_{\alpha}) \geq \kappa_{\mathrm{u}}^{\star}(\kappa_{\alpha})$ which will complete the proof. Fix $P_{X|US}$ and let $P_{UVXY}:=P_{UV}P_{X|US}$ $P_{Y|X}$ and $Q_{UVXY}:=Q_{UV}P_{X|US}P_{Y|X}$. Consider an uncoded transmission scheme in which the channel input $\mathbf{X} \sim f_n(\cdot|\mathbf{u})= P_{X|US}^{\otimes n}(\cdot|\mathbf{u},\mathbf{s})$.  Let the decision rule $g_n$ be specified by the acceptance region $\mathcal{A}_n=\big\{(\mathbf{s},\mathbf{v},\mathbf{y}):D\big(P_{\mathbf{v}\mathbf{y}|\mathbf{s}}||P_{V Y|S}|P_{\mathbf{s}}\big) \leq \kappa_{\alpha}+\eta\big\}$ for some small  $\eta>0$. Then, it follows from \cite[Lemma 2.6]{Csiszar-1982} that for sufficiently large $n$,
\begin{align}
   \alpha_n(f_n,g_n)&= P_{VY|S}^{\otimes n}\left(\mathcal{A}_n^c|\mathbf{s}\right) \leq e^{-n\kappa_{\alpha}},\notag \\
      \beta_n(f_n,g_n)&= Q_{VY|S}^{\otimes n}\left(\mathcal{A}_n|\mathbf{s}\right)\leq e^{-n(\kappa_{\mathrm{u}}^{\star}(\kappa_{\alpha})-O(\eta))}.\notag  
\end{align}
 The proof is complete by noting that $\eta>0$ is arbitrary.

\section{Conclusion}\label{conclu}
This work explored the trade-off between the type I and type II error-exponents for distributed hypothesis testing over a noisy channel. We proposed a separate hypothesis testing and channel coding scheme as well as a joint scheme utilizing hybrid coding, and analyzed their performance resulting in two inner bounds on the error-exponents trade-off. The separate scheme recovers some of the existing bounds in the literature as special cases. We also showed via an example of testing against dependence that the joint scheme strictly outperforms the separate scheme at some points of the error-exponents trade-off. An interesting avenue for future research is the exploration of novel outer bounds that could shed light on  the scenarios where the separate or joint schemes are tight.

\begin{appendices}
\section{Proof that  Theorem \ref{lbbinningts} recovers \cite[Theorem 2]{SD_2020}}\label{genHTsepbndproof}
We  prove that $\lim_{\kappa_{\alpha} \rightarrow 0} \kappa_{\mathrm{s}}^{\star}(\kappa_{\alpha})= \kappa_{\mathrm{s}}$, 
where $\kappa_{\mathrm{s}}$ is the lower bound on the type II error-exponent for a fixed type I error probability constraint and unit bandwidth ratio established in \cite[Theorem 2]{SD_2020}.
Note that $\hat{\mathcal{L}}(0,\omega)=\{P_{UVW}=P_{UV}P_{W|U}, P_{W|U}=\omega(P_U)\}$, $\zeta(0,\omega)=I_P(U;W)$, and $\rho(0,\omega)=I_P(V;W)$. The result then follows from Theorem \ref{lbbinningts}  by noting that  $\hat{\mathcal{L}}(\kappa_{\alpha},\omega)$, $\zeta(\kappa_{\alpha}, \omega)$ and $\rho(\kappa_{\alpha},\omega)$ are continuous  in $\kappa_{\alpha}$ and the fact that $E_{\mathrm{sp}}(P_{SX}, \theta),E_{\mathrm{ex}}\left(R,P_{SX}\right)$ and  $E_{\mathrm{b}}(\kappa_{\alpha},\omega,R)$  are all greater than or equal to zero.

\section{Proof that Theorem \ref{jhtccthm} recovers \cite[Theorem 5]{SD_2020}} \label{corjhtccsteinproof}
We show that $\lim_{\kappa_{\alpha} \rightarrow 0} \kappa_{\mathrm{h}}^{\star}(\kappa_{\alpha})=\kappa_{\mathrm{h}}$, 
 where $\kappa_{\mathrm{h}}$ is as defined in \cite[Theorem 5]{SD_2020}.
Note that  $\zeta'(0, \omega',P_S):= I_P( U; W|S)$, $ \rho(0,\omega',P_S, P_{X|USW})= I_P(Y,V; W|S)$, 
\begin{align}
\hat{\mathcal{L}}_{\mathrm{h}}(0,\omega', P_S, P_{X|USW})&:=\big\{P_{UV\hat WYS}:~ P_{SUVWXY}:= P_S P_{UV}P_{W|US}P_{X|USW}P_{Y|X},~P_{W|US}=\omega'(P_{U}, P_S)\big\}, \notag \\
    \mathcal{L}_{\mathrm{h}}(0)&:=\big\{
   \big(P_{S}, \omega'(\cdot,P_{S}), P_{X|USW},P_{X'|US}\big) :
   I_P( U; W|S)< I_P(Y,V; W|S)   
  \big\}, \notag
\end{align}
and $E_{\mathrm{b}}'(0,\omega',P_S, P_{X|USW})=  I_P(Y,V; W|S)-I_P( U; W|S)$.
The result then follows from Theorem \ref{jhtccthm} via the continuity of $\hat{\mathcal{L}}_{\mathrm{h}}(\kappa_{\alpha},\cdot,\cdot,\cdot)$, $\zeta'(\kappa_{\alpha}, \cdot,\cdot)$, $\rho(\kappa_{\alpha},\cdot,\cdot,\cdot)$, $\mathcal{L}_{\mathrm{h}}(\kappa_{\alpha})$ and $E_{\mathrm{b}}'(\kappa_{\alpha},\cdot,\cdot,\cdot)$ in $\kappa_{\alpha}$.

\section{An auxiliary result}\label{App:auxres}
Here, we prove a result which was used in the proof of Theorem \ref{lbbinningts}, namely Proposition \ref{genchernbnd} given below.
For this purpose, we require a few properties of log-moment generating function, which we briefly review next.
\begin{lemma}[Properties of log-MGF, Theorem 15.3 and Theorem 15.6 in\cite{Polyanskiy-Wu-book}] \label{logmgffact}
The following hold:
\begin{enumerate}[label=(\roman*)]
    \item $\psi_{P_Z,f}(0)=0$ and $\psi'_{P_Z,f}(0)=\mathbb{E}_{P_Z}[f(Z)]$, where $\psi'_{P_Z,f}(\lambda)$ denotes the derivative of $\psi_{P_Z,f}(\lambda)$ with respect to $\lambda$.
    \item $\psi_{P_Z,f}(\lambda)$  is a strictly convex function in $\lambda$. 
    \item $\psi^*_{P_Z,f}(\theta)$ is strictly convex and strictly positive in $\theta$ except  $\psi^*_{P_Z,f}(\mathbb{E}_{P_Z}[Z])=0$.
\end{enumerate}
\end{lemma}
Proposition \ref{genchernbnd} is basically a characterization of the error-exponent region of a hypothesis testing problem which we introduce next.  Let  $P_{X_0X_1} \in \mathcal{P}(\X^2)$ be an arbitrary joint PMF, and consider  a sequence of pairs of $n$-length sequences $(\tilde{\mathbf{x}},\mathbf{x}') $  such that
\begin{align}
   P_{\tilde{\mathbf{x}}\mathbf{x}'}(\tilde x,x') \xrightarrow{(n)} P_{X_0X_1}(\tilde x,x'), ~\forall~ (\tilde x,x') \in \X^2. \label{seqpairconverge} 
\end{align}
Consider the following HT:
\begin{subequations}\label{HTdefchn}
\begin{equation}
 H_0: \mathbf{Y}  \sim  P_{Y|X}^{\otimes n}(\cdot|\tilde{\mathbf{x}}), 
\end{equation}
\begin{equation}
  H_1:\mathbf{Y}  \sim  P_{Y|X}^{\otimes n}(\cdot|\mathbf{x}'). 
\end{equation}
\end{subequations}
 With the achievability of an error-exponent pair $(\kappa_{\alpha},\kappa_{\beta}) $ defined  similar to Definition \ref{deft1t2expdistach}, 
consider the error-exponent region of interest\footnote{As will become evident later, the error-exponent region for the HT in \eqref{HTdefchn} depends on $(\tilde{\mathbf{x}},\mathbf{x}')$ only through its limiting joint type $P_{X_0X_1}$.} 
\begin{align}
&  \mathcal{R}'(P_{X_0X_1}) :=\{(\kappa_{\alpha}, \kappa'(\kappa_{\alpha}, P_{X_0X_1})): \kappa_{\alpha} \in (0,\kappa'^{\star}_{\alpha})\}, \notag
    \end{align}
where $ \kappa'(\kappa_{\alpha},P_{X_0X_1}):= \sup \{\kappa_{\beta}:(\kappa_{\alpha},\kappa_{\beta}) \mbox{ is achievable for HT in \eqref{HTdefchn}}  \} $
and $\kappa'^{\star}_{\alpha}:=\inf \{\kappa_{\alpha}: \kappa'(\kappa_{\alpha},P_{X_0X_1})=0\}$.  
The  following proposition provides a single-letter characterization of $ \mathcal{R}'(P_{X_0X_1})$.
\begin{prop}[Error-exponent region for HT in \eqref{HTdefchn} ] \label{genchernbnd}
\begin{align}
& \mathcal{R}'(P_{X_0X_1})= \underset{\theta \in \mathcal{I}(P_{X_0X_1})}{\cup}\left(\mathbb{E}_{P_{X_0X_1}}\big[\psi^*_{P_{Y|X}(\cdot|X_0),\Pi_{X_0,X_1}}(\theta) \big],~ \mathbb{E}_{P_{X_0X_1}}\big[\psi^*_{P_{Y|X}(\cdot|X_0),\Pi_{X_0,X_1}}(\theta) \big] -\theta \right) \notag
\end{align}
where $\Pi_{\tilde x,x'}(y):=\log \big(P_{Y|X}(y|x')/P_{Y|X}(y|\tilde x)\big)$ for  $(\tilde x,x') \in \X^2$, $\mathcal{I}(P_{X_0X_1}):= \big(-d_{\min}(P_{X_0X_1}),d_{\max}(P_{X_0X_1}) \big)$, $d_{\min}(P_{X_0X_1}) := \mathbb{E}_{P_{X_0X_1}}\big[D\big(P_{Y|X}(\cdot|X_0)||P_{Y|X}(\cdot|X_1)\big)\big]$,  $d_{\max}(P_{X_0X_1}):=\mathbb{E}_{P_{X_0X_1}}\big[D\big(P_{Y|X}(\cdot|X_1)||P_{Y|X}(\cdot|X_0)\big)\big]$
\end{prop}
\begin{proof}
    Let $(\tilde{\mathbf{x}},\mathbf{x}')\in \mathcal{X}^n \times \mathcal{X}^n$ be sequences that satisfy \eqref{seqpairconverge}. For simplicity, we will denote $d_{\max}(P_{X_0X_1})$ and $d_{\min}(P_{X_0X_1})$  by  $d_{\max}$ and $d_{\min}$, respectively. \\
\textbf{Achievability:} We will  show that for $- d_{\min} < \theta < d_{\max}$, 
\begin{align}
\kappa'\left(\mathbb{E}_{P_{X_0X_1}}\left[\psi^*_{{P_{Y|X}(\cdot|X_0)},\Pi_{X_0,X_1}}(\theta)\right],P_{X_0X_1}\right) \geq \mathbb{E}_{P_{X_0X_1}}\left[\psi^*_{{P_{Y|X}(\cdot|X_0)},\Pi_{X_0,X_1}}(\theta)\right]-\theta.\notag 
\end{align}
Consider the Neyman-Pearson test  given by
$g_n(\mathbf{y})=  \ind_{ \big\{ \Pi^{(n)}_{ \tilde{\mathbf{x}}, \mathbf{x}'}(\mathbf{y})\geq n\theta\big\}}$, 
where $ \Pi^{(n)}_{ \tilde{\mathbf{x}}, \mathbf{x}'}(\mathbf{y}):=\sum_{i=1}^n \Pi_{\tilde x_i,x'_i,P_{Y|X}}(y_i)$. Observe that  the type I error probability can be upper bounded for $\theta > -d_{\min}$ and sufficiently large $n$ as follows:
\begin{flalign}
    \alpha_n\left(g_n \right)&= \mathbb{P}_{P_{Y|X}^{\otimes n}(\cdot| \tilde{\mathbf{x}})} \left( \Pi^{(n)}_{ \tilde{\mathbf{x}}, \mathbf{x}'}(\mathbf{Y})\geq n \theta \right) \notag \\
    & \stackrel{(a)}{\leq} e^{-\underset{\lambda  \geq 0}{\sup} ~n \theta \lambda- \psi_{P_{Y|X}^{\otimes n}(\cdot| \tilde{\mathbf{x}}), \Pi^{(n)}_{ \tilde{\mathbf{x}}, \mathbf{x}'}}(\lambda)} \notag \\
    &\stackrel{(b)}{=} e^{-\underset{\lambda \in \mathbb{R}}{\sup}~ n \left(\theta \lambda- \frac 1n \psi_{P_{Y|X}^{\otimes n}(\cdot| \tilde{\mathbf{x}}), \Pi^{(n)}_{ \tilde{\mathbf{x}}, \mathbf{x}'}}(\lambda)\right)}, \label{expchernf} &&
\end{flalign}
where $(a)$ follows from the Chernoff bound,  and $(b)$ holds because for $\theta > - d_{\min}$ and sufficiently large $n$, the supremum in \eqref{expchernf} always occurs at $\lambda \geq 0$.
    To see this, note that 
    the term $l_n(\lambda):=\theta \lambda- n^{-1} \psi_{P_{Y|X}^{\otimes n}(\cdot| \tilde{\mathbf{x}}), \Pi^{(n)}_{ \tilde{\mathbf{x}}, \mathbf{x}'}}(\lambda)$ is a concave function of $\lambda$ by Lemma \ref{logmgffact} (i). Also, denoting its derivative with respect to $\lambda$ by $l_n'(\lambda)$, we have
        \begin{flalign}
     l_n'(0)&=\theta- \frac 1n ~  \mathbb{E}_{{P_{Y|X}^{\otimes n}(\cdot| \tilde{\mathbf{x}})}} \left[\Pi^{(n)}_{ \tilde{\mathbf{x}}, \mathbf{x}'} \right]\label{applylemlogmgf} \\
     &=\theta- \frac 1n \sum_{i=1}^n \mathbb{E}_{{P_{Y|X}(\cdot| \tilde{x}_i)}} \left[\log\left(P_{Y|X}(Y_i|x_i')/P_{Y|X}(Y_i|\tilde x_i)\right) \right] \notag \\
     &\xrightarrow{(n)} \theta+ d_{\min}>0, \label{convergxhntyedist} &&
     \end{flalign}  
     where \eqref{applylemlogmgf} follows from Lemma \ref{logmgffact} (iii), and  \eqref{convergxhntyedist} is due to the absolute continuity assumption, $P_{Y|X}(\cdot|x)\ll P_{Y|X}(\cdot|x')$, $\forall ~(x,x') \in \X^2$ on the channel, and \eqref{seqpairconverge}. Thus, by the concavity of $l_n(\lambda)$, its supremum has to occur at $\lambda \geq 0$.
Simplifying the term within the exponent in \eqref{expchernf}, we obtain
\begin{flalign}
     \frac 1n \psi_{P_{Y|X}^{\otimes n}(\cdot|\mathbf{x}), \Pi^{(n)}_{ \tilde{\mathbf{x}}, \mathbf{x}'}}(\lambda)&=    
    \sum_{\tilde x,x'} P_{\tilde{\mathbf{x}}\mathbf{x}'}(\tilde x,x') \log \left( \mathbb{E}_{{P_{Y|X}(\cdot|\tilde x)}} \left[ \big(P_{Y|X}(Y|x')/P_{Y|X}(Y|\tilde x)\big)^{\lambda}  \right] \right) \label{bndedllratio} \\
    &  \xrightarrow{(n)} \mathbb{E}_{P_{X_0X_1}} \left[ \log \left( \mathbb{E}_{{P_{Y|X}(\cdot|X_0)}} \left[e^{\lambda \Pi_{X_0,X_1}(Y)}\right]\right)\right], \label{expchernsimp} && 
\end{flalign}
where \eqref{expchernsimp} follows from \eqref{seqpairconverge} and the absolute continuity assumption on $P_{Y|X}$. 
Substituting \eqref{expchernsimp} in \eqref{expchernf} and from \eqref{Chernoffexp}, we obtain for arbitrarily small (but fixed) $\delta>0$ and sufficiently large $n$, that
\begin{flalign}
     \alpha_n\left(g_n \right) &\leq e^{-\underset{\lambda \in \mathbb{R}}{\sup}\left( n \left(\theta \lambda-  \mathbb{E}_{P_{X_0X_1}} \left[ \log \left( \mathbb{E}_{{P_{Y|X}(\cdot|X_0)}} \left[e^{\lambda \Pi_{X_0,X_1}(Y)}\right]\right)\right]-\delta\right)\right)} \notag \\
    &= e^{-n\left(\mathbb{E}_{P_{X_0X_1}}\left[\underset{\lambda \in \mathbb{R}}{\sup}\left(\theta \lambda- \mathbb{E}_{{P_{Y|X}(\cdot|X_0)}} \left(e^{\lambda \Pi_{X_0,X_1}(Y)}\right)\right)\right]-\delta\right)} \notag \\
   &= e^{-n\left(\mathbb{E}_{P_{X_0X_1}}\left[\psi^*_{{P_{Y|X}(\cdot|X_0)},\Pi_{X_0,X_1}}(\theta)\right]-\delta\right)}. \label{type1errchnbnd} &&
\end{flalign}
Similarly, it can be shown  that for $\theta < d_{\max}$,
\begin{align}
  \beta_n\left(g_n \right) &\leq  e^{-n\left(\mathbb{E}_{P_{X_0X_1}}\left[\psi^*_{P_{Y|X}(\cdot|X_1),\Pi_{X_0,X_1}}(\theta)\right]-\delta\right)}. \label{type2errchnbnd}
\end{align}
Moreover, for $(\tilde x,x') \in \X^2$, we have
\begin{align}
e^{\psi_{P_{Y|X}(\cdot|x'),\Pi_{\tilde x,x'}}(\lambda)} &= \sum_{y \in \Y} P^{\lambda+1}_{Y|X}(\cdot|x')/P^{\lambda}_{Y|X}(\cdot|\tilde x) =e^{\psi_{P_{Y|X}(\cdot|\tilde x),\Pi_{\tilde x,x'}}(\lambda+1)}.\notag 
\end{align}
It follows that 
\begin{flalign}
   \psi^*_{P_{Y|X}(\cdot|x'),\Pi_{\tilde x,x'}}(\theta)&:= \underset{\lambda \in \mathbb{R}}{\sup} ~\lambda \theta-\psi_{P_{Y|X}(\cdot|x'),\Pi_{\tilde x,x'}}(\lambda) \notag \\
   &= \underset{\lambda \in \mathbb{R}}{\sup}~  \lambda \theta- \psi_{P_{Y|X}(\cdot|\tilde x),\Pi_{\tilde x,x'}}(\lambda+1)  \notag \\
   &=   \psi^*_{P_{Y|X}(\cdot|\tilde x),\Pi_{\tilde x,x'}}(\theta)-\theta.\notag &&
\end{flalign}
Hence,
\begin{align}
    \mathbb{E}_{P_{X_0X_1}}\left[ \psi^*_{P_{Y|X}(\cdot|X_1),\Pi_{X_0,X_1}}(\theta)\right]= \mathbb{E}_{P_{X_0X_1}}\left[\psi^*_{{P_{Y|X}(\cdot|X_0)},\Pi_{X_0,X_1}}(\theta)\right]-\theta. \notag
\end{align}
From this, \eqref{type1errchnbnd} and \eqref{type2errchnbnd}, we obtain for $-d_{\min} < \theta < d_{\max}$ that 
\begin{align}
\kappa'\left(\mathbb{E}_{P_{X_0X_1}}\left[\psi^*_{{P_{Y|X}(\cdot|X_0)},\Pi_{X_0,X_1}}(\theta)\right]-\delta,P_{X_0X_1} \right) \geq \mathbb{E}_{P_{X_0X_1}}\left[\psi^*_{{P_{Y|X}(\cdot|X_0)},\Pi_{X_0,X_1}}(\theta)\right]-\theta-\delta.\notag 
\end{align}
Then, the proof of achievability is completed by noting that $\delta>0$ is arbitrary and $\kappa'(\kappa_{\alpha},P_{X_0X_1})$ is a continuous function of $\kappa_{\alpha}$ for a fixed $P_{X_0X_1}$.\\
\textbf{Converse:} 
Let $\mathcal{I}_n(\tilde x,x'):=\{i \in [n]: \tilde x_{i}=\tilde x \mbox{ and }x_{i}'=x'\}$. For any $\theta \in \mathbb{R}$ and decision function $g_n$, we have from \cite[Theorem 14.9]{Polyanskiy-Wu-book}  that
\begin{flalign}
  \alpha_n(g_n)+ e^{-n \theta}\beta_n(g_n) &\geq \mathbb{P}_{P_{Y|X}^{\otimes n}(\cdot|\tilde{\mathbf{x}})} \left(\log \left(P_{Y|X}^{\otimes n}(\mathbf{Y}|\mathbf{x}')/P_{Y|X}^{\otimes n}(\mathbf{Y}|\tilde{\mathbf{x}}) \right) \geq n \theta\right). \notag
  \end{flalign}
 Simplifying the RHS above, we obtain
 \begin{flalign} 
  &\mathbb{P}_{P_{Y|X}^{\otimes n}(\cdot|\tilde{\mathbf{x}})} \left(\log \left( P_{Y|X}^{\otimes n}(\mathbf{Y}|\mathbf{x}')/P_{Y|X}^{\otimes n}(\mathbf{Y}|\tilde{\mathbf{x}}) \right) \geq n \theta\right) \notag \\
&=\mathbb{P}_{P_{Y|X}^{\otimes n}(\cdot|\tilde{\mathbf{x}})} \Bigg(\sum_{\tilde x,x'} \sum_{ i \in \mathcal{I}_n(\tilde x,x')}  \log \left( P_{Y|X}(Y_i|x_i')/P_{Y|X}(Y_i|\tilde  x_i) \right) \geq n \theta\Bigg)  \notag \\
 &= \mathbb{P}_{P_{Y|X}^{\otimes n}(\cdot|\tilde{\mathbf{x}})} \left(\sum_{\tilde x,x'} \sum_{ i \in \mathcal{I}_n(\tilde x,x')}  \log \left( P_{Y|X}(Y_i|x_i')/P_{Y|X}(Y_i|\tilde  x_i) \right) \geq  \sum_{(\tilde x,x') \in \X^2} n P_{\tilde{\mathbf{x}}\mathbf{x'}}(\tilde x,x') \theta\right)  \notag \\
     & \stackrel{(a)}{\geq }  \mathbb{P}_{P_{Y|X}^{\otimes n}(\cdot|\tilde{\mathbf{x}})} \left( \bigcap_{\tilde x,x'}  \left( \sum_{ i \in \mathcal{I}_n(\tilde x,x')}  \log \left( P_{Y|X}(Y_i|x_i')/P_{Y|X}(Y_i|\tilde  x_i) \right) \geq   n P_{\tilde{\mathbf{x}}\mathbf{x'}}(\tilde x,x') \theta\right)\right)  \notag \\
       &  \stackrel{(b)}{= }  \prod_{(\tilde x,x') \in \X^2}  \mathbb{P}_{P_{Y|X}^{\otimes n}(\cdot|\tilde{\mathbf{x}})} \left( \sum_{ i \in \mathcal{I}_n(\tilde x,x')}  \log \left( P_{Y|X}(Y_i|x_i')/P_{Y|X}(Y_i|\tilde  x_i)\right) \geq   n P_{\tilde{\mathbf{x}}\mathbf{x'}}(\tilde x,x') \theta\right), \notag &&
\end{flalign} 
where 
\begin{enumerate}[label=\emph{(\alph*)}]
    \item follows since 
    \begin{align}
       & \bigcap_{\tilde x,x'} \left\{  \sum_{ i \in \mathcal{I}_n(\tilde x,x')}  \log \left(P_{Y|X}(Y_i|x_i')/P_{Y|X}(Y_i|\tilde x_i)\right) \geq   n P_{\tilde{\mathbf{x}}\mathbf{x'}}(\tilde x,x') \theta\right\} \notag \\ 
       & \qquad \qquad \qquad  \subseteq \left\{\sum_{\tilde x,x'} \sum_{ i \in \mathcal{I}_n(\tilde x,x')}  \log \left( P_{Y|X}(Y_i|x_i')/P_{Y|X}(Y_i|\tilde x_i) \right) \geq  \sum_{(\tilde x,x') \in \X^2} n P_{\tilde{\mathbf{x}}\mathbf{x'}}(\tilde x,x') \theta\right\}; \notag 
    \end{align}
    \item is due to the independence of the events $\big\{\sum_{ i \in \mathcal{I}_n(\tilde x,x')}  \log \big(P_{Y|X}(Y_i|x_i')/P_{Y|X}(Y_i|\tilde x_i) \big) \geq   n P_{\tilde{\mathbf{x}}\mathbf{x'}}(\tilde x,x') \theta\big\}$ for different $(\tilde x,x') \in \X^2$.
\end{enumerate}
Define $b_{\tilde x,x'}(\theta):=\min_{\substack{\tilde Q_{\tilde x} \in \mathcal{P}(\Y): \mathbb{E}_{\tilde Q_{\tilde x}}\left[\log \left(P_{Y|X}(Y|x')/P_{Y|X}(Y|\tilde x)\right)\right] \geq \theta}}~D\big(\tilde Q_{x}||P_{Y|X}(\cdot|\tilde x) \big)$. Then, for arbitrary $\delta >0$, $\delta'>\delta$ and sufficiently large $n$, we can write 
\begin{flalign}
    \alpha_n+ e^{-n \theta}\beta_n     & \stackrel{(a)}{\geq}  \prod_{(\tilde x,x') \in \X^2}  e^{-nP_{\tilde{\mathbf{x}}\mathbf{x'}}(\tilde x,x')~\left(b_{\tilde x,x'}(\theta)+ \delta \right)}\notag \\
  & \stackrel{(b)}{\geq} \prod_{(\tilde x,x') \in \X^2} e^{-nP_{\tilde{\mathbf{x}}\mathbf{x'}}(\tilde x,x')\left(\psi^*_{P_{Y|X}(\cdot|\tilde x), \Pi_{\tilde x,x'}}( \theta)+ \delta\right)} \notag \\
  & \stackrel{(c)}{=}  e^{-n \left( \mathbb{E}_{P_{X_0X_1}}\left[\psi^*_{{P_{Y|X}(\cdot|X_0)}, \Pi_{X_0,X_1}}( \theta)\right]+ \delta'\right)}, \notag &&
  \end{flalign}
where $(a)$ follows  from \cite[Theorem 15.9]{Polyanskiy-Wu-book};
 $(b)$ follows since $b_{\tilde x,x'}(\theta)=\psi^*_{P_{Y|X}(\cdot|\tilde  x), \Pi_{\tilde x,x'}}( \theta)$ from \cite[Theorem 15.6]{Polyanskiy-Wu-book} and \cite[Theorem 15.11]{Polyanskiy-Wu-book}; and 
   $(c)$  is due to \eqref{seqpairconverge}. The equation above implies that
\begin{align}
   \limsup_{n \rightarrow \infty} \min \left\lbrace-\frac{1}{n}\log\alpha_n, -\frac{1}{n}\log\beta_n+\theta \right\rbrace \leq \mathbb{E}_{P_{X_0X_1}}\left[\psi^*_{{P_{Y|X}(\cdot|X_0)}, \Pi_{X_0,X_1}}( \theta)\right]+\delta'. \notag
\end{align} 
Hence, if  $ -\log (\alpha_n)/n 
  >\mathbb{E}_{P_{X_0X_1}}\big[\psi^*_{{P_{Y|X}(\cdot|X_0)}, \Pi_{X_0,X_1}}( \theta) \big]+\delta'$ for all sufficiently large $n$, 
then
\begin{align}
   \limsup_{n \rightarrow \infty} -\frac{1}{n} \log\beta_n  \leq \mathbb{E}_{P_{X_0X_1}}\left[\psi^*_{{P_{Y|X}(\cdot|X_0)}, \Pi_{X_0,X_1}}( \theta) \right]-\theta+\delta'.\notag
\end{align}
Since $\delta$ (and $\delta'$) is arbitrary,  this implies via the continuity of $ \kappa'\left(\kappa_{\alpha},P_{X_0X_1} \right)$ in $\kappa_{\alpha}$ that 
\begin{align}
 \kappa'\left(\mathbb{E}_{P_{X_0X_1}}\left[\psi^*_{{P_{Y|X}(\cdot|X_0)},\Pi_{X_0,X_1}}(\theta)\right],P_{X_0X_1} \right) \leq \mathbb{E}_{P_{X_0X_1}}\left[\psi^*_{{P_{Y|X}(\cdot|X_0)},\Pi_{X_0,X_1}}(\theta)\right]-\theta.\notag 
\end{align}
To complete the proof, we need to show that $\theta$ can be restricted to lie in $ \mathcal{I}(P_{X_0X_1})$. Towards this, it suffices to show the following:
\begin{enumerate}[label=(\roman*)]
    \item 
   $ \mathbb{E}_{P_{X_0X_1}}\big[\psi^*_{{P_{Y|X}(\cdot|X_0)},\Pi_{X_0,X_1}}\left(-d_{\min}\right) \big]=0$,
\item 
   $ \mathbb{E}_{P_{X_0X_1}}\big[\psi^*_{{P_{Y|X}(\cdot|X_0)},\Pi_{X_0,X_1}}\left(d_{\max}\right) \big]= d_{\max}$, and
\item $\mathbb{E}_{P_{X_0X_1}}\big[\psi^*_{{P_{Y|X}(\cdot|X_0)},\Pi_{X_0,X_1}}(\theta) \big]$ and $\mathbb{E}_{P_{X_0X_1}}\big[\psi^*_{{P_{Y|X}(\cdot|X_0)},\Pi_{X_0,X_1}}(\theta) \big]-\theta$ are convex functions of $\theta$.
\end{enumerate}
We have
\begin{flalign}
&\mathbb{E}_{P_{X_0X_1}}\left[\psi^*_{{P_{Y|X}(\cdot|X_0)},\Pi_{X_0,X_1}}\left(-d_{\min}\right) \right]\notag \\
    &:=\sup_{\lambda \in \mathbb{R}} -\lambda ~\mathbb{E}_{P_{X_0X_1}}\big[\kl{P_{Y|X}(\cdot|X_0)}{P_{Y|X}(\cdot|X_1)}\big]-  \mathbb{E}_{P_{X_0X_1}}\left[\psi_{{P_{Y|X}(\cdot|X_0)},\Pi_{X_0,X_1}}(\lambda) \right]   \notag\\
    & \leq  \sum_{\tilde x,x'}P_{X_0X_1}(\tilde x,x')~\left[ \sup_{\lambda_{\tilde x,x'} \in \mathbb{R} } -\lambda_{\tilde x,x'} D\left(P_{Y|X}(\cdot|\tilde x)||P_{Y|X}(\cdot|x')\right)-  \psi_{{P_{Y|X}(\cdot|\tilde  x)},\Pi_{\tilde x,x'}}(\lambda_{\tilde x,x'}) \right] \notag \\
    & =0, \label{equshowdiv} &&
\end{flalign}
where \eqref{equshowdiv} follows since each term inside the square braces in the penultimate equation is zero, which in turn follows from Lemma \ref{logmgffact} (iii).
Also,
\begin{flalign}
\mathbb{E}_{P_{X_0X_1}}\left[\psi^*_{{P_{Y|X}(\cdot|X_0)},\Pi_{X_0,X_1}}\left(-d_{\min}\right) \right] &=\sum_{\tilde x,x'}P_{X_0X_1}(\tilde x,x')~\psi^*_{{P_{Y|X}(\cdot|\tilde x)},\Pi_{\tilde x,x'}}\left(-d_{\min}\right)\geq 0,\label{equshowdiv2} 
\end{flalign}
where \eqref{equshowdiv2} follows from the non-negativity of $\psi^*_{{P_{Y|X}(\cdot|\tilde  x)},\Pi_{\tilde x,x'}}$ for every $(\tilde x,x')\in \X^2$ stated in Lemma \ref{logmgffact} (iii).
Combining \eqref{equshowdiv} and \eqref{equshowdiv2} proves $(i)$.
We also have
\begin{flalign}   \mathbb{E}_{P_{X_0X_1}}\left[\psi^*_{{P_{Y|X}(\cdot|X_0)},\Pi_{X_0,X_1}}\left(d_{\max}\right) \right]- d_{\max} &=\mathbb{E}_{P_{X_0X_1}}\left[\psi^*_{P_{Y|X}(\cdot|X_1),\Pi_{X_0,X_1}}(d_{\max})\right] =0, \label{previousineq} 
\end{flalign}
where  the final equality follows similarly to the proof of $(i)$. This proves $(ii)$.
Finally, (iii) follows from Lemma \ref{logmgffact} (iii) and the fact that a weighted sum of convex functions is convex provided the weights are non-negative, thus completing the  proof.
\end{proof}

\end{appendices}
\bibliographystyle{IEEEtran}
\bibliography{ref}

\end{document}